\documentclass[times,twocolumn,final,authoryear]{elsarticle}

\usepackage{jasr}
\usepackage{framed,multirow}

\usepackage{amssymb}
\usepackage{latexsym}
\usepackage[toc]{appendix} 

\usepackage[switch]{lineno}

\newcommand\sw{SW}
\newcommand\m{MP}

\usepackage{url}
\usepackage{xcolor}
\definecolor{newcolor}{rgb}{.8,.349,.1}

\usepackage[citebordercolor=white]{hyperref}

\journal{Advances in Space Research}

\begin{document}

\verso{Ahmed \textit{et al.}}
  
\begin{frontmatter}

\title{Characteristics and development of the main phase disturbance in geomagnetic storms ($\mathrm{Dst}\leq-50 ~\mathrm{nT}$)}

\author[1,2]{O. Ahmed}
\cortext[cor1]{email:osmanr18@ymail.com;oahmedindris@stu.kau.edu.sa (O. Ahmed)}
\cortext[cor1]{email:derouichmoncef@gmail.com (M. Derouich)}
\author[1]{B. Badruddin}
\cortext[cor1]{Corresponding author: }
  
  \ead{badr.physamu@gmail.com; bzahmad@kau.edu.sa}

\author[1]{M. Derouich}

\address[1]{Astronomy and Space Science Department, Faculty of Science, King Abdulaziz University, P.O. Box 80203, Jeddah 21589, Saudi Arabia}
\address[2]{Department of Physics, Faculty of Natural and computational Sciences, Debre Tabor University, P.O. Box 272, Debre Tabor, Ethiopia}


\begin{abstract}
In this study, we present geomagnetic storms (GSs) 
selected from three solar cycles, spanning the years 1995 to 2022. We studied the development of the main phase of storms with in disturbance storm time (Dst) amplitudes ranging from $\mathrm{Dst}=-64$ nT to $\mathrm{Dst}=-422$ nT. 
In order to determine the solar wind (SW) parameters that mainly influence the main phase development of a GS, which can best describe the SW-magnetosphere coupling, we divided our selected GSs into four groups based on main phase duration. Superposed epoch analysis was performed on the selected geomagnetic indices, SW plasma and field parameters, and their derivatives separately for each group.
To that end, the dynamics of GS main phase development is mainly guided by interplanetary driver’s magnetic field southward component, (-Bz).
It has been determined that there is a temporal difference between the peak values of Bz and Dst. As a result, Dst is delayed from Bz by 1 -- 4 hours, which is crucial for space weather forecasting. Our findings support the theory that the arrival of high-speed magnetic field structures in front of the Earth’s magnetosphere is responsible for the development of GS, which begins a few hours later.
The peak of Dst has a direct relationship with the amplitude of storm sudden commencement (SSC) and an inverse relationship with the duration of SSC. The inter-relationship between the peaks of the three indices (Dst, AE, and ap) during GS, is also obtained. Dst is found to be more closely related to ap than AE.
To determine the best fit SW parameter to the geomagnetic activity indices, we used a linear correlation between the peak values of individual geomagnetic indices and SW plasma and field parameters and their derivatives.
An electric field related function involving speed and IMF (v$^{\frac{4}{3}}$Bz) when coupled with a viscous term ($\rho^{\frac{1}{2}}$) (v$^{\frac{4}{3}}$Bz$\rho^{\frac{1}{2}}$) correlates very well with the intensity of the GS (Dst$ _\mathrm{min} $ or $ \Delta$Dst) and the magnitude of (ap$_\mathrm{max}$) and (AE$_\mathrm{max}$) during storms. However, a related function (v$^{\frac{4}{3}}$B$\rho^{\frac{1}{2}}$) represents slightly better the peak of AE$_\mathrm{max}$ during the storms.
We strongly encourage higher-resolution observations to confirm and reach a thorough conclusion. 
\end{abstract}


\end{frontmatter}


\section{Introduction}\label{sec:intro}
The activity of the sun is the main reason to the explosion of flares and the ejection of charged materials called coronal mass ejections (CMEs) and the later has direct effect for the Earth's disturbance.
Earthward CMEs are the primary causes of magnetic field disturbance on Earth \citep[e.g.,][]{2014JGRA..119.5117W, 2016JGRA..121.7423W}. 
A popular concept is that the main phase (MP) of a GS begins when a large southward interplanetary magnetic field (IMF) Bz carried by a SW structure strikes the magnetopause \citep[e.g.,][]{1994JGR....99.5771G,  Vichare2005SomeCO, RAWAT20101364} and it is a crucial component in the description of GSs \citep{1961PhRvL...6...47D, 1966JGR....71..155F, 1997JGR...10214215Y, 2015RAA....15...85R, 2018cosp...42E2795S} despite the fact that the development of \m\ of GSs are not similarly related to the changes of \sw\ parameters during intense GSs \citep[e.g.,][]{2022Ap&SS.367...10B}. In addition to its role to develop the GS the intensity of GSs is most strongly influenced by the IMF's southward component Bz \citep{1989JGR....94.8835G, Srivastava2004SolarAI, 2008JASTP..70..245G}.  The evolution of AE activity during GS development (Dst value) was compared to IMF Bz, SW velocity, and dynamic pressure. During the GS MP, AE rises fast as the southward IMF Bz, westward vBz, and SW pressure rise \citep{2000AdSpR..26..111O}. \cite{2018NPGeo..25...67G} used ACE SW data to study the relationship between AE and Bz.
\\
Earlier studies indicate that, in addition to southward IMF Bz, the dawn-dusk electric field (Ey) is taken to be the primary initiator for the development of GSs.
The strength of a GS is well determined by the SW electric field Ey \citep[e.g., see][ and references therein]{1975JGR....80.4204B, 2000JGR...105.7707O, wang2003empirical, 2023SpWea..2103314M}. \\
ICMEs and CIRs are two large-scale structures with enhanced magnetic field in the interplanetary (IP) space mainly responsible for GSs. ICME generated GS are transient and stronger, while CIR generated storms are comparatively weaker and last longer. CMEs are associated with active regions on the Sun, ejected from the solar atmosphere and propagating in IP space evolve in the space \citep[e.g.,][]{2006SSRv..124..145G}. An ICME is the CME event in the IP space. This structure reaches and interacts with the Earth's magnetosphere, causing a GS if circumstances are met.  Magnetic cloud (MC) refers to a subset of ICMEs with specific properties (high magnetic field with turning IMF, northward to southward or southward to northward, and low plasma beta within) \citep[e.g.,][]{1981JGR....86.6673B}.  
An ICME/MC moving with high speed in IP space relative to the ambient solar wind can produce a shock wave and a sheath region ahead of it. This thick sheath region ahead of CME ejecta (ICME) occurs as a result of the compression of the ambient SW plasma and magnetic field. The magnetic field in the sheath region is generally intense and turbulent, whereas the magnetic field in the ICME/MC is likewise powerful but smooth (for example, flux rope). 
Thus, by the time a CME structure evolves and propagates in IP space and interacts with the magnetosphere, it could have the following probable structures: (a) CME ejecta without magnetic cloud structure (referred to as ICME), (b) CME ejecta with magnetic cloud structural properties (referred to as MC), (c) ICME with sheath region ahead of it (referred to as SH+ICME structure), and (d) MC with sheath region ahead of it (referred to as SH+MC structure). CIRs, on the other hand, are formed when high-speed SW (emitted by coronal holes) interacts with slower SW in the IP space. 
The structure so generated is known as a corotating interaction region (CIR). A well-developed CIR may include a forward shock, stream interface, and reverse shock \citep[e.g.,][]{1999SSRv...89...21G}.
Depending on the appropriate plasma/field conditions in these structures, a GS may be formed during the passage of an ICME, MC, SH, SH+ICME, SH+MC region/structure, or CIR (Table \ref{tabl1}).\\
Simple superposed epoch analysis (SEA) \citep[e.g.,][]{1998P&SS...46.1015B, singh2009geoeffectiveness, Mustajab2013RelativeGO, 2016Ap&SS.361..253B} and double superposed epoch analysis (DSEA) \citep[e.g.,][]{2010AnGeo..28.2177Y} have been utilized effectively to study the average behavior of GS and their development due to (shock/sheath, ICME/MC/Ejecta/CIR, etc.) in relation to SW properties. 
Recently, \citep{2018SoPh..293..126P} used SEA technique to analyze magnetic storms, comparing the reaction of SW parameters, v, Bz, and the product vBz. They concluded that the parameter vBz is more important than v and Bz independently.
More recently, \citep{2020Ge&Ae..59..639D, 2021Univ....7..138Y, 2022JGRA..12730423P, 2023SpWea..2103314M} have used the DSEA approach to analyze the dynamics of geomagnetic disturbances.  \\
\cite{2021Univ....7..138Y} underlined the significance of correctly identifying IP drivers (SH/CME/MC/CIR, etc.) and supported their claim by studying SW characteristics and geomagnetic indices (Dst, Kp, and AE) during GS generated by various drivers.
The DSEA approach was used to investigate the dynamics of geomagnetic indices (Dst, ap, and AE) with SW parameters during large GSs (Dst $\leq-50$ nT) caused by various sources (CIR/SH/MC/ICME/ejecta) \citep{2020Ge&Ae..59..639D}. They found that sheath (SH) storms have the highest average value in the indices studied, while ejecta storms have the lowest. They also discovered a link between ring current and Dst variance. \cite{2022JGRA..12730423P} used the SEA technique to study the evolution of field-aligned currents (FAC) and ionospheric equivalent currents throughout the MP of GSs (Dst $\leq-50$ nT). 
They concluded that the response in currents found in their investigation agrees with the findings \citep{2002JGRA..107.1440H, 2007GeoRL..34.2105P} that magnetospheric and auroral activity respond faster at storm onset in SH storms than MC storms. \cite{2023GeoRL..5003151P} examined the time evolution of the AE index with concurrent developments in the field aligned current, both during SH and MC storms. 
They noted a very strong correlation between AE and FAC.\\
SEA has been applied in the past to study the average response of IP structures (shock/SH/MC/CME(ejecta)/CIR etc.) on geomagnetic activity and their relative geoeffectiveness. Such studies have also invoked, in order to study the dynamics of geomagnetic disturbance, the variation in SW plasma and field parameters and their various derivatives/functions. Most of these SEA based studies were with reference (zero epoch) to arrival time of IP structures or with reference to GS onset, in general. We have adopted a unique approach by grouping the GS on the basis of MP durations of different bins and/or GS based on MP steps (one, two, three, multiple step GS). This approach makes it possible to study the average behavior of GS of different MP durations and/or GS with different rate of their development (faster/slower) and also developing in different steps. The division in such groups makes the estimation of intensity of GS (Dst$_\mathrm{min}$) and corresponding evaluation of SW plasma parameters during MP more accurate, as compared to, if GS of all MP durations (wider durations) are superposed together. This division overcomes to some extent, the disadvantage of simple SEA, that is, at points away from the epoch (e.g., at Dst$_\mathrm{min}$) average gets less accurate\citep[][]{2023SpWea..2103314M}. Another uniqueness of our approach in SEA averages is that we have compared the MP dynamics of GS of different duration/step-decreases with variation of one, two and three parameter CFs simultaneously. In addition to study the dynamics of GS during the MP of GS, we have utilized the superposed results to discuss about the time lag between post-disturbed period and peak in southward IMF. Another aim of this work is to obtain the best possible SW-magnetosphere CFs which represents not only the magnitude of GS as measured by Dst$_\mathrm{min}$ but also magnitude of geomagnetic disturbance in mid-latitude (ap$_\mathrm{max}$) and polar (AE$_\mathrm{max}$) regions representing substorm intensity. This aim has been approached with the use of linear correlation analysis. Correlation analysis has also been adopted with the aim to obtain the inter-relationship between the magnitude attained in three geomagnetic indices (Dst, ap, AE) representing geomagnetic disturbances in three different domains (equatorial, mid-latitude and polar) during the MP of GSs.\\
There have been numerous potential SW-magnetosphere CFs proposed, notably since the space age, when SW observations became available. Previous studies has made great progress in discovering acceptable CFs involving one/two/multiple parameters \citep[e.g., see][]{1981SSRv...28..121A, 1990P&SS...38..627G, 2007JGRA..112.1206N, 2014JGRA..119..751B}. 
Four SW parameters and their various combinations with distinct powers over these parameters, which appear in most of the CFs are SW velocity (v) and its density ($\rho$), IMF (B) and its north-south component (Bz).\\
Early research on establishing a relation of geomagnetic indices with single SW parameters or their suitable combinations started mainly after space age. Using daily Kp average data, a direct relation between Kp with SW velocity (v) was suggested \citep[][]{snyder1963solar}. Later such attempts explored the role of IMF strength B and its southward component (-Bz or Bs) \citep[e.g.,][]{1971JGR....76.5189A, 1973P&SS...21.2139B}.
Some simple combinations/variants involving  southward IMF (Bs), total IMF (B), SW velocity (v),  density ($\rho$) and/or SW pressure (P) duskward electric field (vBs) and total electric field (vB), were also explored time to time in different solar conditions (SH/ICME/CIR etc.), associating them with different geomagnetic parameters (Dst/Kp/ap/AE etc.). Among them, vBs is most studied CF \citep[e.g.,][and references therein]{1975JGR....80.4204B, 1989JGR....94.8835G, 2007GeoRL..34.2105P, 2011SpWea...9.7005R, 2016Ap&SS.361..253B, 2018SoPh..293..126P, 2022Ap&SS.367...10B, 2022JGRA..12730423P, 2023SpWea..2103314M}. Some simple variants of v, Bz or B explored in earlier studies are; Bzv$^2$ \citep{1975P&SS...23...75M, 1982JGR....87.2558H}, Bv$^2$ \citep{Crooker1977OnTH}, Bz$^2$v \citep{1982JGR....87.2558H}, B$^2$v 
\citep{1981GeoRL...8..179B, 1982JGR....87.2558H}. \\
\cite{1982JGR....87.2558H} studied correlation of three functions vBs, v$^2$Bs snd vBs$^2$ and tested their ability to predict geomagnetic activity as measured by AL index. Their study yielded best correlation coefficient (0.97) with v$^2$Bs, 0.92 with vBs and 0.82 with vBs$^2$. \cite{2013P&SS...85..123B} studied the SW-magnetosphere coupling on different time scales (yearly, half-yearly, 27 day, daily, 3-hour and hourly) using four geomagnetic indices (Dst, ap, aa and AE) over a long period of time. They analyzed the relationship of these indices, at different time scales with candidate SW CFs (B, v, Bz, Ey(-vBz), vB, v$^2$Bz and v$^2$B. They suggested that two functions that are highly correlated with all four geomagnetic indices at different time scales are IP electric field vB and Bv$^2$. 
In addition to vBs and vB, other merging/electric field-related CFs were tested in some studies invoking IMF clock angle ($\theta$). In such CFs, the amplitudes of electric-field-related functions were; vB \citep{1983JGR....88.9125D, 1979GeoRL...6..577K, 1983JGR....88.5727W}; v$^{\frac{4}{3}}$B$\rho^{\frac{1}{6}}$ \citep{1982P&SS...30..359V}; $\rho^{\frac{1}{2}}$v$^2$Bs \citep{2006JGRA..111.4221T}; v$^{\frac{4}{3}}$B \citep{2007JGRA..112.1206N}. \citep[See Table of candidate CFs in][]{1981SSRv...28..121A, 1990P&SS...38..627G, 2007JGRA..112.1206N, 2008JGRA..113.4218N, 2014JGRA..119..751B}.\\
Most of the studies are confined to single geomagnetic parameter (e.g., Dst/Kp/AE/AL/AU etc.). However, many studies use multiple indices in their work \citep[e.g.,][]{2007JGRA..112.1206N, 2017SoPh..292..140V, 2020Ge&Ae..59..639D, 2021Univ....7..138Y, 2023SpWea..2103314M, 2023AdSpR..71.1137B}. Studies were performed to obtain suitable CFs that relate to the intensity of GS during ICME generated CIR generated or both. 
Furthermore, CFs during a long train of continuous periods (which includes both geomagnetically quiet and disturbed conditions) have previously been investigated \citep[][]{2007JGRA..112.1206N, 2014ApJS..213...21V}. Our attempt in this work is to find a suitable CF that correlates well with the magnitudes of three geomagnetic parameters (equatorial Dst, mid-latitude ap, and polar AE indices) during the geomagnetically disturbed conditions ($\mathrm{Dst}\leq-50$ nT). This CF is expected to be useful in estimating the geomagnetic disturbance level from SW data with good accuracy, not only in the equatorial region but also in the mid-latitude and polar regions. \\
We have also studied the inter-relation between the peak values in three geomagnetic indices (Dst, ap, AE) during the MP of GSs. Some studies in recent past have studied inter-relationship between geomagnetic indices (Dst, ap, AE) \citep[e.g.,][]{2016STP.....2d..11B, 2017SoPh..292..140V} as well as between these indices and SW electric field \cite[e.g.,][]{2023AdSpR..71.1137B} during the MP of the GS generated by the IP structures (SH/ICME/CIR etc).
Throughout the MP of the GS a better correlation between (Dst, Kp) than (Dst, AE) was detected both during storms driven by ICMEs and CIRs. The correlation coefficient for ICME-generated storms between (Dst, Kp) and (Dst, AE) is reported to be higher for (Dst, Kp) \citep[see][]{2016STP.....2d..11B, 2018AdSpR..61..348B, 2023AdSpR..71.1137B}.
\cite{2023AdSpR..71.1137B} investigated the interrelationship between multiple indices (Dst, Kp), (Dst, AE), and (Kp, AE) and found a correlation coefficient which is lowest for (Dst, AE) during GS caused by SH events. They studied the relationship between different parameters during ICME and CIR events also. (Dst, Ap) and (Ap, AE) correlation coefficient was determined during CIR passage by \cite{2017SoPh..292..140V} and their correlation between (AE, Ap) is rather high compared to (Dst, Ap). It has been suggested by \citep{2020Ge&Ae..59..639D} that the pattern in the dynamics of the AE and ap indices is similar during GSs. \\
After selecting 57 strong GSs (Dst $\leq -50$ nT), the solar IP source of individual GS was identified. Then invoking the criteria of MP duration/steps taken to reach the Dst$_\mathrm{min}$,  which is related to rate of the GS development, we classified these selected GS into four duration-categories. In order to compare the GS development with the variations in various SW plasma and field parameters during the development of GS with different durations, we selected various plasma/field parameters (and their functions). We applied SEA to study the average behavior of not only the equatorial Dst index but also simultaneous changes in mid-latitude (ap) and polar (AE) indices, during the development of GS with four different main phase duration/rates development. Such plots, provide insight about similarities and distinctions in the dynamical behavior of three geomagnetic indices together with the simultaneous dynamical changes in SW plasma and field parameters (and their various functions). This aim has been realized by a systematic use of suitable parameters/functions of single, two and three SW plasma/field parameters and using correlation analysis between geomagnetic and SW parameters.
Such comparison between time evolution of geomagnetic indices and simultaneous variations in plasma/field parameters will be useful to compare the dynamics of SW parameters (and their functions) with the dynamics of geomagnetic disturbances in three domains of terrestrial magnetosphere (equatorial, mid-latitude and polar). It will help us to isolate the SW parameter/function which better follow the development in geomagnetic activity and ultimately lead to isolate the parameter/function that better represents the intensity of geomagnetic disturbance as seen in three geomagnetic parameters.\\
Although various CFs have previously been proposed to infer the strength of the GS using SW plasma and field parameters, more work is needed to identify a suitable function that can be used to better represent not only the intensity of GS as measured in the equatorial region (Dst), but also the peaks of geomagnetic disturbances in mid-latitude (ap) and polar regions (AE) regions during the occurrence of GS.\\
Because the physical mechanisms for the generation and development of the three geomagnetic indices (Dst, ap, and AE) are not identical, an attempt is also made to determine the strength of the inter-relationship between them (as inferred from correlation coefficients), which may be useful in understanding the coupling between different geomagnetic domains of the magnetosphere and differences in their dynamical behavior.\\
For SW-magnetosphere coupling and possible GS prediction from space-based observations, it is important to understand the lead/lag time between (a) the onset of GS and the onset of change in relevant SW plasma and field parameters/functions, as well as (b) the time of maximum geomagnetic disturbance and the time of maximum value of SW plasma and field parameters.\\
We have adopted two approaches; SEA and correlation analysis, with specific aims. SEA results are used to study the dynamics of GS, after grouping them on the basis of durations/steps in the MP, and comparing their dynamics with the changes/variations in probable CFs.
Then, using the same set of individual GS and implementing correlation analysis, we tried to determine a suitable function which may be useful to predict not only the intensity of the GS (as measured by (Dst$_\mathrm{min}$) but also to assess the magnitude of geomagnetic activity in mid-latitude region (ap$_\mathrm{max}$) and polar regions (AE$_\mathrm{max}$) during the intense geomagnetic activity generated by the same IP structure.
Correlation analysis has also been employed for the study of inter-relationship between the peaks in (ap, AE, Dst) during the MP of GSs. \\
Section \ref{sec:data} discusses data and analysis selection, grouping, and analyzing techniques. The criteria for classifying our data are specifically explained. Section \ref{sec:res} discusses the results; based on of superposed analysis (\ref{sec:sea}); the relationship between peak Dst and Bz (\ref{sub:DstBz});
relation between geomagnetic indices and SW plasma and field parameters and their derivatives (\ref{sub:correlation}), in Sections \ref{sub:correndex} we discuss inter-relation between geomagnetic indices, and \ref{SSC} discuss SSCs of Dst and Bz.  
Section \ref{sec:conc} presents the summary of results and conclusions.

\section{Data and analysis}\label{sec:data}
\subsection{Classifying the groups based on \m\ duration}\label{subsec:dec}

We have classified our selected 57 GS events in to four (MP durations) groups, explained in Table \ref{tabl1}.
In order to accomplish this, we have taken into account two factors for categorizing those groups: the first is the \m\ time span, which has been taken to be from the onset to the peak of GS (Dst$_\mathrm{min}$), and the second is based on the number of dips (steps) to reach Dst$_\mathrm{min}$ during its \m\ period.
We classified 14 events as d$ _{1} $ (one step decline) based on this criterion, a quick decrease takes 8 hours or less to reach the minimal Dst value.
We have labeled 13 events as d$_2$ (two step decrease) which takes 9 -- 12 hours to reach its minimum. Another 13 events classified as d$_3$ (three step decrease)
which takes 13 -- 22 hours to reach its Dst peak, and the rest 17 events are classified as d$_4$ (multiple step decrease) as shown in the seventh column of Table \ref{tabl1}; the decline in these events is very slow, taking more than 22 hours to reach minimum Dst.
\subsection{Data selection}\label{subsec:selection}

In this study, we chose 57 selected GS events measured with Dst (ranging from $\mathrm{Dst}=-64$ nT to $\mathrm{Dst}=-422$ nT) at 1-hour resolution from 1995 to 2022, spanning three solar cycles (solar cycles 23, 24, and 25).
We have retrieved 1-hour resolution geomagnetic indices, SW plasma and field parameters from Web-based data browse and retrieval tool for the NSSDC OMNI data set\footnote{\url{https://omniweb.gsfc.nasa.gov/ow.html}} \citep{King2005SolarWS}. 
The 57 events were chosen after accessing all storms from 1995 to 2022 and meeting the following criteria:
\begin{itemize}
\item The threshold Dst on the depth of a dip is $\leq-50$ nT
\item Smoothness of the drop and recovery profiles is taken into consideration, and some high amplitude multiple dip structures are excluded.
\item Full or nearly full recovery profiles are considered and selected. 
\end{itemize} 
We chose 57 events based on the above criteria, despite the fact that the criteria are highly subjective for different observers. 
Individual GS and their sources were identified with a careful scrutiny of timings using available data set\footnote{\url{https://izw1.caltech.edu/ACE/ASC/DATA/level3/icmetable2.htm}}, \footnote{\url{http://www.iki.rssi.ru/omni/catalog/}}, \footnote{\url{https://space.ustc.edu.cn/dreams/wind_icmes/index.php}} and tabulated in Table \ref{tabl1}.
In our study, we have considered three geomagnetic indices, SW plasma and field parameters and their functions for studying of their relationship with the intensity of geomagnetic disturbance. 
We have retrieved hourly averaged of four IMF related parameters namely; IMF magnitude B (nT), the north-south IMF Bz (nT), Sigma in IMF magnitude $\sigma$B$_\mathrm{m}$ (nT) and Sigma in IMF vector $\sigma$B$_\mathrm{v}$ (nT), two IP SW plasma parameters (proton density $\rho$ (cm$^{-3}$) and velocity v (kms$^{-1}$)), three geomagnetic activity indices (equatorial and low latitude geomagnetic index Dst (nT), linear mid-latitude geomagnetic activity index ap (nT); which is derived from the quasi-logarithmic Kp, and high latitude geomagnetic activity index AE (nT)) and ten derived parameters [flow pressure P (nPa), dawn-to-dusk electric field Ey (mVm$ ^{-1} $), the ratio of plasma thermal to magnetic pressure (plasma $\beta$), and the products (functions) of B, Bz, v and $\rho$, called SW-magnetosphere coupling functions (CFs) (vB, vBz$^2$, vEy, v$^{\frac{4}{3}}$Bz, v$^{\frac{4}{3}}$Bz$\rho$, v$^{\frac{4}{3}}$Bz$\rho^{\frac{1}{2}}$, v$^{\frac{4}{3}}$B$\rho^{\frac{1}{2}}$). In this way, a total of 19 parameters which are considered to be indicators of the development and intensity of GS \m.\\ 
\subsection{Selecting probable candidate coupling functions}\label{coupling}

Although our selection of SW parameters and their derivative functions is not exhaustive, we have considered for our study, some candidate CFs; single parameter (Bz, B, v and $\rho$), simple expression (functions) involving two SW parameters (vBz$^2$, v$^2$Bz) electric field-related (merging) functions involving two terms (vBz, vB, v$^{\frac{4}{3}}$Bz), electric field-related (merging) functions coupled with a viscous term involving three terms (v$^{\frac{4}{3}}$Bz$\rho$, v$^{\frac{4}{3}}$Bz$\rho^{\frac{1}{2}}$, v$^{\frac{4}{3}}$B$\rho^{\frac{1}{2}}$).
This selection of functions, although not exhaustive, considers some of the successfully employed functions one or more geomagnetic conditions (quiet/moderately distributed/strongly distributed) and better connected with one or more geomagnetic activity indices representing geomagnetic disturbances in one or more magnetic domains (equatorial/mid-latitude/polar). In addition, we have considered some variants of earlier suggested electric field related term coupled with a viscous term. For isolating a CF, whose amplitude can be related fairly accurately with the strength of magnetic disturbance in all three geomagnetic latitudinal domains (equatorial, mid-latitude and polar), we adopted a systematic approach. \\
I. We isolated two (best and second best) related (single) SW parameters whose amplitude best relates the amplitudes of geomagnetic disturbances in three latitudinal domains (represented by Dst$_\mathrm{min}$, ap$_\mathrm{max}$ and AE$_\mathrm{max}$).\\
II. These two parameters were then coupled with other SW parameters with varying powers over them, and two (best and second best) related CFs in involving two parameters were obtained.\\
III. These two parameters (both electric field related) we then coupled with some viscous terms, and a best CF is obtained whose amplitude can used to predict the intensity of geomagnetic disturbances in three geomagnetic domain with fairly good accuracy.\\

\subsection{Data analysis procedures}
In this work, we have applied two data analysis techniques; SEA and linear regression analysis.
SEA is a very useful technique for the cause and effect study. It provides genuine effects after removing small scale (unwanted) fluctuations as a result of averaging effect. This analysis can be used to detect a signal (genuine effect) in the presence of noise (unrelated variation) whenever the noise sums incoherently while the signal is re-enforced by superposition. SEA has been successfully utilized not only to demonstrate the average behavior of certain phenomena (effects) caused by their sources, but also to compare the development of a phenomenon with the simultaneous variation in source-related parameters and their functions. \\
For analysis purpose, zero epoch in the SEA procedure is taken as the onset time of MP of the GS. In order to gain more insight into the MP dynamics, the SEA technique is applied not only for the GS of all selected MP durations together, but also for the group of GS divided on the basis of durations of MP in different time bins, and, whether total MP developments happen, in one, two, three or multiple steps.\\
Regression analysis provides a good supplement to SEA. Linear regression analysis is applied to obtain the best SW parameter/function that represents the magnitude in three geomagnetic indices (Dst, ap and AE) at the time of strong geomagnetic disturbances.
Linear regression analysis is done (a) to search for a best SW-magnetosphere CFs to predict the intensity of GS and (b) to find the inter-relationship between different geomagnetic indices.\\
In this way, we have also analyzed the correlation between storm indices with SW plasma and field parameters (and their derivatives) \c\ by applying linear regression analysis. Based on the correlation analysis we get information on the parameters that are well correlated with amplitude of geomagnetic indices (Dst, AE \& ap).\\ Since southward Bz is crucial parameter for GS \citep{1974JGR....79.1105R}, the time delay between extreme value of geomagnetic index Dst (Dst$_\mathrm{min}$) and southward IMF Bz have been determined. Moreover, observed SSCs in relation to Bz are analyzed as well. The inter-relationship between storm indices have also been discussed.\\
\begin{table*}
\centering
\footnotesize{
\caption{List of the selected 57 GS events profile ($\mathrm{Dst}\leq-64$ nT) with onset to peak time profiles with the duration of \m\ storm, the peak Dst value and the amplitude of Dst ($\Delta $Dst).
}\label{tabl1}
\begin{tabular}{l cccccccc}
\hline
\textbf{No of GS}  & \textbf{GS start date}& \textbf{Recovery Start}  &\textbf{Decrease}  &\textbf{$\Delta$Dst}&\textbf{Dst$_\mathrm{min}$}&\textbf{Group}&\textbf{Sources}\\
\textbf{events}&\& \textbf{time (UT)}&\textbf{date} \& \textbf{time(UT)}&\textbf{time(hrs)}&\textbf{(nT)}&\textbf{(nT)}&&\\
\hline
1 & 1995-04-06T16:00&1995-04-07T19:00 &26&158 & -149&d$_3$&CIR\\
2 & 1995-10-18T12:00 &1995-10-19T00:00&11& 136 &-127& d$_3$&SH+MC\\
3 & 1997-01-10T02:00 &1997-01-10T10:00& 7 &  101 &-78& d$_1$&SH+MC\\
4 & 1997-04-21T09:00 &1997-04-22T00:00&14& 106 &-107&d$_2$&MC\\
5 & 1997-05-01T14:00 &1997-05-02T01:00&10& 69 &-64&d$_3$&CIR\\
6 & 1997-05-15T02:00 &1997-05-15T13:00&10& 133 &-115&d$_3$&SH+MC\\
7 & 1997-05-26T11:00 &1997-05-27T07:00&17& 85 &-73&d$_3$&SH+MC\\
8 & 1997-06-08T05:00 & 1997-06-09T05:00&22& 70 &-84&d$_4$&MC\\
9 & 1997-09-02T23:00 & 1997-09-03T23:00&23& 111 &-98&d$_4$&SH+ICME\\
10 & 1997-10-10T17:00 & 1997-10-11T04:00&10&116 &-130&d$_2$&SH+MC\\
11 & 1997-11-06T23:00 & 1997-11-07T05:00& 5 & 105 &-110&d$_1$&SH\\
12 & 1998-01-06T14:00 & 1998-01-07T05:00&14& 84 &-77&d$_2$&SH+MC\\
13 & 1998-03-10T10:00 & 1998-03-10T21:00&10& 131 &-116&d$_2$&CIR\\
14 & 1998-06-25T18:00 & 1998-06-26T05:00 &10&  129 &-101&d$_2$&SH+ICME\\
15 & 1998-08-05T23:00 & 1998-08-06T12:00 & 12 & 151 &-138&d$_3$&SH+ICME\\
16 & 1998-08-26T08:00 & 1998-08-27T10:00 & 25 & 186 &-155&d$_4$&SH+ICME\\
17 & 1998-09-25T00:00 & 1998-09-25T08:00 & 7 & 198 &-207&d$_3$&SH+MC\\
18 & 1998-11-13T00:00 & 1998-11-13T22:00 & 21 & 132 &-131&d$_4$&MC\\
19 & 1999-01-13T02:00 & 1999-01-14T00:00 & 21 & 113 &-112&d$_4$&SH+ICME\\
20 & 1999-09-22T16:00 & 1999-09-23T00:00 & 7 & 200 &-173&d$_2$&SH+ICME\\  
21 & 2000-01-11T08:00 & 2000-01-11T23:00&13& 93 &-80&d$_2$&CIR\\
22 & 2000-04-06T16:00 &2000-04-07T01:00&8& 287 &-292&d$_1$&SH\\ 
23 & 2000-07-15T15:00 &2000-07-16T01:00&9& 307 &-300&d$_3$&SH+MC\\
24 & 2000-08-12T01:00 &2000-08-12T10:00&8& 214 &-234&d$_1$&SH+MC\\
25 & 2000-09-17T19:00 &2000-09-18T00:00&4& 229 &-201&d$_1$&SH+MC\\  
26 & 2001-03-31T03:00 &2001-03-31T09:00&5& 413 &-387&d$_1$&SH+ICME\\
27 & 2001-04-11T15:00 &2001-04-12T00:00&8& 269 &-271&d$_1$&SH+MC\\
28 & 2001-04-18T01:00 &2001-04-18T06:00& 5 & 106 &-114&d$_1$&SH\\
29 & 2001-04-21T22:00 &2001-04-22T16:00&17& 114 &-102&d$_3$&SH+MC\\
30 & 2001-08-17T12:00 &2001-08-17T22:00&9& 149 &-105&d$_1$&SH\\
31 & 2001-10-03T06:00 &2001-10-03T15:00&8& 120 &-166&d$_1$&MC\\
32 & 2001-11-05T18:00 &2001-11-06T07:00&12& 314 &-292&d$_3$&SH+ICME\\
33 & 2001-11-24T06:00 &2001-11-24T17:00&10& 223 &-221&d$_3$&SH\\
34 & 2002-09-07T12:00 &2002-09-08T01:00&12& 153 &-181&d$_2$&SH\\
35  & 2003-11-19T22:00 &2003-11-20T22:00&23& 420 &-422&d$_4$&SH+MC\\
36  & 2004-04-03T14:00 &2004-04-04T01:00&10& 107 &-117&d$_2$&SH\\
37  & 2004-08-30T00:00 &2004-08-30T23:00&22& 147 &-129&d$_4$&SH\\
38  & 2005-01-21T17:00 &2005-01-22T07:00&11& 120 &-97&d$_4$&ICME\\
39  & 2005-05-15T03:00 &2005-05-15T09:00& 5 & 299 &-247&d$_1$&SH+MC\\
40  & 2006-12-14T14:00 &2006-12-15T08:00& 17 &155 &-162&d$_4$&SH+MC\\
41  & 2009-07-21T22:00 &2009-07-22T07:00& 8 & 88 &-83&d$_2$&ICME\\
42  & 2011-08-05T19:00 &2011-08-06T04:00& 8 & 140 &-115&d$_2$&SH\\
43  & 2011-10-24T18:00 &2011-10-25T01:00& 7 & 167 &-147&d$_2$&SH+MC\\
44  & 2012-04-23T04:00 &2012-04-24T05:00& 24 & 138 &-120&d$_4$&SH+ICME\\
45  & 2012-07-14T18:00 &2012-07-15T19:00& 22 & 154 &-139&d$_4$&SH+ICME\\
46  & 2012-11-12T16:00 &2012-11-14T08:00& 39 &  127 &-108&d$_4$&SH+MC\\
47  & 2013-03-17T06:00 &2013-03-17T21:00& 14 &  147 &-132&d$_4$&SH+ICME\\
48  & 2013-05-31T17:00 &2013-06-01T09:00& 15 & 140 &-124&d$_2$&CIR\\
49  & 2013-06-05T13:00 &2013-06-07T03:00& 37 & 70 &-78&d$_4$&SH+MC\\
50  & 2014-02-26T22:00 &2014-02-28T00:00& 25 & 102 &-97&d$_3$&CIR\\
51  & 2015-01-07T07:00 &2015-01-07T12:00& 4 & 116 &-107&d$_1$&SH+MC\\
52  & 2015-03-17T05:00 &2015-03-17T23:00& 17 & 279 &-234&d$_3$&SH+MC\\
53  & 2015-06-21T18:00 &2015-06-23T05:00& 34 & 240 &-198&d$_4$&SH+ICME\\
54  & 2015-12-31T02:00 &2016-01-01T01:00& 22 & 124 &-116&d$_4$&SH+MC\\
55  & 2016-10-13T02:00 &2016-10-13T18:00& 15 & 131 &-110&d$_1$&SH+MC\\
56  & 2017-05-27T21:00 &2017-05-28T08:00& 10 & 168 &-125&d$_1$&SH+MC\\
57  & 2018-08-25T08:00 &2018-08-26T07:00& 22 & 194 &-175&d$_4$&SH+MC\\ 
\hline
\multicolumn{8}{l}{\footnotesize d$_1$, Duration of one step decrease less than 8 hours. } \\  
\multicolumn{8}{l}{\footnotesize d$_2$, Two step decrease duration between 9--12 hours.}\\
\multicolumn{8}{l}{\footnotesize d$_3$, Three step decrease duration between 13--22 hours.}\\
\multicolumn{8}{l}{\footnotesize d$_4$, Multiple step decrease duration $\geq 23$ hours.} \\
\end{tabular}}
\end{table*}

\section{Results and Discussion}\label{sec:res}
Based on the development time of the \m\ disturbances, as indicated in Section \ref{subsec:dec}, we have categorized the selected 57 GS events plotted with 1-hour resolution, as $\sim$25\% are one stage/step development, $\sim$23\% two stage developments, $\sim$23\% three stage developments and $\sim$30\% multiple stage developments. In general, $\sim$75\% of our GS events evolve through more than one stage steps to reach the Dst minimum, which is slightly lower than the result of \citep{2020EGUGA..2220900M}, where $\sim$87\% of their data evolve through multiple stages of \m\ development. 

\subsection{Superposed epoch analysis}\label{sec:sea}
\begin{figure*}\centering
\includegraphics[width = 0.89\textwidth, height= 7.6cm]{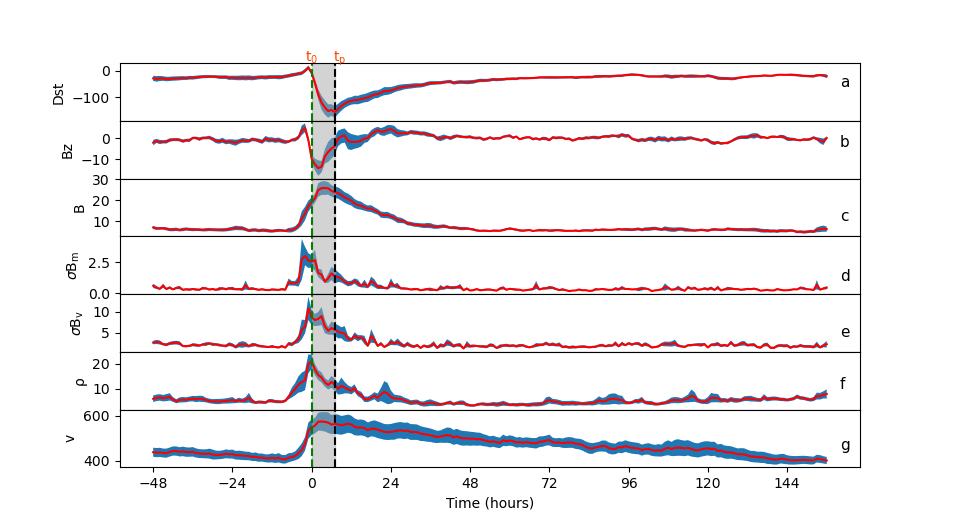}\label{i}(i)
\includegraphics[width = 0.89\textwidth, height= 7.6cm]{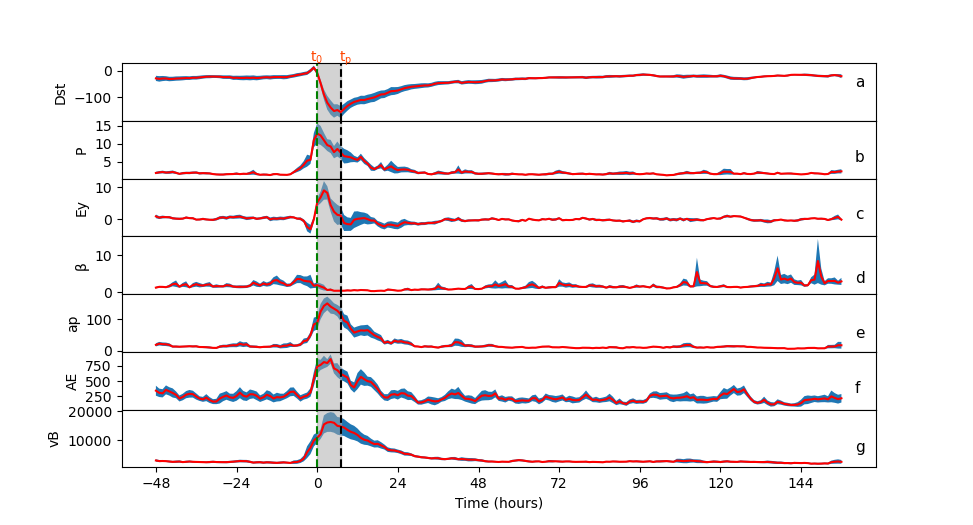}\label{ii}(ii)
\includegraphics[width = 0.89\textwidth, height= 7.6cm]{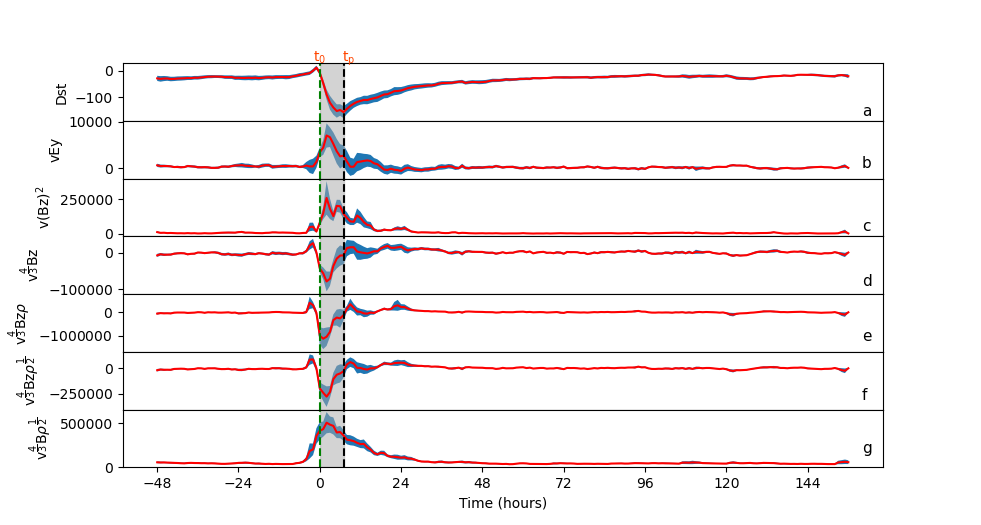}\label{iii}(iii)
\caption{\small Selected one step (d$_1$) hourly resolution superposed temporal profiles of geomagnetic indices, SW plasma and field parameters and their derivatives [Dst (nT), ap (nT), AE (nT), B (nT), Bz (nT), $\sigma\mathrm{B_m}$ (nT), $\sigma\mathrm{B_v}$ (nT), P (nPa), $\rho$ ($\mathrm{cm}^{-3}$), $\beta$, $\mathrm{v}$ ($\mathrm{kms}^{-1}$), $\mathrm{Ey}$ $(\mathrm{mVm}^{-1})$, $\mathrm{vB}$ $(10^{-6}\mathrm{Tms}^{-1})$, $\mathrm{vEy}$ $(\mathrm{Vs}^{-1})$, $\mathrm{vBz^2}$ $(\mathrm{10^{-15}T^2ms}^{-1})$, v$^{\frac{4}{3}}\mathrm{Bz}$ $[10^{-5}\mathrm{T}(\mathrm{\frac{m}{s}})^{-\mathrm{a}}]$, v$^{\frac{4}{3}}\mathrm{Bz}\rho$ $(10^{-11}\mathrm{T}\mathrm{m}^{\mathrm{b}}\mathrm{s}^{\mathrm{a}})$, v$^{\frac{4}{3}}$Bz$\rho^{\frac{1}{2}}$ $(10^{-8}\mathrm{T}\mathrm{m}^{\mathrm{c}}\mathrm{s}^{\mathrm{a}})$, v$^{\frac{4}{3}}$B$\rho^{\frac{1}{2}}$ $(10^{-8}\mathrm{T}\mathrm{m}^{\mathrm{c}}\mathrm{s}^{\mathrm{a}})$] subjected to SEA. Where $\mathrm{a}=\frac{-4}{3}$, $\mathrm{b}=\frac{-5}{3}$ and $\mathrm{c}=\frac{-1}{6}$. 
The two vertical broken lines indicate: the onset of storm \m\ at t$_0$ (green) and the peak of storm at minimum Dst t$_\mathrm{p}$, respectively. The shaded gray area represents the storm \m duration from onset to minimum Dst. \\
}\label{d1}
\end{figure*}

\begin{figure*}\centering
\includegraphics[width = 0.89\textwidth, height= 7.6cm]{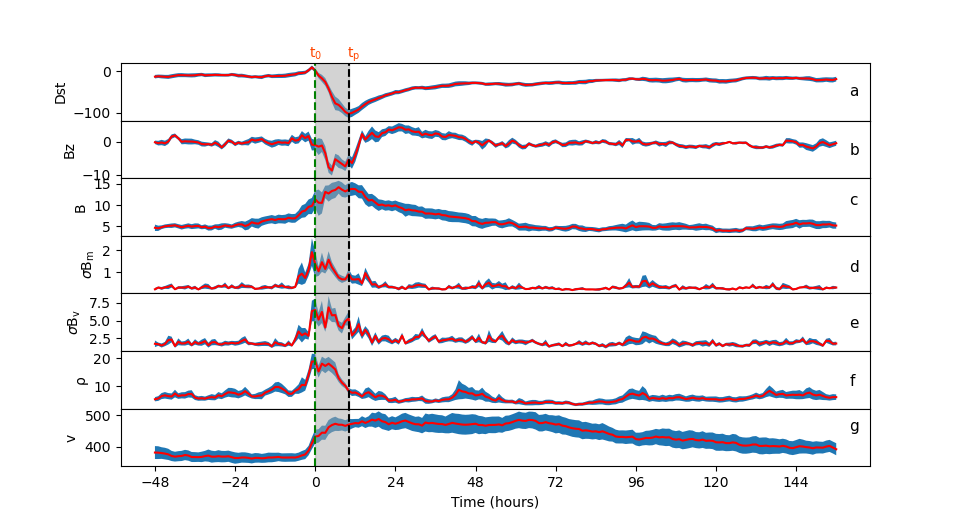}(i)
\includegraphics[width = 0.89\textwidth, height= 7.6cm]{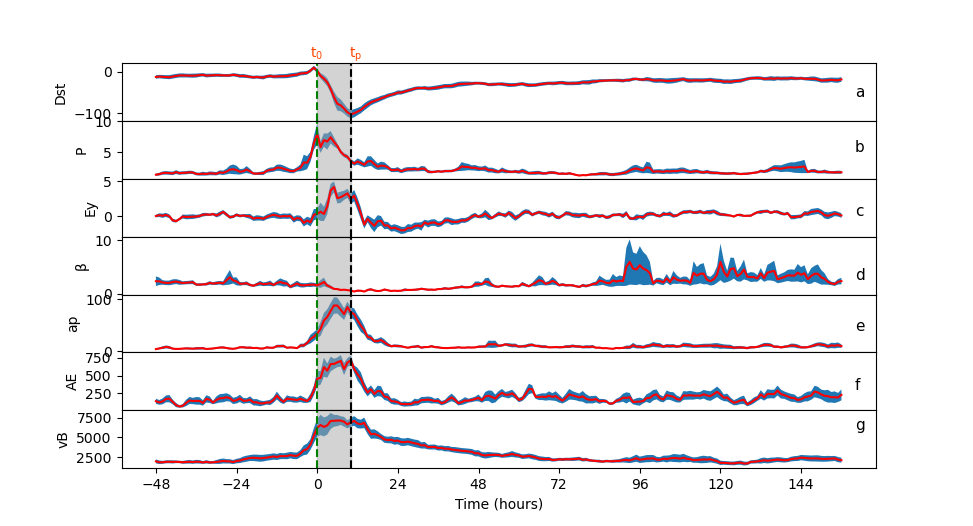}(ii) 
\includegraphics[width = 0.89\textwidth, height= 7.6cm]{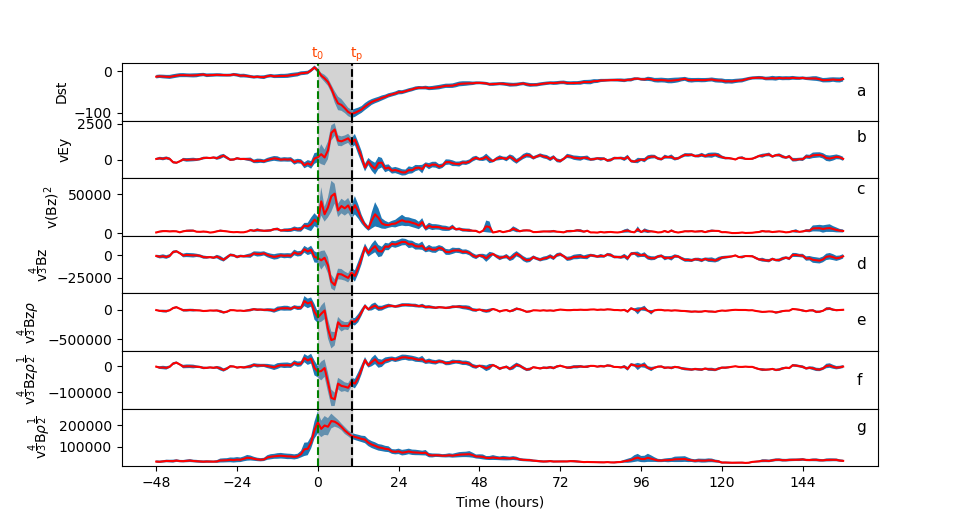}(iii)
\caption{\small Same as Figure \ref{d1} for SEA results for 13 selected two step drop GS events (d$_2$).}\label{d2}
\end{figure*}

\begin{figure*}\centering
 \includegraphics[width = 0.89\textwidth, height= 7.6cm]{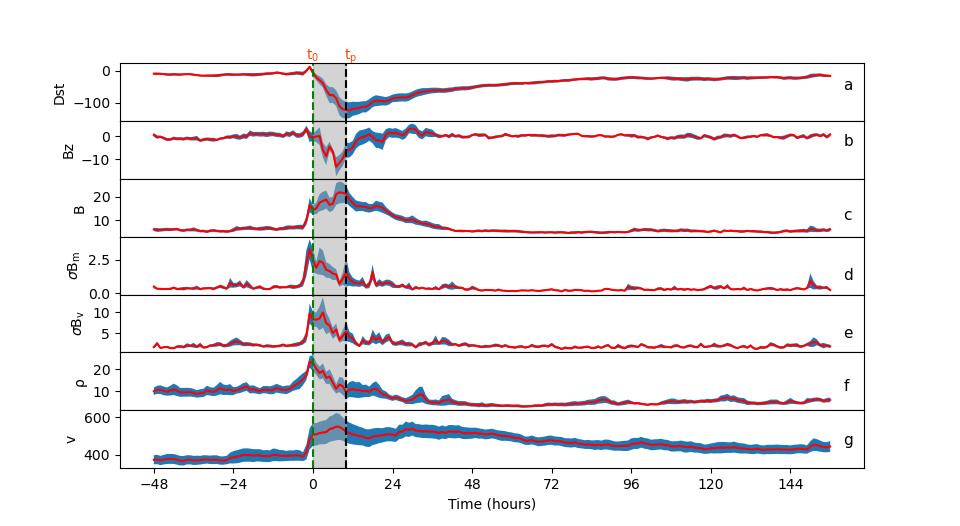}(i)
\includegraphics[width = 0.89\textwidth, height= 7.6cm]{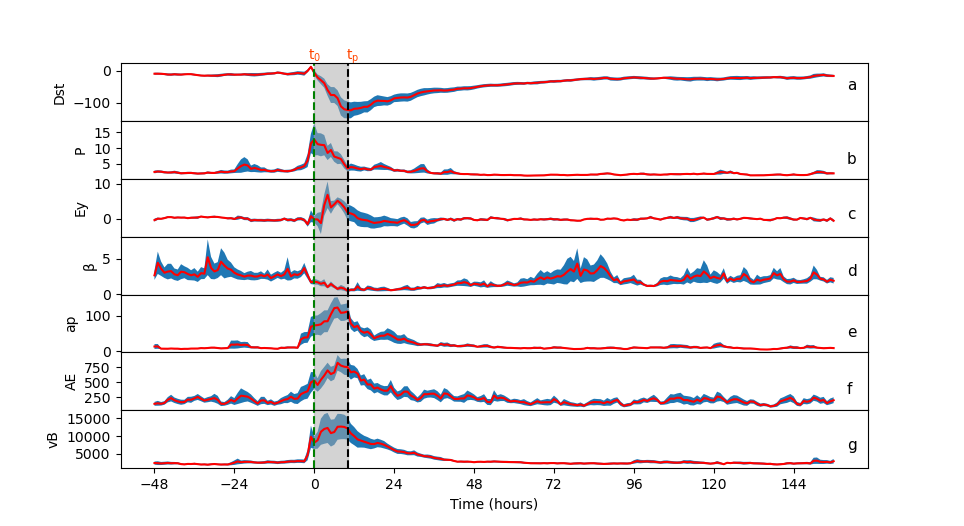}(ii)
\includegraphics[width = 0.89\textwidth, height= 7.6cm]{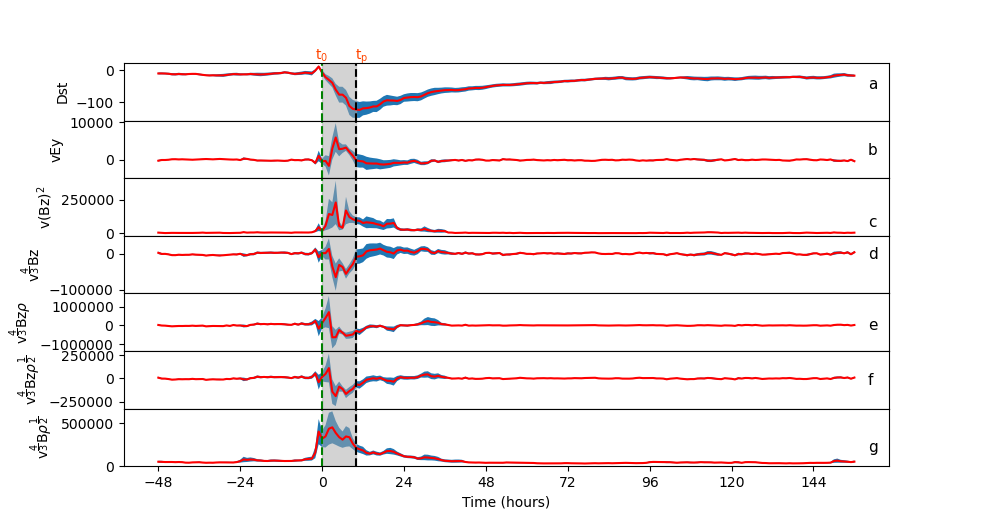}(iii)
\caption{\small Same as Figure \ref{d1} for SEA results for a group of 13 three step decrease GS events (d$_3$).} \label{d3}
\end{figure*}

\begin{figure*}\centering
\includegraphics[width = 0.89\textwidth, height= 7.6cm]{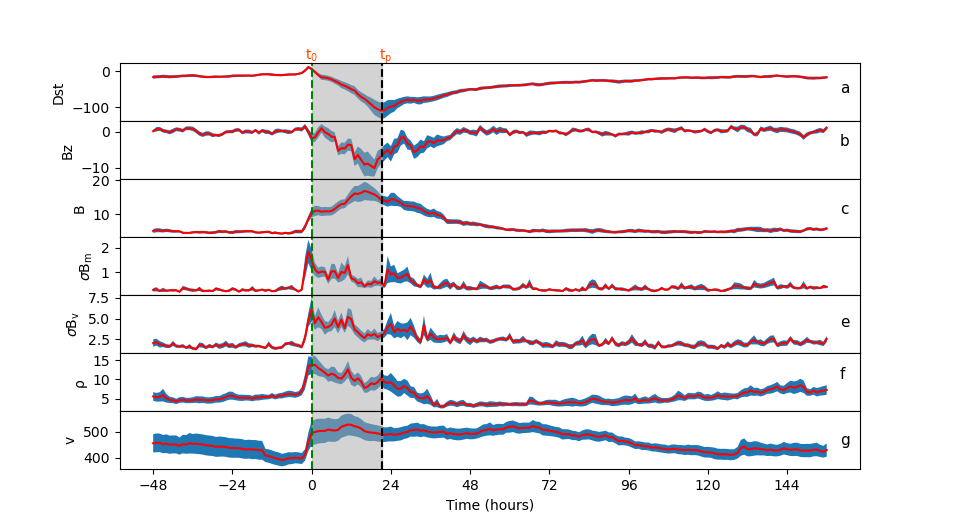}(i)
\includegraphics[width = 0.89\textwidth, height= 7.6cm]{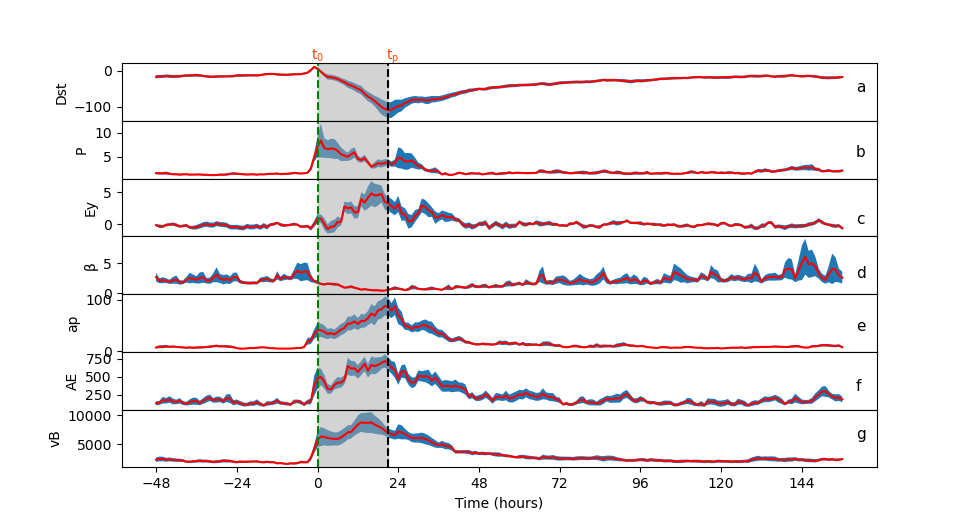}(ii)
\includegraphics[width = 0.89\textwidth, height= 7.6cm]{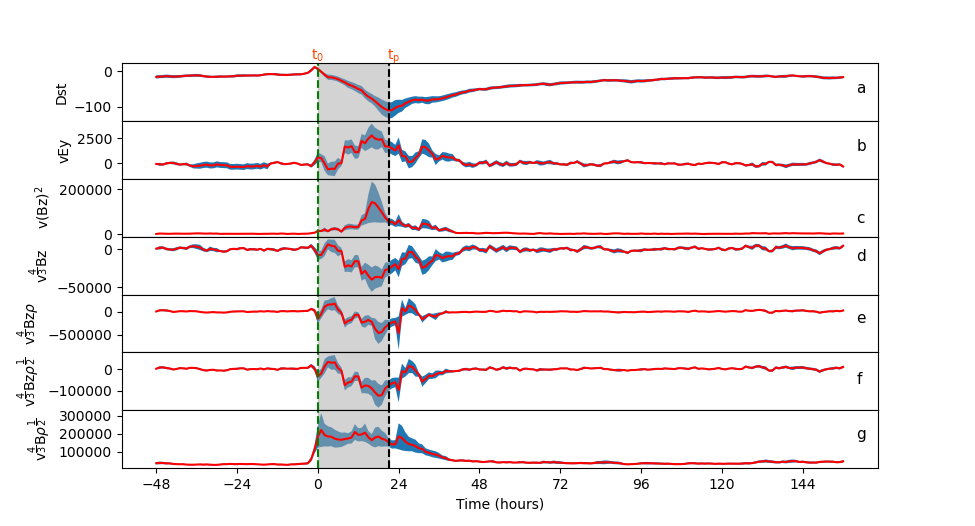}(iii)
\caption{\small Same as Figure \ref{d1} for SEA results for a group of 17 multiple step decrease GS events (d$_4$).}\label{d4}
\end{figure*}

\begin{figure*}\centering
\includegraphics[width = 0.89\textwidth, height= 7.6cm]{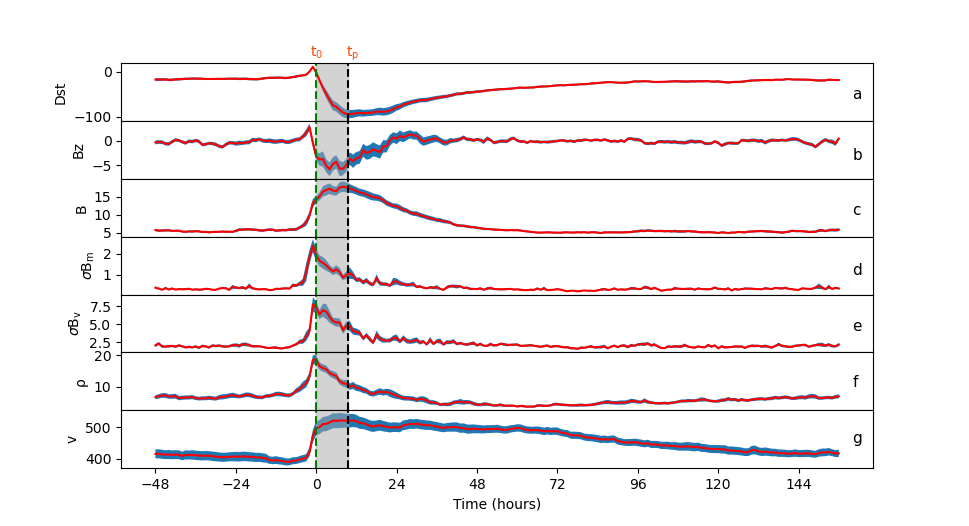}(i)
\includegraphics[width = 0.89\textwidth, height= 7.6cm]{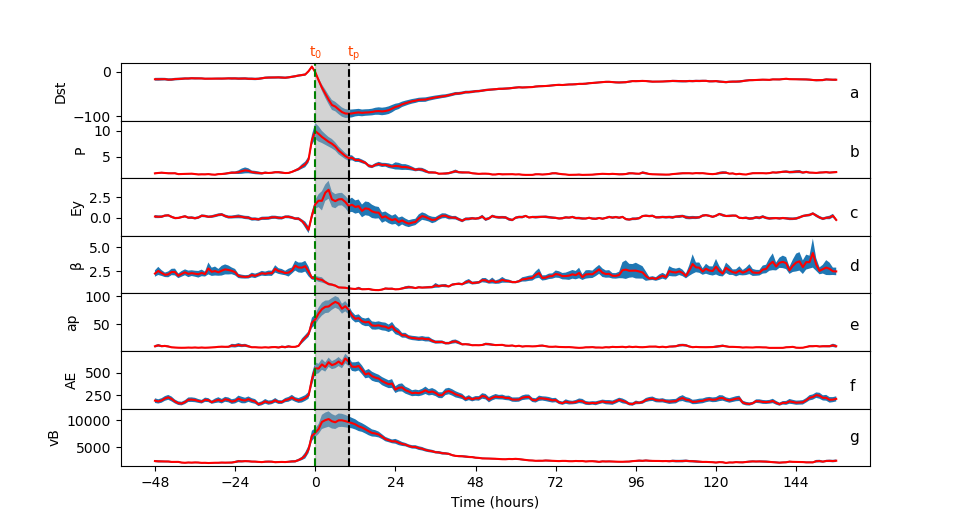}(ii)
\includegraphics[width = 0.89\textwidth, height= 7.6cm]{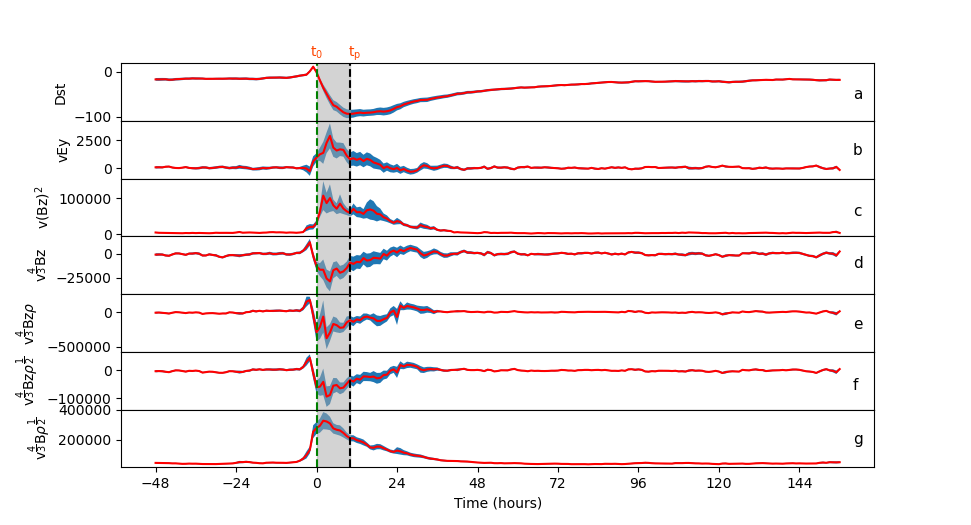}(iii)
\caption{\small Same as Figure \ref{d1} for SEA results for all selected 57 GS events}\label{d5}
\end{figure*}

\begin{figure}
\centering
\includegraphics[height=8.2cm, width=9.4cm]{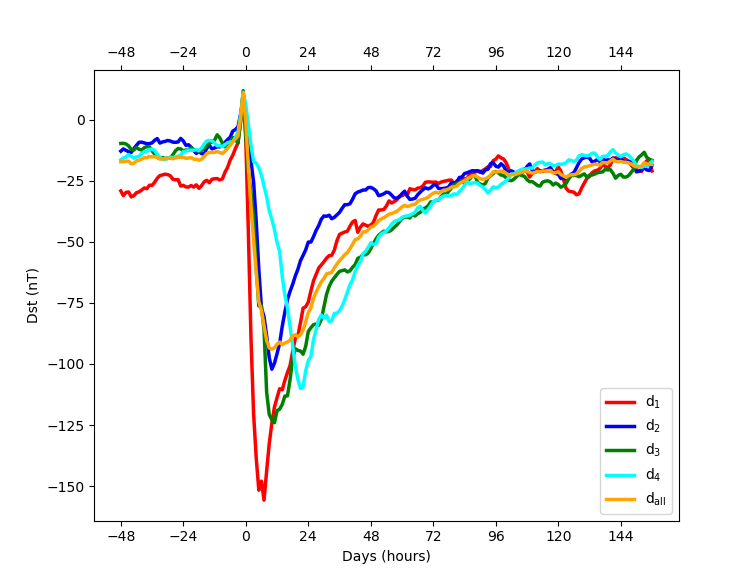}	
\caption{\small The line plot depicts the intensity of peak Dst for various groups of storms. The \m\ dynamics of Dst are illustrated for the very quick decrease (group d$_1$ red colored), fast decrease (group d$_2$ blue colored), slow decrease (group d$_3$ green colored), very slow decrease (group d$_4$ cyan colored), and the whole data set (orange colored).}
\label{Dst_group}
\end{figure}

For various purposes, we chose the aforementioned geomagnetic indices, SW plasma and field parameters and their product CFs, and they were subjected to SEA, in order to investigate the dynamics of GS \m\ developments in relation to simultaneous time variation of 
SW plasma and field parameters and selected derivatives during GSs of varying durations.
For the reason of visibility and to avoid squeezed view of figures, we have grouped the 19 selected parameters and their derived CFs into three panel sets (i,ii and iii) of superposed graphs, each set has 7 panels (a, b, c, d, e, f \& g) of parameters labeled vertically with a common shared x-axis as time (hours) from top to bottom (a, b, c, d, e, f \& g) shown in Figures (\ref{d1} - \ref{d5}). For the purpose of getting detail information to the dynamics of all the above mentioned SW plasma and field parameters and their derived CFs, 
we have plotted the geomagnetic index Dst at the top panel for each group of the figures, which helps to bring us a more detail analysis of their dynamics with the geomagnetic disturbances.\\
Study of the development of GS, its distinct phases (e.g. main and recovery phases) often evoke SEA procedure on selected GS events with a targeted epoch time (e.g., GS start time) without taking into account the total time of development of GS (MP) to reach its peak level of disturbance (Dst$_\mathrm{peak}$). Although, such studies were successful in their goal, we adopted somewhat improved approach by dividing the selected GSs on the basis of their MP durations/steps to attain its peak. This approach helps us to give us better averaged amplitude (by dividing them into suitable groups of selected bins of MP durations), compared to averaged amplitude of GSs of all possible MP durations. Another advantage of our approach is that it gives us a good chance to compare the average behavior of the development of one-step, two-step and multiple step GSs separately with simultaneous development in suitable CFs, and also to compare their time variation response in GS development.\\
A DSEA approach was used in certain superposed analysis-based studies \citep[e.g.,][]{1997JGR...10214215Y, 2010AnGeo..28.2177Y, 2023SpWea..2103314M} to explore GS evolution. \cite{2023SpWea..2103314M} comment in their paper that the single point SEA approach provides accurate behavior of the average storm around the epoch (e.g., GS MP onset). However, when one moves away from the epoch, the accuracy of the average Dst$_\mathrm{min}$ (e.g., at the end point of MP) decreases \citep[see also,][]{1997JGR...10214215Y, 2021Univ....7..138Y}. 
In our SEA based analysis, with MP onset set as zero epoch, we could minimize this possible compromise to accuracy around Dst$_\mathrm{min}$ by superposing the GS events after dividing them into groups within a limited  range of  MP durations; thus avoiding the compromise on the accuracy in determining  the average intensity of GS (e.g., Dst$_\mathrm{min}$).\\
We employed the method of SEA on hourly  resolution data of three geomagnetic indices (Dst, ap and AE) and selected SW plasma and field parameters (and their derivatives); zero epoch is taken as the onset of GS (Figures \ref{d1} - \ref{d5}). In these figures t$_0$, denoted by a vertical dashed green line, represents the onset (sharp decrease of Dst) of the GS, and t$_\mathrm{p}$, indicated by a black dashed line, represents the peak time of the storm at Dst$_\mathrm{min}$. 
The gray area represents the \m\ of GSs. Using this method, it is possible to identify patterns in the dynamics of the geomagnetic indices with simultaneous variations IP SW-plasma parameters, \c\ during storms with different \m\ durations. This also helps to observe distinction and similarities in GS of different durations.
For Figures from (\ref{d1} - \ref{d5}), we used hourly data 48 hours prior to the onset time of disturbance and the plot extends to 156 hours after the onset which includes the recovery phase from the storm peak time until it reaches the pre-onset level. Such extended plot helps to observe the post disturbances in detail and sporadic transient enhancement (SSC) prior to MP onset time  for all groups superposed analysis. It is also useful to see the time lag between the storm indices and IP parameters. However, in this study our interest is concentrated on the \m\ development of GSs. The SSC enhancements before the MP onset observed in Dst and IMF Bz are discussed in detail at Section \ref{SSC}.
In all the superposed analyzed Figures (\ref{d1} - \ref{d5}) the red line is our data points and the cyan shades are the standard error of mean (SEM), where $\mathrm{SEM}=\frac{\mathrm{STDEV.S}}{\sqrt{\mathrm{N}}}$ as STDEV.S is standard deviation and N is the number of GS events for each group. Standard deviation is calculated from the mean as $\mathrm{STDEV.S}=\sqrt{\frac{(\mathrm{x_{i}-x_{m}})^2}{\mathrm{N}-1}}$, where x$_\mathrm{i}$ \& x$_\mathrm{m}$ are data point and sample mean, respectively.\\
Figure \ref{d1} depicts a set of 14 one step dip events used in SEA at a resolution of one hour. When we zoom out and examine the storm \m\ development of each individual events, we notice one step of dip known as one step decrease events, as described in Section \ref{subsec:dec}.\\
Figure \ref{d1}(i) displays the following data in the panels from top to bottom: (a) Dst index; (b) north-south component of IMF Bz; (c) IMF magnitude B; (d) sigma in IMF magnitude $\sigma$B$_\mathrm{m}$; (e) sigma in IMF vector $\sigma$B$_\mathrm{v}$; (f) plasma density  $\rho$ and (g) solar wind velocity v. It takes 8 hours to complete the \m\ of GS development from the moment of storm onset (epoch) to the lowest Dst value. The MP of GS  has a single step visible dip; it begins as a rapid reduction. SSC is evident just before the MP onset (t$_0$); a discussion about SSC and its relation with Bz as given in detail in Section \ref{SSC}. 
All the parameters plotted in Figure \ref{d1}(i) start increasing significantly a few hours before the onset time (t$_0$) indicating arrival of shock-sheath turbulent region indicated by enhanced $\sigma$B$_\mathrm{m}$ and $\sigma$B$_\mathrm{v}$. SSC is also considered as a geomagnetic signature of shock.
More specifically $\sigma$B$_\mathrm{v}$ \& proton density $\rho$ are at their peak levels at the GS onset time, while $\sigma$B$_\mathrm{m}$ reaches its peak a few hours earlier than MP onset, indicating that shock/sheath magnetic field turbulent regions passing during onset time of GS. 
At the GS onset time, enhanced B, v, $\sigma$B$_\mathrm{m}$, $\sigma$B$_\mathrm{v}$ \& $\rho$ indicates that a high density, high speed turbulent shock/sheath magnetic structure is passing. 
When the GS begins to recover, the Bz which was negative (southward) during the MP, turns positive (northward). Moreover, at this time $\sigma$B$_\mathrm{m}$, $\sigma$B$_\mathrm{v}$ and density decreased significantly to near normal level IMF B and plasma velocity v remain high, indicating that high-speed quiet magnetic ejecta/cloud structure is still passing. \\ 
Figure \ref{d1}(ii) consists of another seven panels plotted with a resolution of one hour. This superposed plot results of the seven parameters are (a) Dst index, (b) solar wind flow pressure P, (c) duskward electric field Ey, (d) plasma $\beta$, (e) ap index, (f) AE index and (g) product vB, from top to bottom. Increased P at the commencement of GS is a sign that highly pressured plasma has passed. 
The down-dusk electric field (Ey) is enhanced during the MP on an average, it start increasing three hours earlier than MP onset time. 
Ey has been suggested to determine the SW coupling \citep[e.g.,][and references therein]{1975Sci...189..717B, 2000JGR...105.7707O}.\\
The ratio of thermal to magnetic pressures $ \beta $ begins to fall before the GS onset time. The decrease of $ \beta $ indicates that magnetic field pressure takes precedence over thermal pressure. 
The dynamics of the geomagnetic indices ap, AE, and IP field-plasma derivative vB are similar but ap follows vB better than AE during MP of the GS. 
The geomagnetic index ap and product vB begin 3 hours before the Dst onset time, while AE 2 hours ahead of Dst epoch. The other geomagnetic index ap peak occurs one hour after the Dst peak. As shown in Table \ref{tabl4}, AE and vB peaks 3 hours before the Dst peak time.\\
As seen in Figure \ref{d1}(iii) it has seven more panels with a 1-hour resolution of same 14 events. They are essentially different CFs involving the parameters v, B, Bz, Ey and $\rho$. The idea to use fractional powers for SW plasma parameters was motivated by some earlier publications \citep[e.g.,][]{1982P&SS...30..359V, 1986ASSL..126..119M, 1989JGR....94.8835G, 2006JGRA..111.4221T, 2007JGRA..112.1206N}. The superposed result of seven panels from top to bottom are (a) Dst index and various derivatives (functions) (b) vEy (Bzv$ ^2 $), (c) vBz$^2$, (d) v$^{\frac{4}{3}}$Bz, (e) v$^{\frac{4}{3}}$Bz$\rho$, (f) v$^{\frac{4}{3}}$Bz$\rho^{\frac{1}{2}}$ and (g) v$^{\frac{4}{3}}$B$\rho^{\frac{1}{2}}$. 
All the derived CFs onset time started 2 to 5 hours earlier than the MP onset. However, the GS MP end time lags behind the other CFs by a few hours. The morphology of v$^{\frac{4}{3}}$Bz, v$^{\frac{4}{3}}$Bz$\rho$ and  v$^{\frac{4}{3}}$Bz$\rho^{\frac{1}{2}}$ are identical to that of Dst dynamics and exhibits visible SSC features (see also Table \ref{tabl4}). \\
Figure \ref{d1}(i - iii) shows that high density turbulent IMF shock/sheath structure has developed ($\sim$5 -- 6) hours before the Dst onset time. IMF Bz points south 3 hours before Dst decrease begins. For this set of events, based on the superposed plot, the peak of Dst is $\sim$5 hours later than the peak of Bz.\\
Figure \ref{d2}(i - iii) is similar to the corresponding parameters as Figure \ref{d1}(i - iii), which represents average profile for 13 two step decrease GS events utilized for SEA. Zero epoch is again taken as the onset time of the GS MP. Two step decreases in Dst is evident during the \m\ even in the superposed plot with the first faster dip accounting for $\sim$60\% of the \m\ duration and the second slower dip accounting for $\sim$30\% of the \m\ duration.\\
According to Figure \ref{d2}(i), Bz starts turning to southward slightly earlier than the MP onset and remains strongly southward for a considerable period.
The IMF magnitude B begins to increase earlier than to and remains at its maximum for considerably long time, even though Dst begins to recover while B remains at high. The disturbed IMF levels represented by enhanced $\sigma$B$_{\mathrm{m}}$, $\sigma$B$_{\mathrm{v}}$ \& proton density $\rho$ are at their maximum during the storm onset time, although their level remains enhanced for few more hours. These three parameters started increasing for four to five hours before the onset time t$_0$. 
All SW plasma and field parameters begin a few hours before the storm's MP onset time. The IP plasma speed v begins three hours before Dst decrease.\\	
Characteristic features of different parameters in \ref{d2}(ii) show that, with the exception of $ \beta $, all parameters begin to increase a few hours before the MP onset time. Although enhanced during the MP of the storm, the dynamical behavior of ap and AE indices during this phase is not exactly similar to Dst; ap attains its peak earlier (and remains enhanced for a longer duration), AE enhancement is even faster (and the peak intensity is broaden). These differences in dynamics of three geomagnetic indices during developments in MP of GS is possibly related to activity domain they represent and the physical mechanism operating with in the magnetospheric regions, during their development, are not exactly similar.\\
As shown in Figure \ref{d2}(iii), all product CFs begin to change three to six hours before storm onset time and remain enhanced during the MP although with some step variation in this period. For more details about time variations in these parameters see Figure \ref{d2}(iii).\\	
Figure \ref{d3}(i - iii) is plotted in the same pattern with corresponding parameters as previous two. It represents superposed results for 13 three-step decrease GS events. Even after superposed averaging Dst has primarily three steps of dip in the \m\, with the first and second dips accounting more than 75\% of the \m\ duration.\\
According to Figure \ref{d3}(i), the IMF Bz progresses in three steps, to reach maximum (-ve) Bz value in three steps. Interestingly almost similar is the increase pattern in IMF B. Bz begins flipping to southward 2 hours earlier than the MP onset and peaks 4 hours earlier than the corresponding Dst peak values.
The IMF magnitude B begins to increase earlier than MP onset and remains high for a quite long time. The $\sigma$B$_{\mathrm{m}}$, $\sigma$B$_{\mathrm{v}}$ \& proton density $\rho$ reach their maximum around the storm onset time. These three parameters increased for two to five hours until they reached their peak (See Table \ref{tabl4}).  
The IP plasma speed v begins to increase before MP onset and peaks of v occurred $\sim$4 hours before the Dst$_\mathrm{min}$. All SW plasma and field parameters begin a few hours before the MP onset time.\\
According to Figure \ref{d3}(ii) changes in all plotted parameters begin 2 to 4 hours before the MP onset time. 
With the exception of three step dips in its \m\, Figure \ref{d3}(iii) contains the same CFs as in Figure \ref{d1}(iii).
All product CFs begin one to four hours before MP onset time and peak some time with in the MP duration. All the plotted CFs develop during the \m\, in steps.\\
Figure \ref{d4}(i - iii) represents 17 multi-step decrease GS events underwent SEA. MP has a long duration ($\sim$22 hours) and slowly progressing MP in average profile.\\
According to Figure \ref{d4}(i), Bz has multiple steps decreases, which reaches its (-ve) peak and starts recovering, during the later parts of MP. Bz begins flipping to south 3 hours earlier than the Dst onset and peaks 2 hours earlier than the Dst$_\mathrm{min}$ peak value. The IMF magnitude B begins to increase earlier than t$_0$ and remains high for several hours. The parameters $\sigma$B$_\mathrm{m}$, $\sigma$B$_\mathrm{v}$ \& proton density $\rho$ begin to enhance 1 to 7 hours earlier than storm onset time.
$\sigma$B$_\mathrm{m}$ and $\sigma$B$_\mathrm{v}$ are at their maximum during the storm onset time. When the Dst begins to fall, both begins to decrease, although it remains significantly high during initial half period of the \m. 
These three parameters increased for one to seven hours until they reached their peak (See Table \ref{tabl4}). All SW plasma and field parameters begin a few hours before the MP onset time. Another noticeable feature of this plot is that Ey is much enhanced, although fluctuating during the MP.\\
The parameters plotted in \ref{d4}(ii) are same as those in \ref{d1}(ii), with the exception of slower multiple step dips in Figure \ref{d4}(ii) during the \m\ of GSs. All the parameters begin to change 1 to 2 hours before the MP onset time. 
With the exception of slower multiple step dips in its \m\, as shown in Figure \ref{d4}(iii), corresponding parameters as the same as for Figure \ref{d1}(iii). Changes in plotted functions start one to four hours before MP onset time. Their variations during the MP are not smooth but proceed in mult-steps. \\
Figure \ref{d5}(i - iii) is similar in plotted to parameters of Figure \ref{d1}(i - iii), which represents the average profile of all 57 GS events underwent SEA. Most of the features discussed in Figures (\ref{d1}-\ref{d4}) are visible in the averaged plot, with difference in amplitudes, duration in the MP, their time variations as it is the SEA averaged profile of all the four groups of GS.\\	
According to the results of the SEA, one step dip storms are more intense, with an averaged Dst$ _\mathrm{min}$ of $-155$ nT (red colored), followed by three step dip storms with Dst$ _\mathrm{min}$ of $-124$ nT (green colored), multiple step decrease storms with Dst$_\mathrm{min}$ of $-110$ nT (cyan colored), and two step decrease storms with Dst$ _\mathrm{min}$ of $-102$ nT (blue colored), (see Figure \ref{Dst_group} and Table \ref{tabl3}). We can generalize with exceptions, that GSs with the shorter \m\ duration, are in general more intense, while storms with the longer \m\ duration are less intense.
For all groupings of storm events, Bz reaches its peak between the time of start and the time of Dst minimum. Our findings are similar with the previous studies \citep[e.g.,][]{2015RAA....15...85R}.

\begin{table*}

\caption{\small Onset and peak time lags between geomagnetic index Dst and SW plasma and field parameters, their derivatives as well as other geomagnetic indices. From $2^\mathrm{nd}$ to $6^\mathrm{th}$ column represent onset time lags and the last five columns shows peak time lags. The negative onset timings implying that the geomagnetic index Dst starts decreasing earlier than those parameters, and negative peak time lag shows that Dst reaches its minimum earlier than the other parameters; while zero values implying that storm onset time and peak time coincided with other parameters. \label{tabl4}}
\centering
\begin{tabular}{l cccccccccc}
\hline\hline
\textbf{Parameters}  & \multicolumn{5}{c}{\underline{\textbf{Onset time lag (hour)}}}  &\multicolumn{5}{c}{\underline{\textbf{Peak time lag (hour)}}}\\
&d$_1$&\textbf{d$_2$}&\textbf{d$_3$}&\textbf{d$_4$}&\textbf{d$_5$}&\textbf{d$_1$}&\textbf{d$_2$}&d$_3$&d$_4$&d$_5$\\
			\hline\hline
			Bz &3 &6 &2 &3 &1 &5 & 5& 4&2&2\\ 	
			B &3 & 4&1 &1 &3 & 3& 3&3 &5&2\\ 			
			$\sigma$B$_{\mathrm{m}}$ &6 &4 &3 & 1&6 & 9& 11& 12&22&11\\ 		
			$\sigma$B$_{\mathrm{v}}$ &5 & 5& 2& 2& 6& 8& 6& 8&21&11\\  		
			$\rho$ &6 &4 &5 &7 &6 &7 & 10& 12&20&10\\ 	
			v &3 &3 &1 &1 &1 & 4& -9& 4&10&3\\
			P &5 &4 & 4& 1&3 & 7&10 & 11&20&10\\  		
			Ey &3 &3 &2 & 2&3 &5 & 5&7 &5&6\\ 		
			$\beta$ &11 &-3 &2 & 2&2 &-10 & -4&1 &2&-9\\ 		
			ap &3 &4 &3 & 2&4 & -1& 5&4 &1&4\\
			AE &2 &1 &4 & 1&1 &3 & 0&4 &1&1\\ 		 		  		
			vB &3 & 4&4 &1 &3 &3 & 4&3 &8&6\\ 		
			vEy &2 &3 &1 &1 &3 & 5& 5&7 &5&6\\ 		
			vBz$^2$ &2 &3 &2 &1 &2 &5 & 5& 7&5&8\\
			v$^{\frac{4}{3}}$Bz &5 &3 &2 &4 &5 &5 & 5&7 &5&8\\  		
			v$^{\frac{4}{3}}$Bz$\rho$ &4 &6 &2 &2 &3 & 5& 6& 8&20&7\\  		
			v$^{\frac{4}{3}}$Bz$\rho^{\frac{1}{2}}$ &4 &5 &2 & 2&3 & 5& 5&7 &3&7\\ 		
			
			v$^{\frac{4}{3}}$B$\rho^{\frac{1}{2}}$ &5 & 5&4 & 4&4 & 6& 6&8 &-3&7\\
			\hline\hline
			\multicolumn{10}{l}{\footnotesize d$_5$, represents total 57 events \m\ duration.}\\
			\multicolumn{10}{l}{\footnotesize d$_1$, d$_2$, d$_3$ \& d$_4$ are adopted from Table \ref{tabl1}}\\
		\end{tabular}
	\end{table*}
	
		\subsection{Relation between peaks of Dst and Bz}\label{sub:DstBz}
	
\begin{figure}
		\includegraphics[height=7.2cm, width=0.51\textwidth]{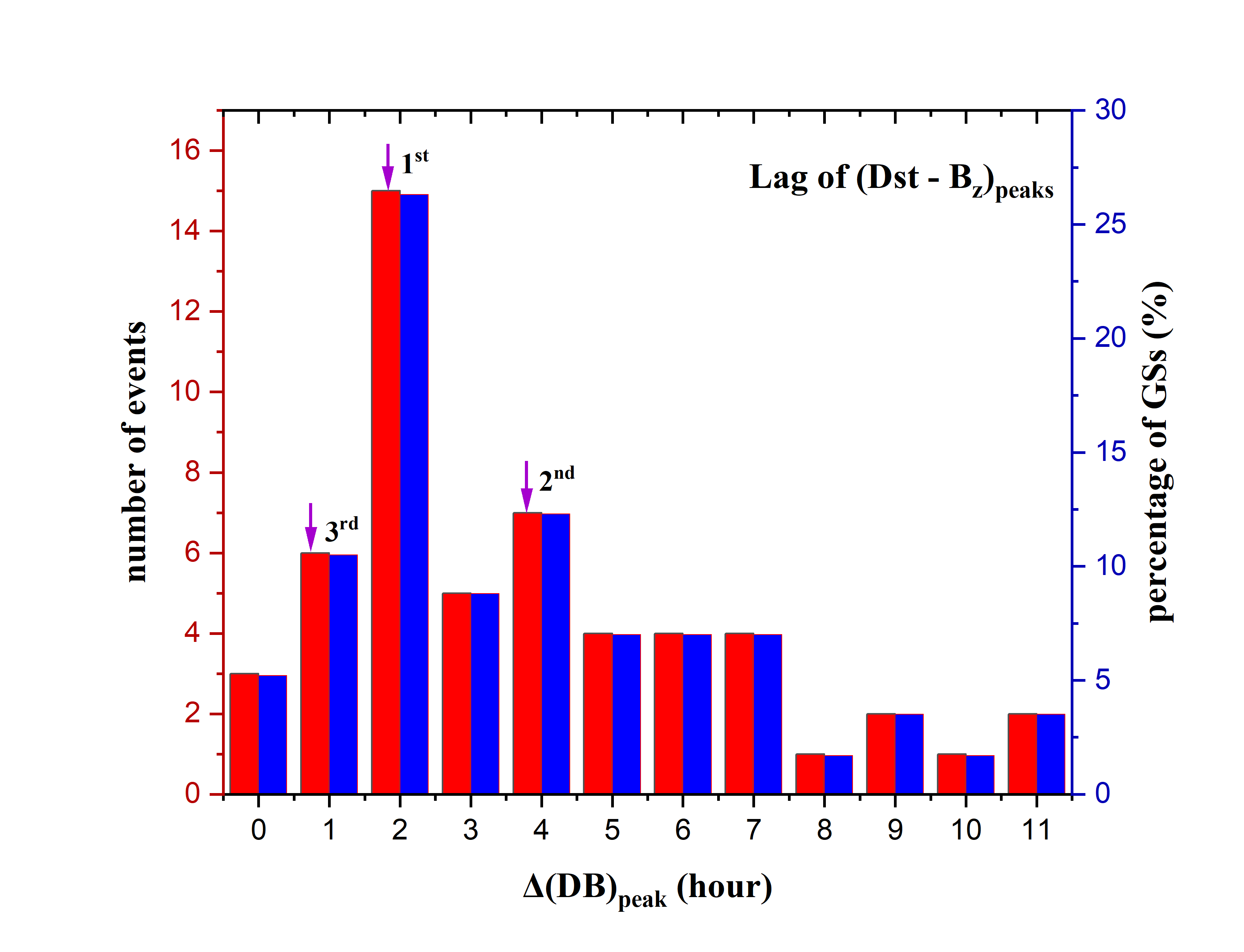}
		\caption{\small The distribution of the number of GSs (left axis) and the percentage of GSs on the right axis are shown in the histogram plot as functions time lag $\Delta$(DB)$_\mathrm{peak}$. The majority of storms are seen to have delays of between one and four hours between the Dst and Bz peaks.}\label{fig:dstbz}
	\end{figure}

Simplest and most studied single SW merging term/parameter in SW-magnetosphere coupling studies is the southward component of the IMF (-Bz), starting from early 1960's \citep[e.g.,][]{1961PhRvL...6...47D, 1971JGR....76.5189A, 1972JGR....77.2964T}, is explored even now in different SW conditions and using different geomagnetic activity indices/parameter \citep[e.g.,][]{2023AdSpR..71.1137B}.	
The IMF Bz is considered to be the main driver for the \m\ GS development \citep[e.g., see][and references therein]{1994JGR....99.5771G, 2018cosp...42E2795S} which reaches its peak (-ve) value a few hours ahead of GS peak value (Dst$ _\mathrm{min}$). 
This time lag, even on average basis, is evident in all five (Figures \ref{d1}- \ref{d5}) plots discussed earlier.
The IMF Bz turns southward (negative) value near simultaneously at the (about 1 hour earlier) onset of MP of GS and remains enhanced during the MP.
In Figures \ref{d1}(i) to \ref{d5}(i) the peak value of southward Bz are ahead of the Dst$ _\mathrm{min}$ values by a few hours. The peak value Dst has a time lag with (-Bz) peak value is less than five hours as seen from SEA plots. Consistent with \citep{2005GeoRL..3218103G} who reported delay of about 2 hours for individual studied events. This SW parameter also appears commonly in most of the earlier proposed CFs in one form or the other. Thus, a discussion of its time lag with GS MP onset time (start time os sudden and fast decrease in Dst values) and with time of most intense geomagnetic activity (time of Dst$ _\mathrm{min}$) during the development of GS both on average and individual event basis is useful for the understanding of SW-magnetosphere coupling.\\
Figure \ref{fig:dstbz} shows a histographic representation of the time lag $\Delta$(DB)$_\mathrm{peak}$ (hours) values between Dst and Bz, with the left side of the y-axis colored red for the number of GS events for a given delay time and the right side colored blue represents the percentage. The peak of 26\% of the GS events lags two hours from Bz peak of 15 events, and the second one is four hours with seven events accounting for 12\% of the total events. Furthermore, the majority of the events lag by one to four hours (with an average of 1 hour 30 minutes), which is close to earlier results \citep{2005GeoRL..3218103G} by 30 minutes, accounting for 58\% of the total events.\\
We also scrutinized this time lag for individual events in different groups (d$ _{1} $, d$ _{2} $, d$ _{3} $ and d$ _{4} $).
According to our findings, in general, one and two step decrease GS events have a shorter time lag than three and multiple step dip GSs which is consistent with earlier studies \citep[e.g.,][]{1989JGR....94.8835G} they reported time lag of 1 hour for moderate storms and even shorter period for intense storms, which is short duration compared to our result.
It is known that the temporal lag between GS indices and \sw\ parameters for different GSs is didfferent \citep[e.g.,][]{2022Ap&SS.367...10B} however, 18 minutes to 4 hours lag was reported in their study between SYM-H and Bz which is in agreement with our findings. Other previous studies also support this result, e.g., \cite{2015RAA....15...85R, 2018cosp...42E2795S} reported that two to three hours prior to the \m\ of GSs, there is a delay between peaks of Bz and the Dst that can be utilized to forecast the intensity of a GS. Recently, \citep{2023GeoRL..5003151P} calculated the time lag  between different SW parameters and the magnetospheric response for different driver sources as CIR/SH/MC as 40 min, 40 min and 60 min respectively. According to our findings, in general, one and two step decrease GS events have a shorter onset time lag than three and multiple step dip GSs.
%
	\begin{table}
		
		\caption{\small For each of the 57 storm events, the time lag between the geomagnetic index Dst and the southward IMF Bz is recorded. The first column contains the time lag in hours, the second column contains the number of GSs, and the third column contains the percentage of events for a given time lag.\label{tab3}}
		\centering
		\begin{tabular}{l ccc}
			\hline\hline
			$\Delta$(DB)$_\mathrm{peak}$ & GS events& Percentage \\
			(hour)&(number) &(\%)\\
			\hline\hline
			negative &3 & 5.3\\
			0 & 3 &5.3\\
			1 & 6 &10.5\\
			2 & 15 &26.3\\
			3 & 5 &8.8\\
			4 & 7 &12.3\\
			5 & 4 &7.0\\
			6 & 4 &7.0\\
			7 & 4 &7.0\\
			8 & 1 &1.7\\
			9 & 2 &3.5\\
			10& 1 &1.7\\
			11& 2 &3.5\\
			\hline\hline
			\multicolumn{3}{l}{\footnotesize $\Delta$(DB)$_\mathrm{peak}$, is the time lag between Dst and Bz.}
		\end{tabular}
	\end{table}

	\subsection{Relation between geomagnetic indices and SW plasma and field parameters}\label{sub:correlation}
\begin{figure*}
	\centering
		\includegraphics[height=3.6cm, width=5.7cm]{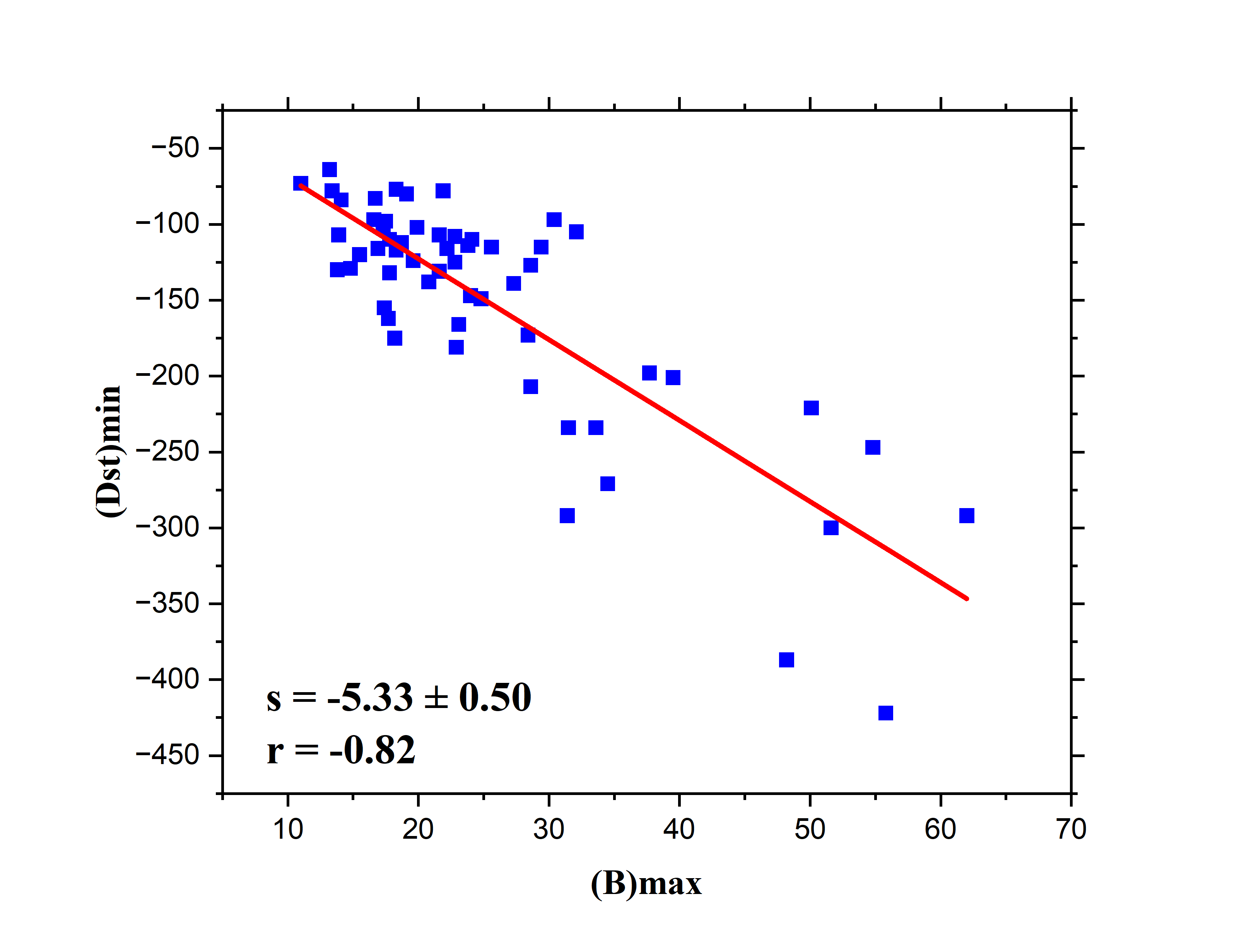}
		\includegraphics[height=3.6cm, width=5.7cm]{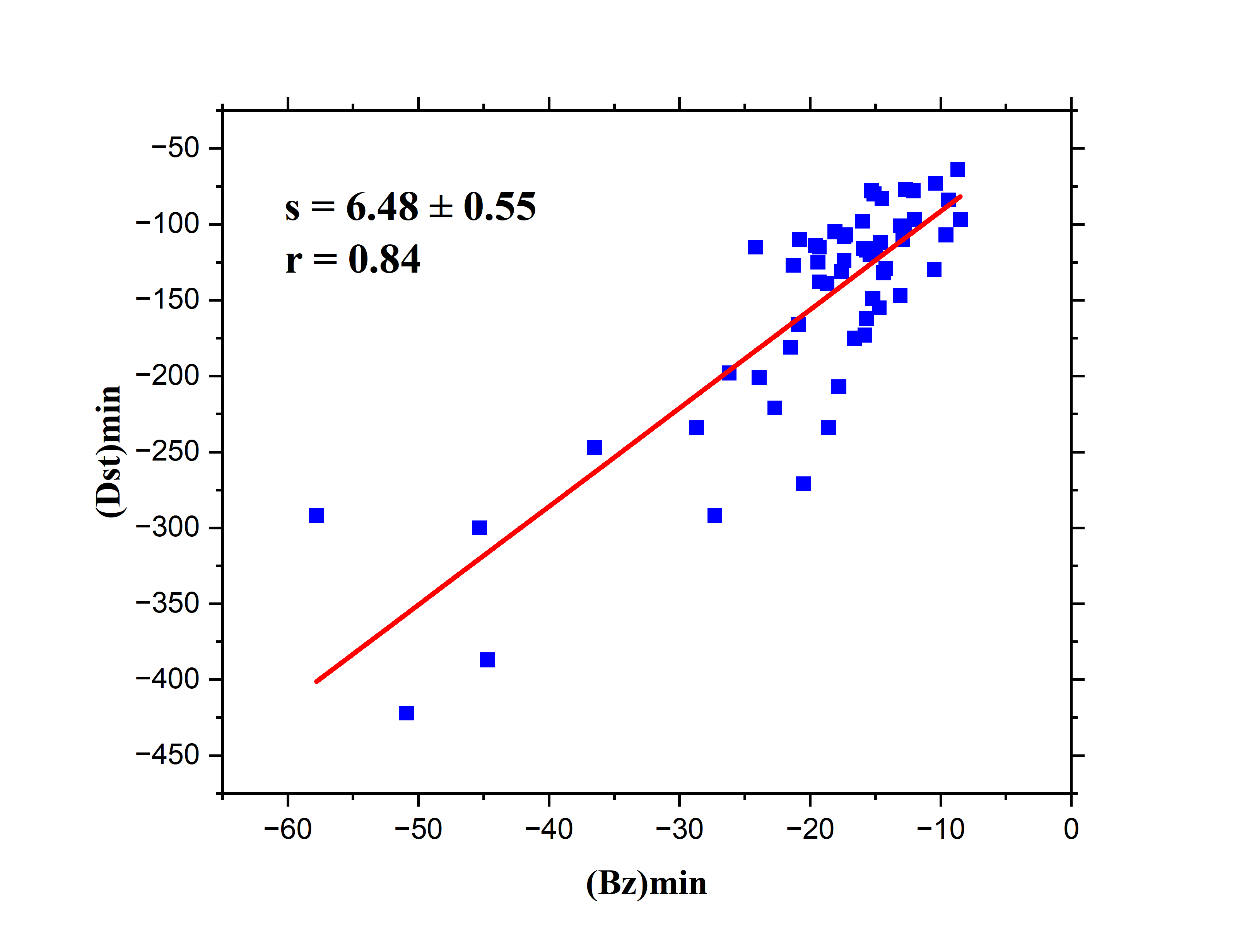}
		\includegraphics[height=3.6cm, width=5.7cm]{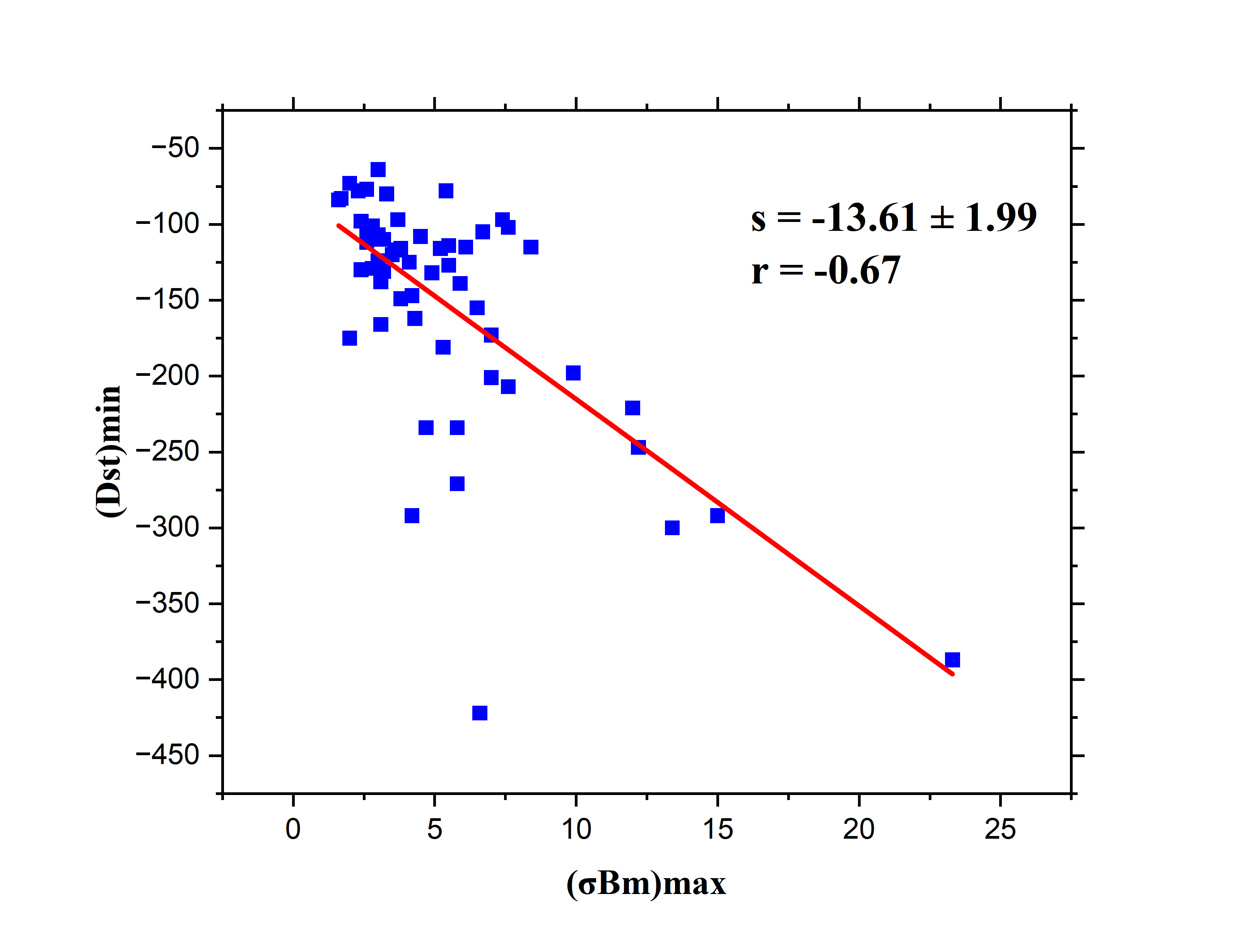}
		\\[\smallskipamount]
		\includegraphics[height=3.6cm, width=5.7cm]{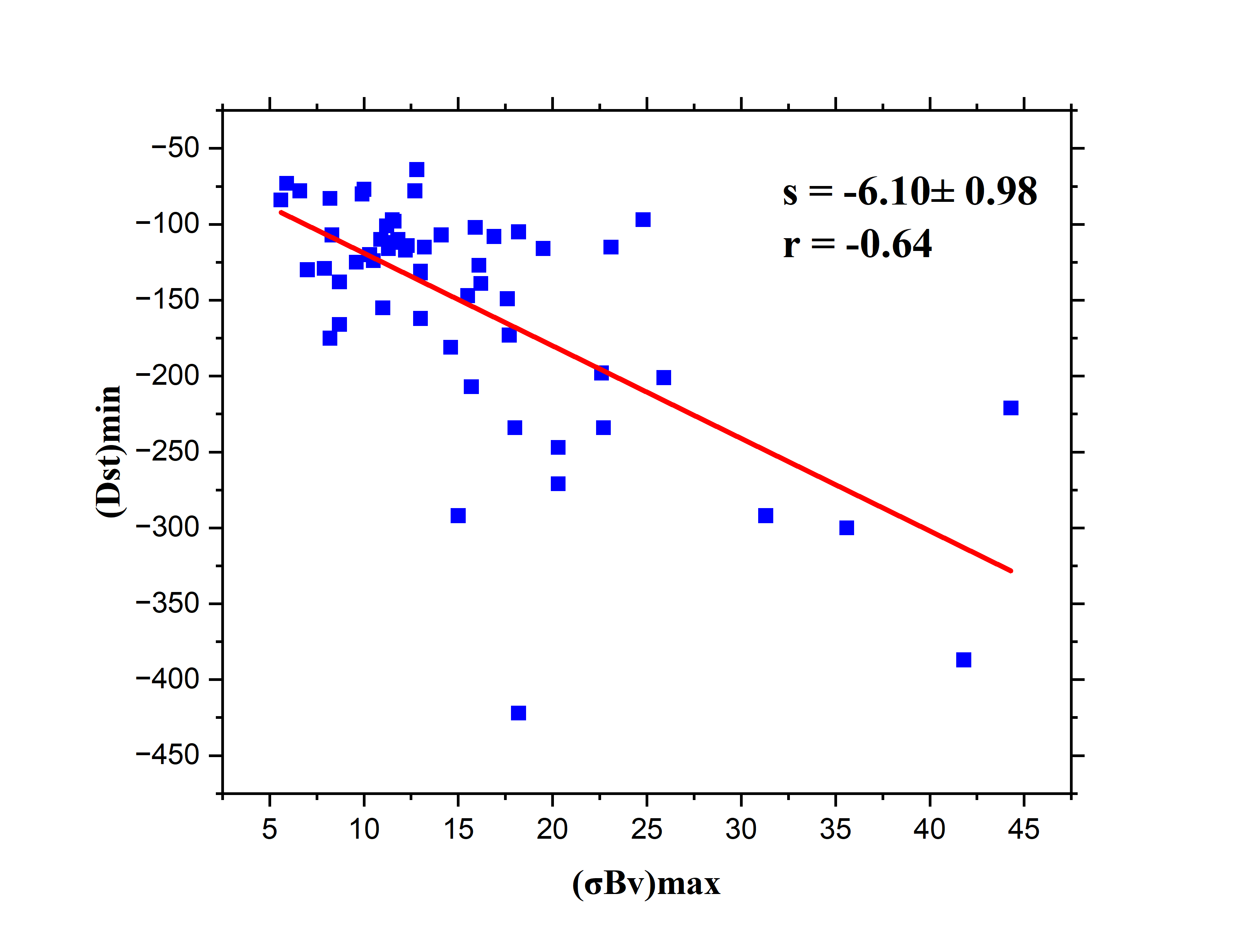}
		\includegraphics[height=3.6cm, width=5.7cm]{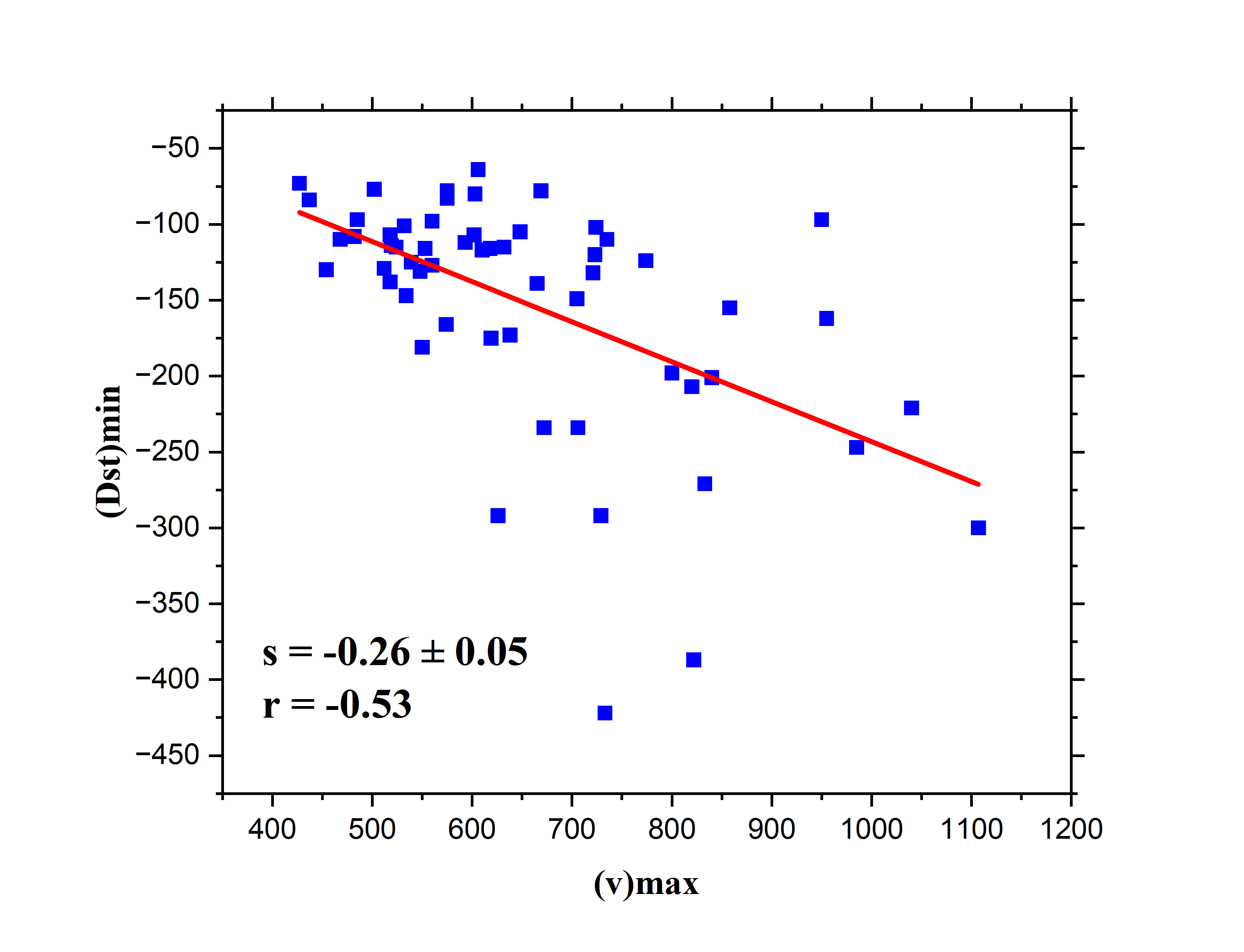}
		\includegraphics[height=3.6cm, width=5.7cm]{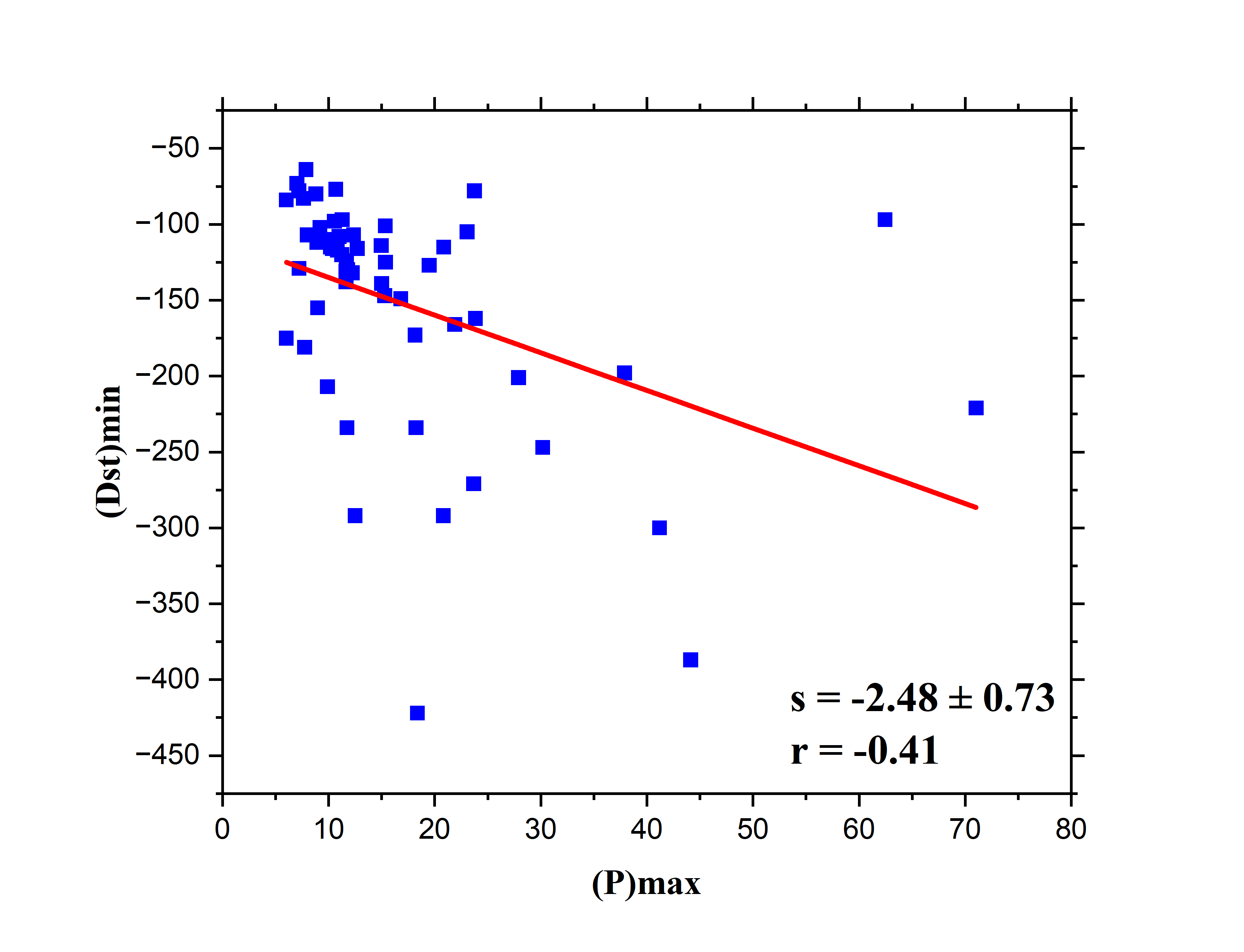}
		\\[\smallskipamount]
		\includegraphics[height=3.6cm, width=5.7cm]{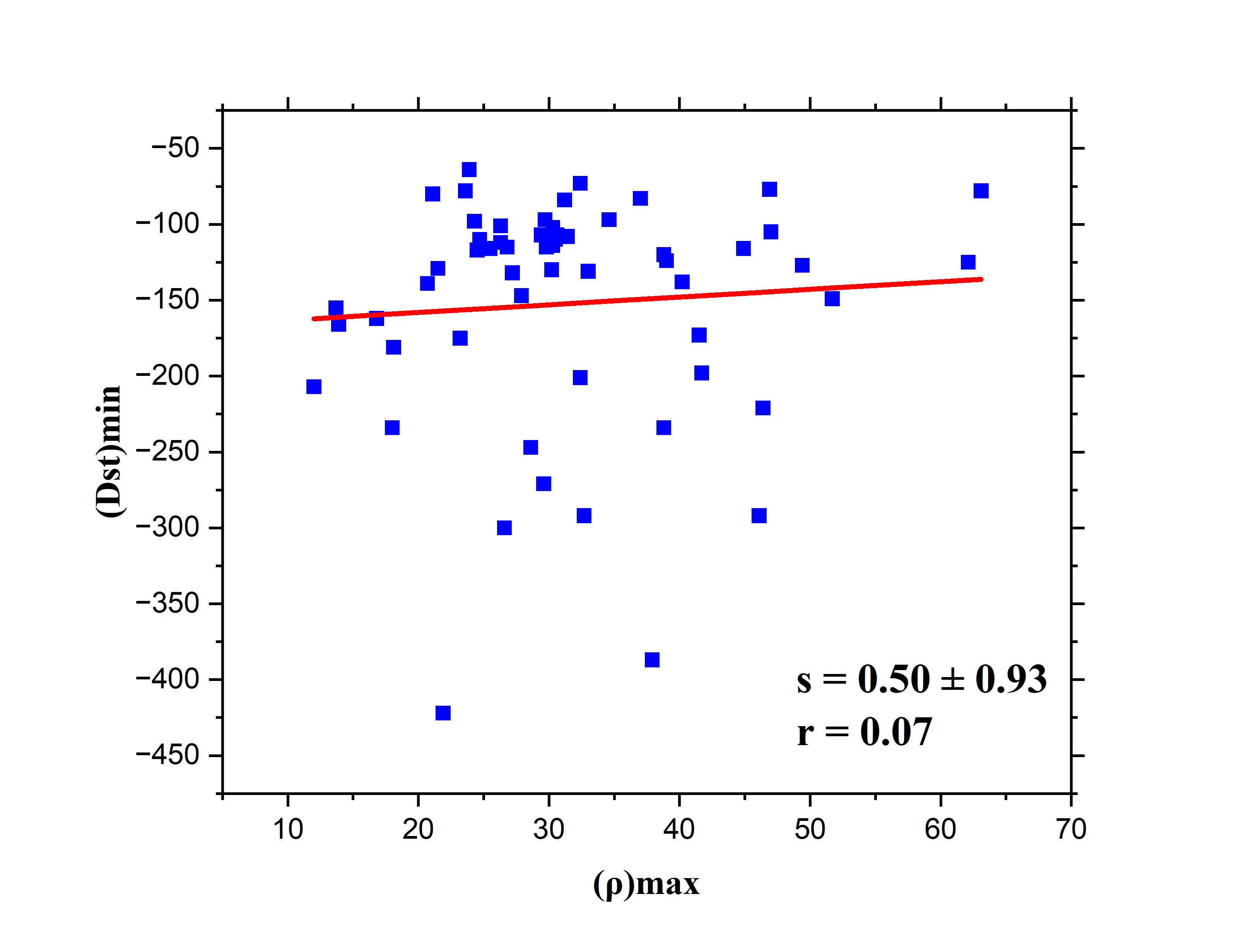}
		\includegraphics[height=3.6cm, width=5.7cm]{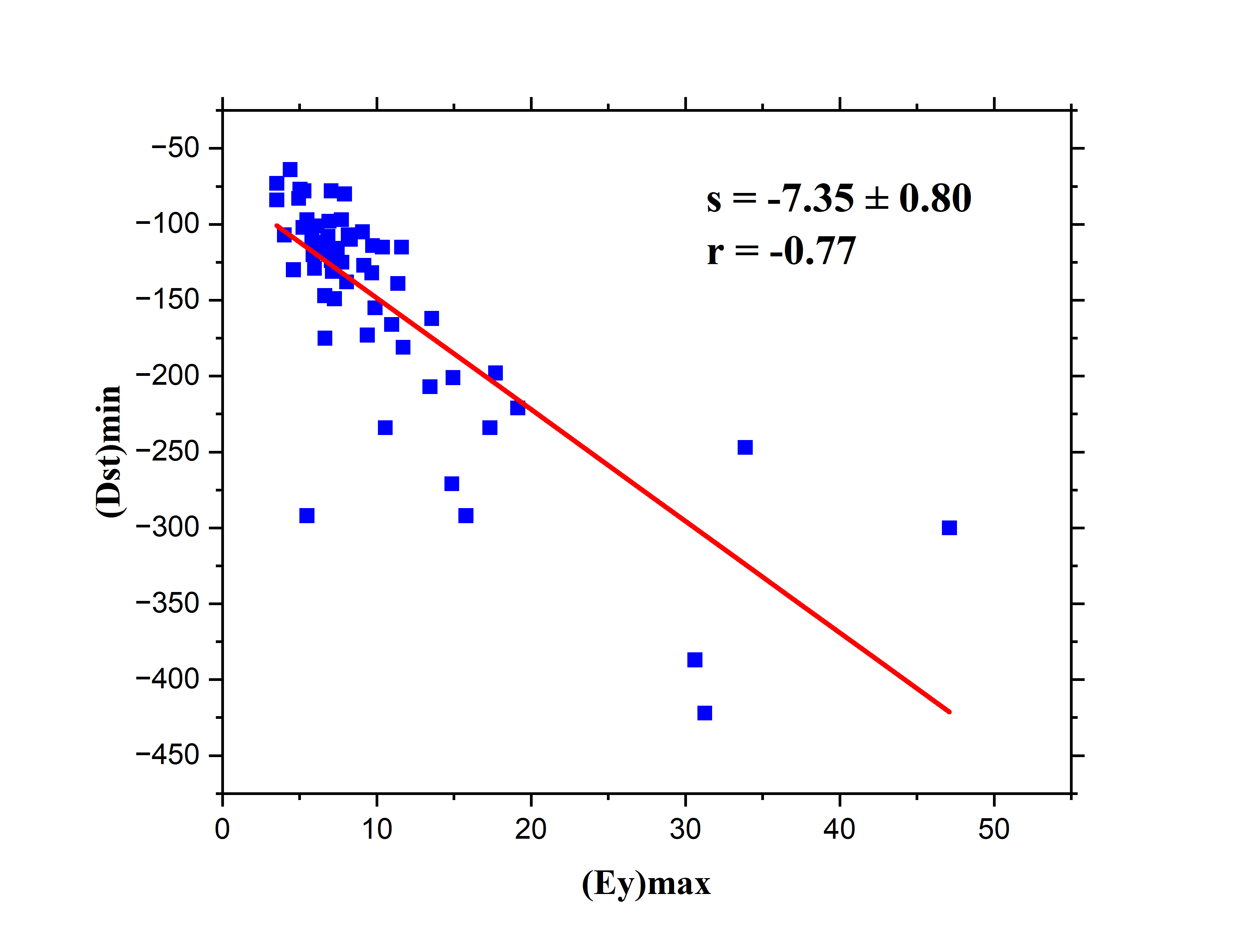}
		\includegraphics[height=3.6cm, width=5.7cm]{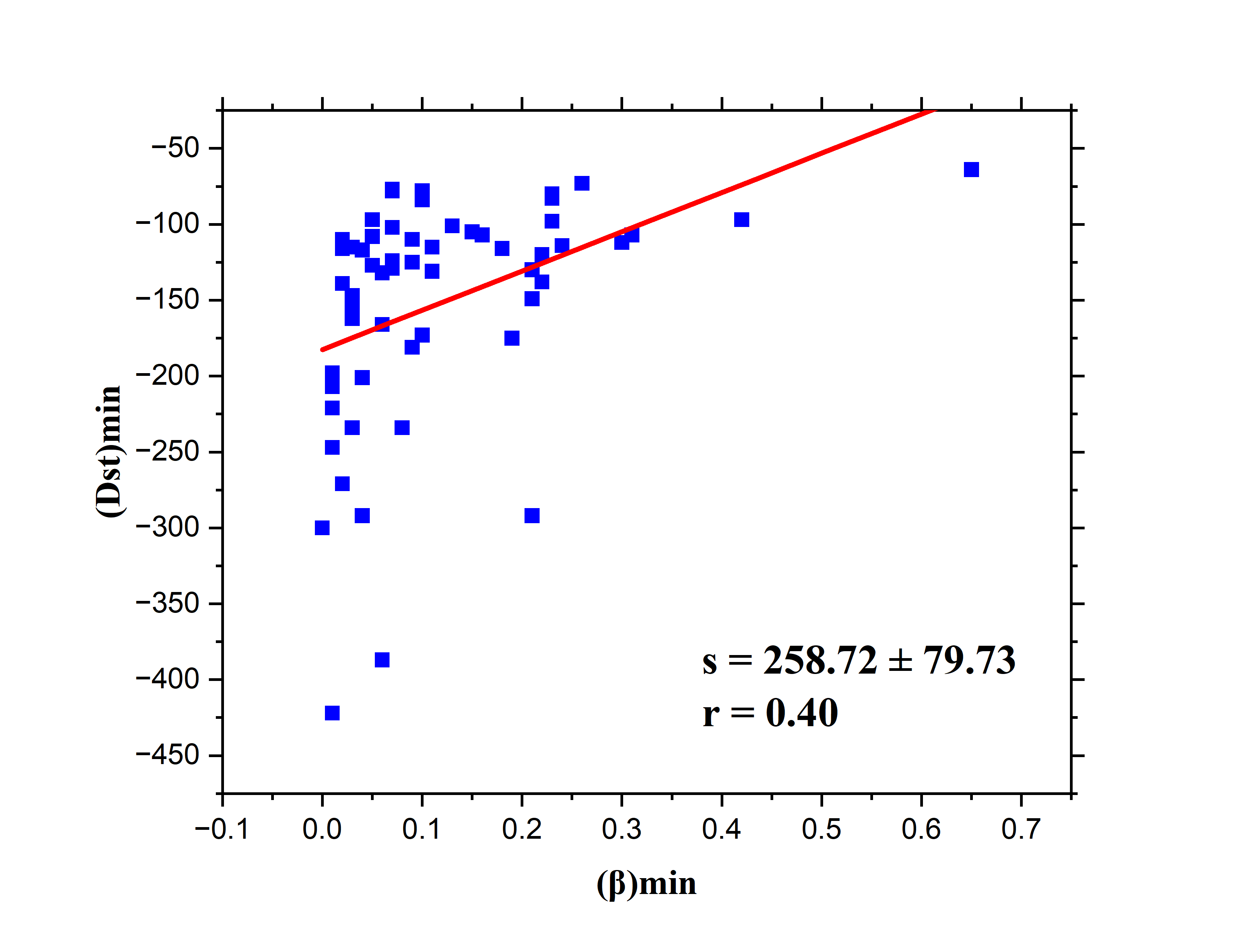}
		\\[\smallskipamount]
		\includegraphics[height=3.6cm, width=5.7cm]{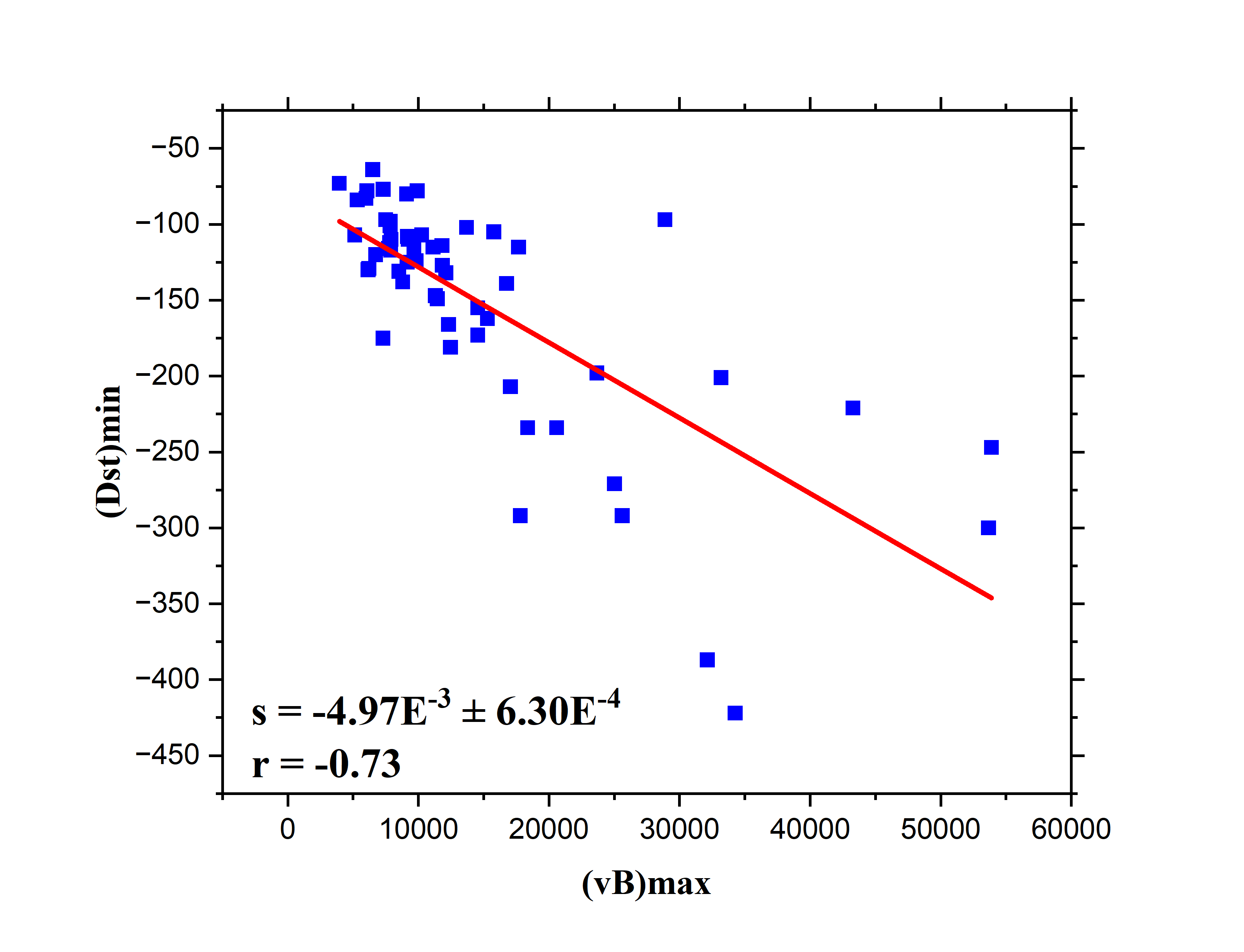}
		\includegraphics[height=3.6cm, width=5.7cm]{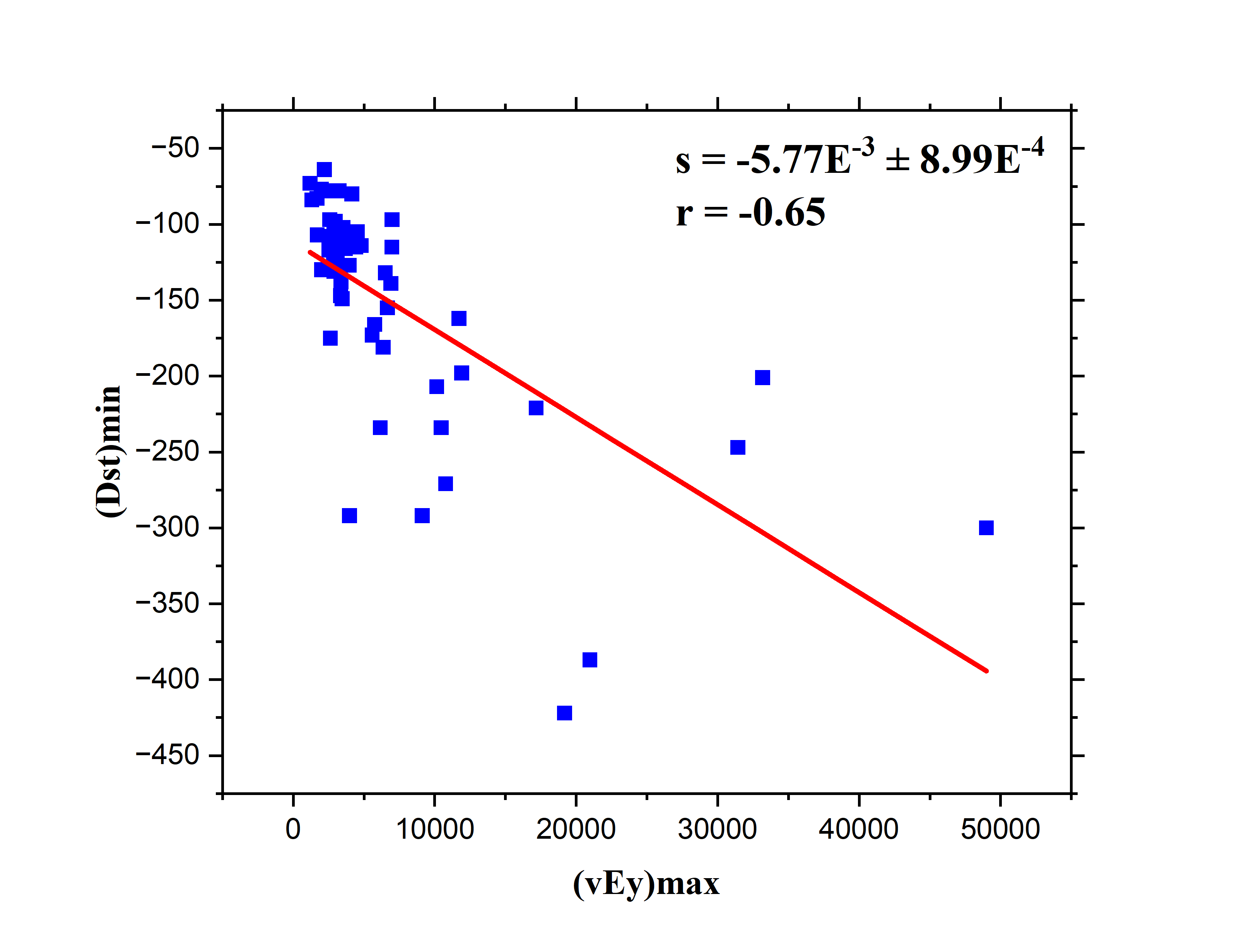}
		\includegraphics[height=3.6cm, width=5.7cm]{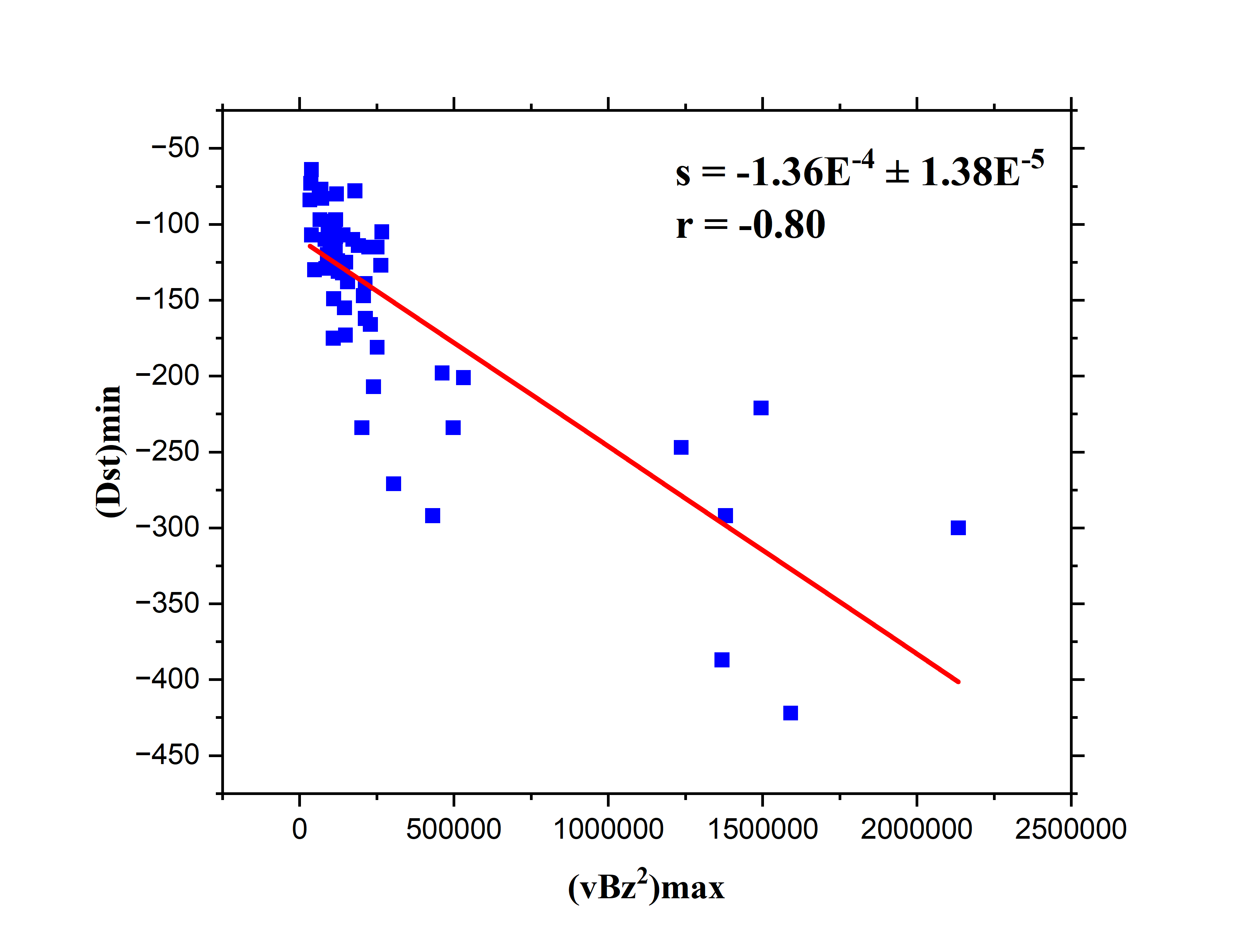} 
		\\[\smallskipamount]    
		\includegraphics[height=3.6cm, width=5.7cm]{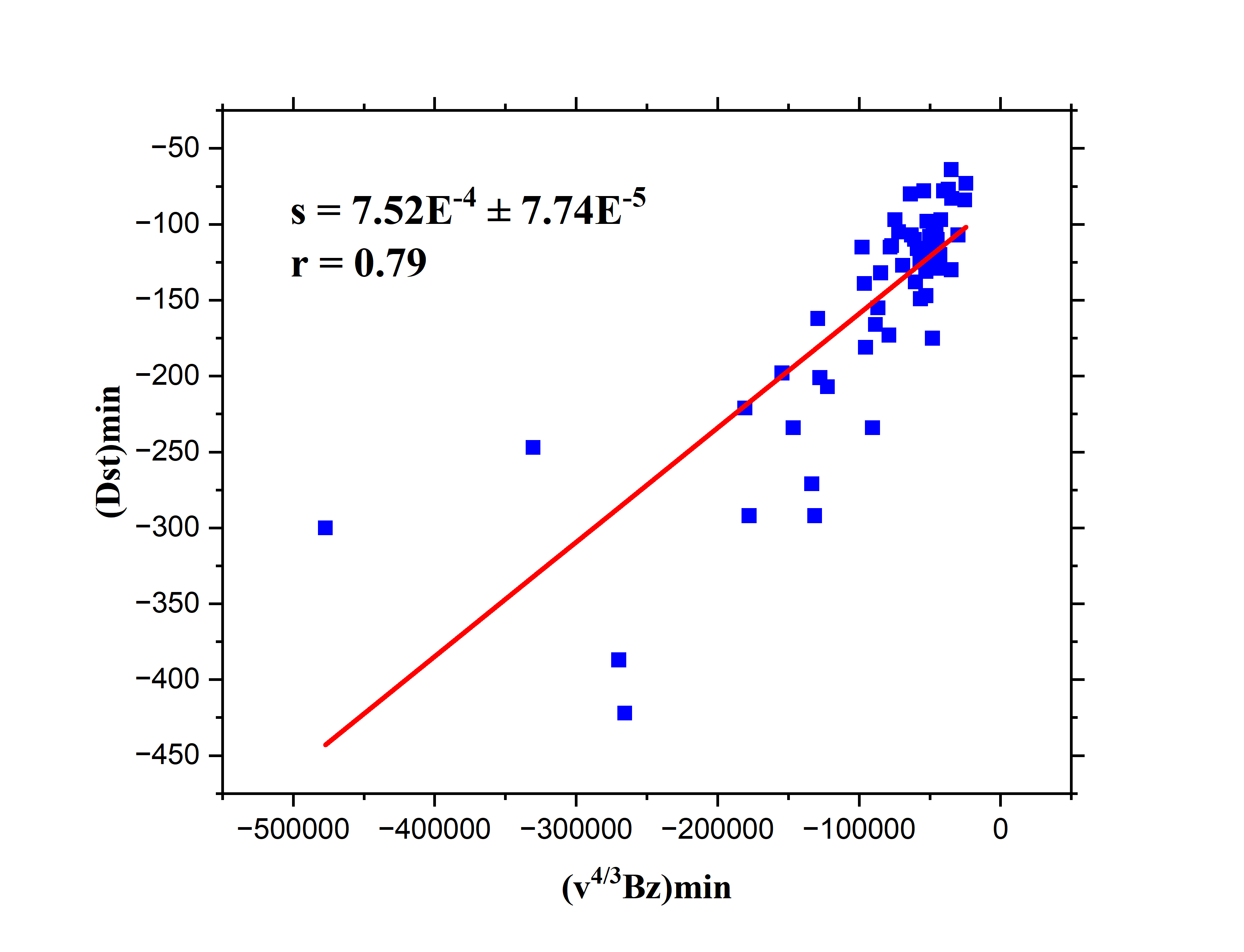}
		\includegraphics[height=3.6cm, width=5.7cm]{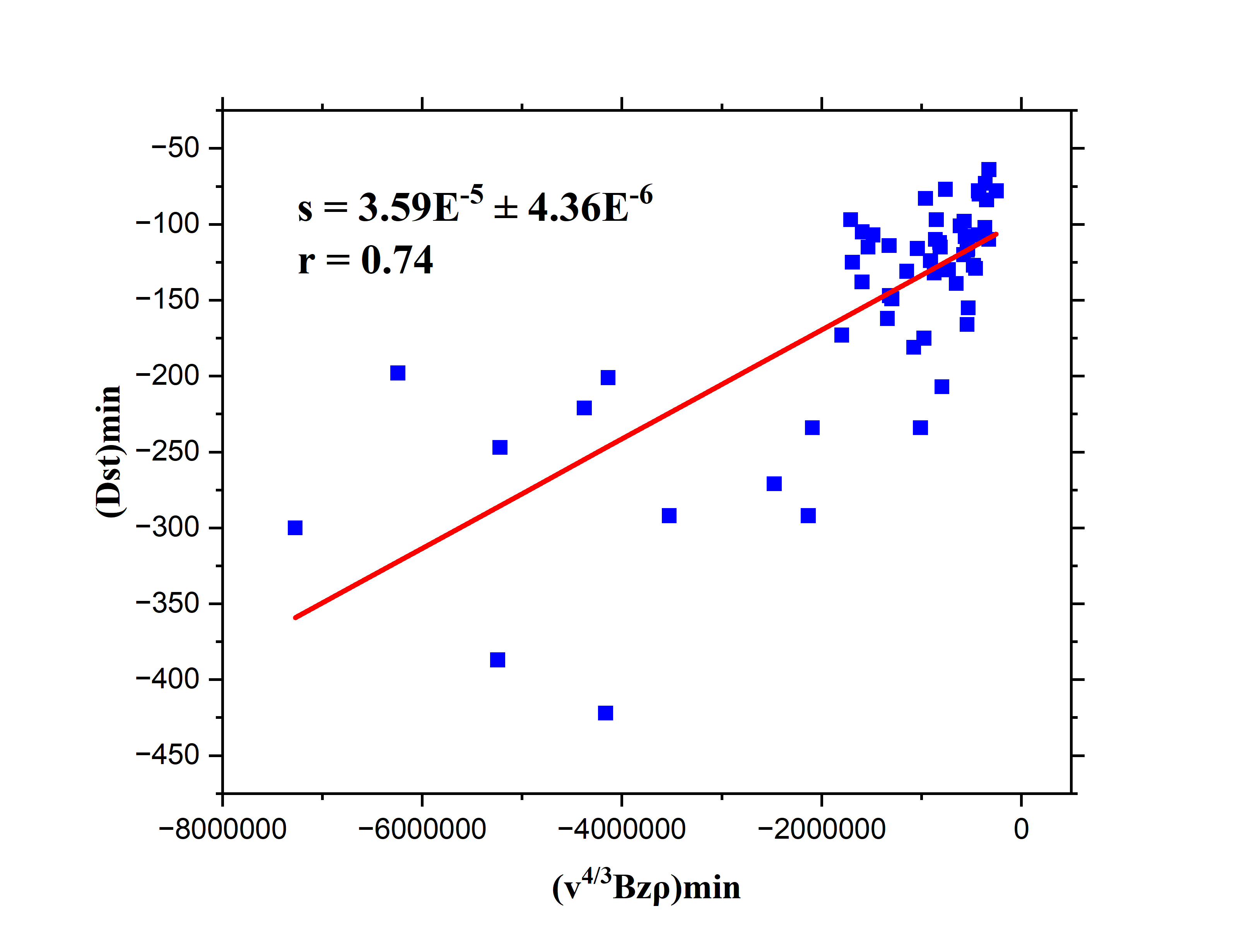}
		\includegraphics[height=3.6cm, width=5.7cm]{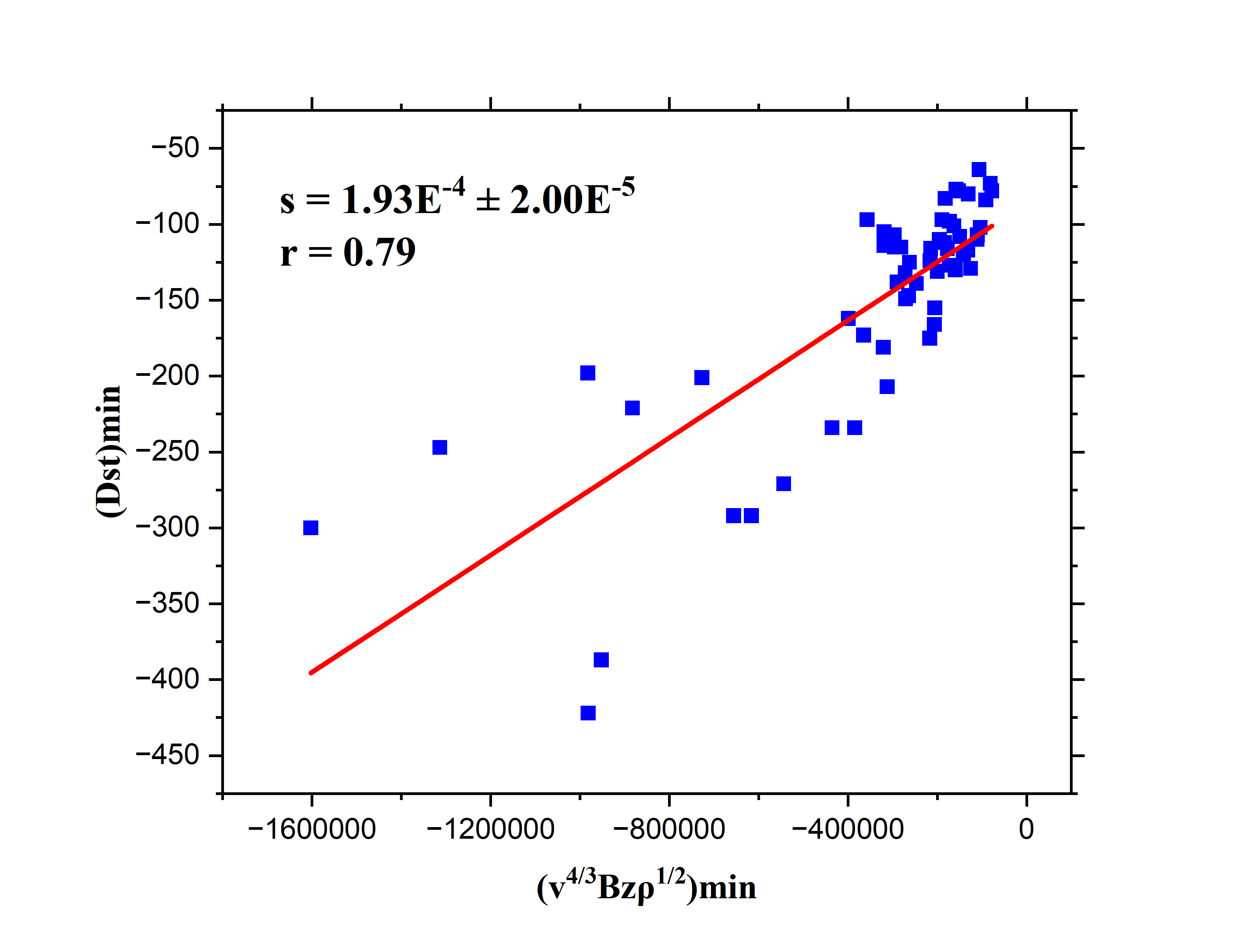}
		\\[\smallskipamount]
		\centering
		\includegraphics[height=3.6cm, width=5.7cm]{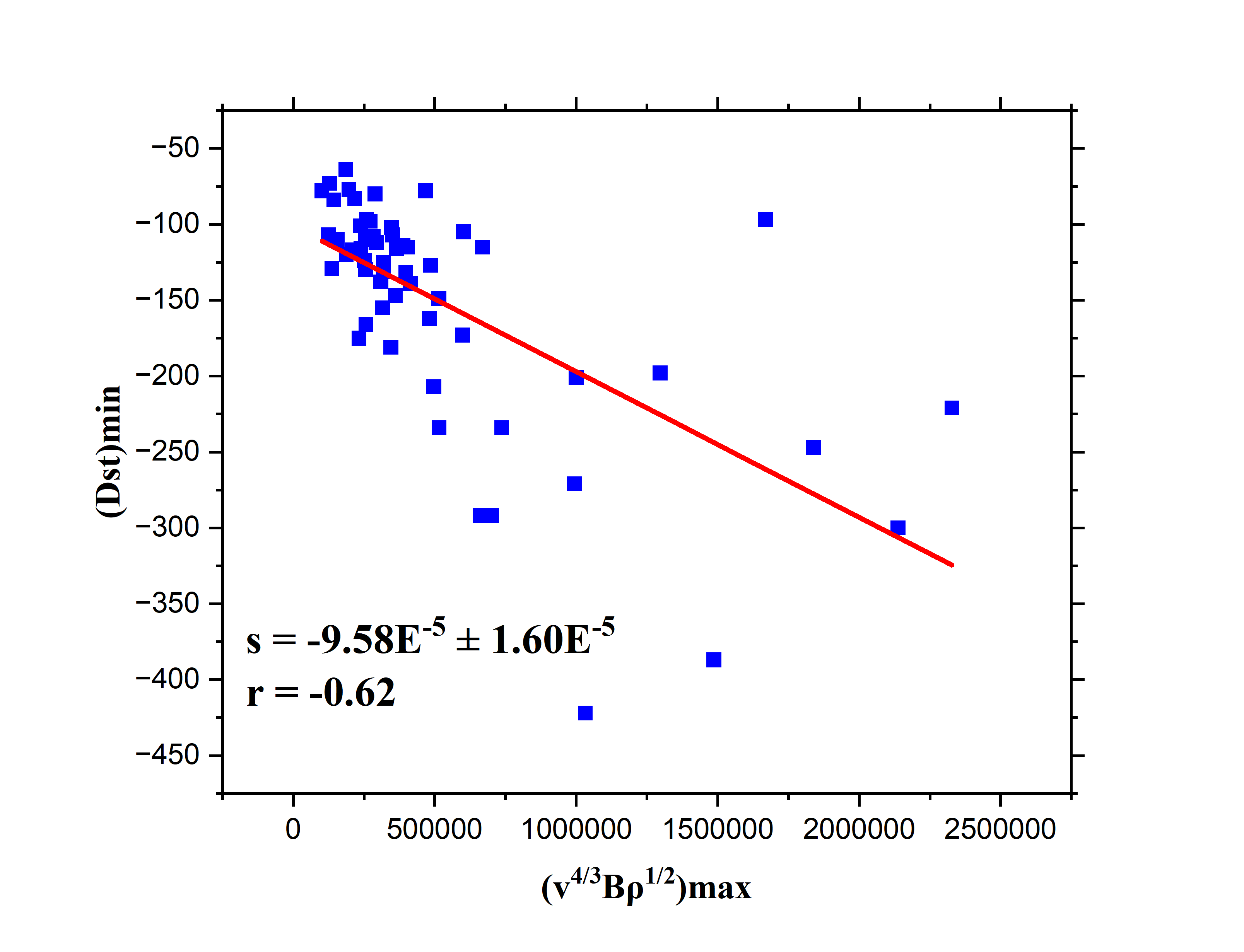}
\caption{\footnotesize The scatter plots depict the best linear fit between the peak values of Dst during GS with the peak values of selected SW plasma and field parameters and derived CFs, from $y=sx+i$, where s is slope and i is y-intercept, r is measure of linear correlation between Dst and other parameters called Pearson's correlation coefficient ($r=\frac{\Sigma(x_i-\overline{x})(y_i-\overline{y})}{\sqrt{\Sigma(x_i-\overline{x})^2\Sigma(y_i-\overline{y})^2}})$, where  $x_{i}=$ values of the values of the x-variable in a sample
			$\overline{x}=$	mean of the values of the x-variable\\
			$y_{i}=$ values of the y-variable in a sample,   
			$\overline{y}=$	mean of the values of the y-variable\\ Dst is strongly related with the single parameter Bz with the best fit equation Dst$_\mathrm{min}=[6.48\pm0.55$](Bz$_\mathrm{min}$)+($-26.70\pm12.03$).}\label{fig:Dstcorr}		
	\end{figure*}

\begin{figure*}
	\centering
		\includegraphics[height=3.8cm, width=5.9cm]{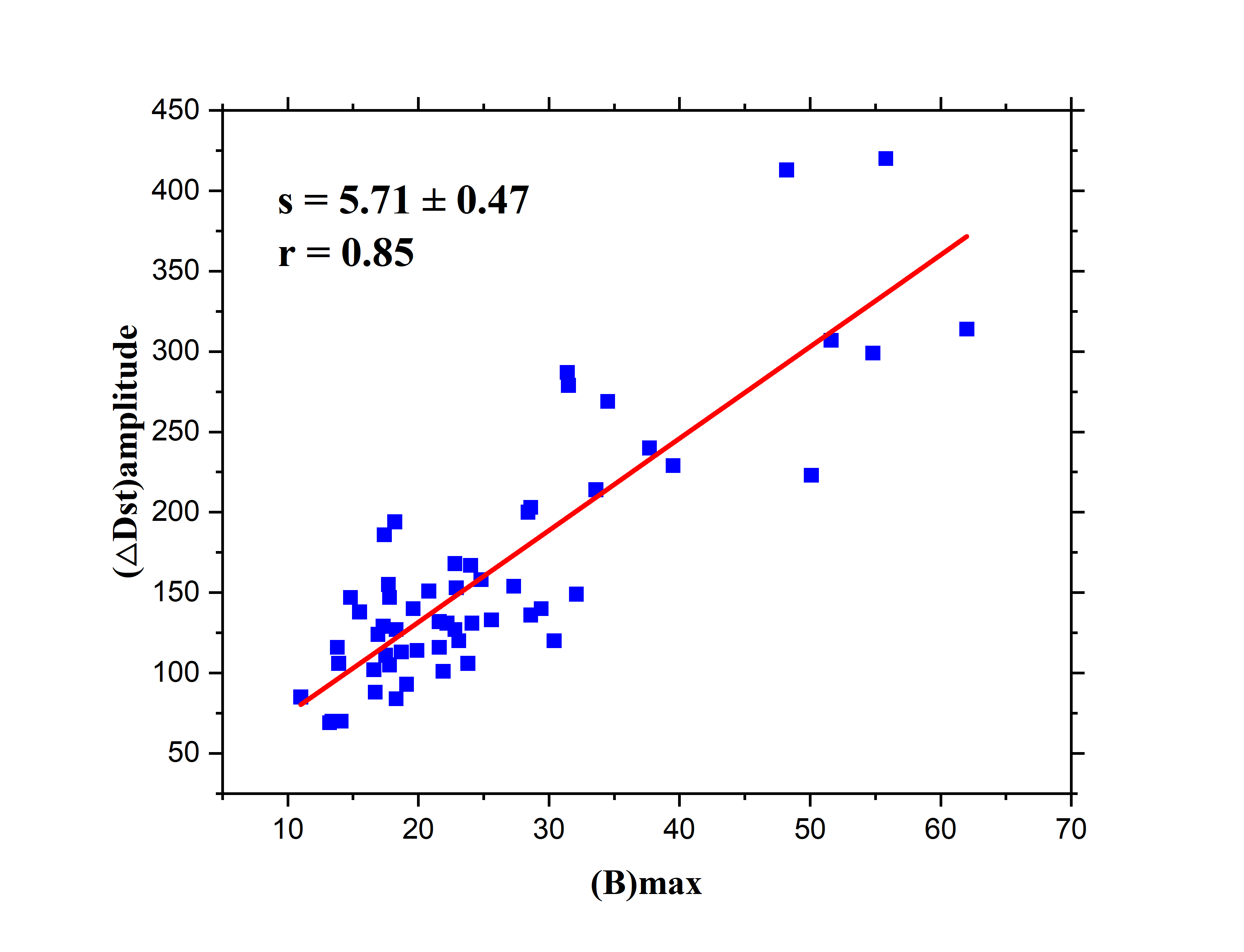}
		\includegraphics[height=3.8cm, width=5.9cm]{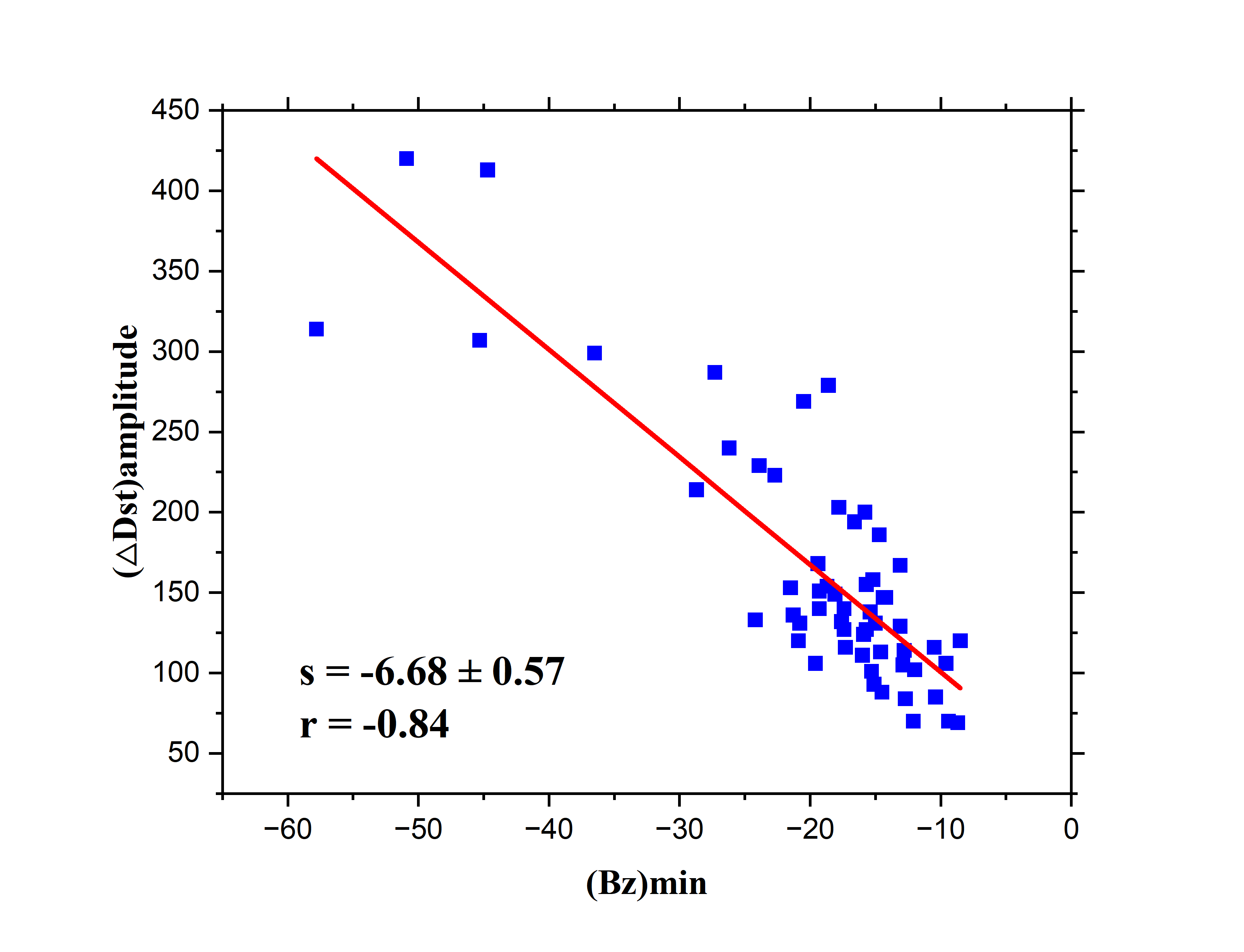}
		\includegraphics[height=3.8cm, width=5.9cm]{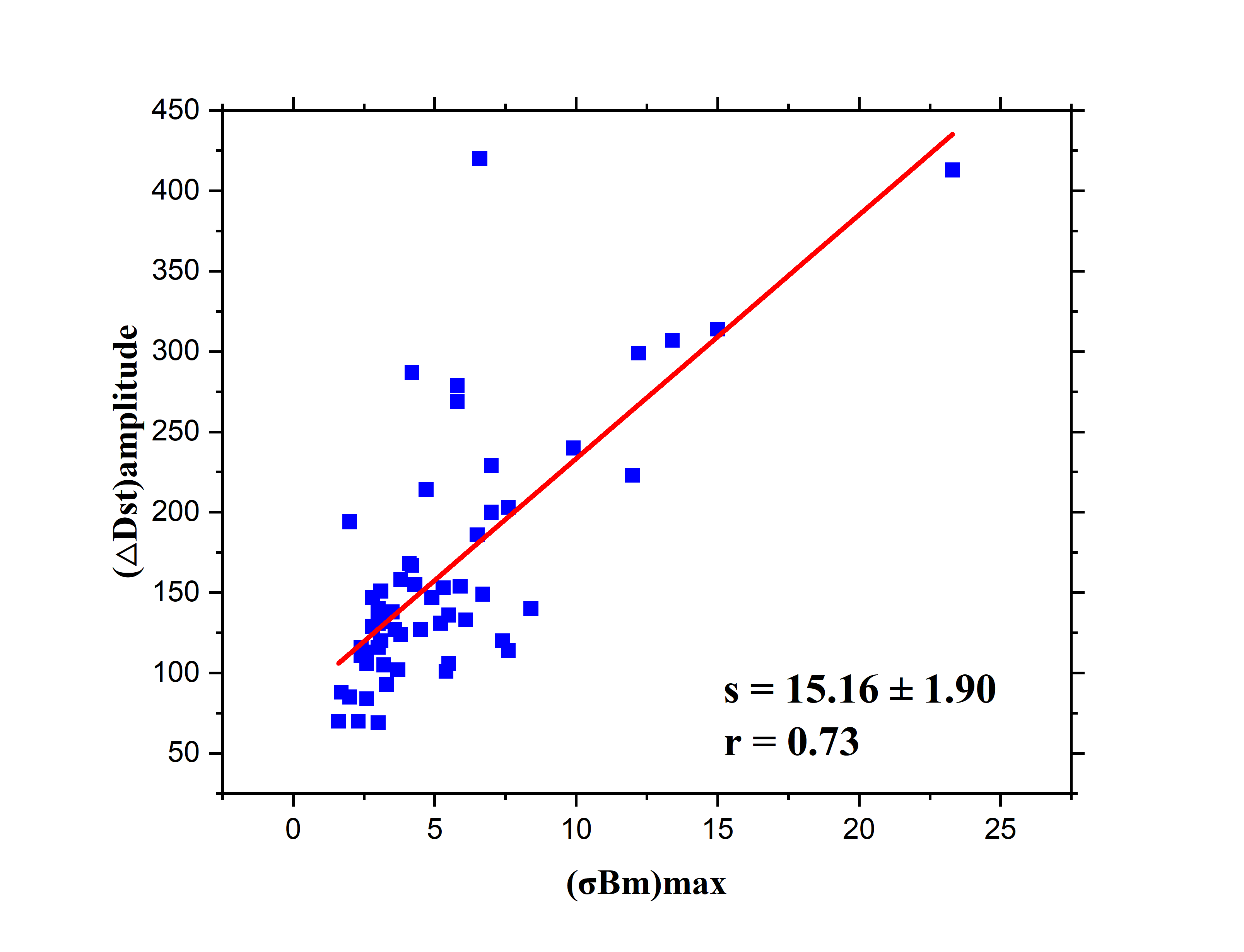}
		\\[\smallskipamount]
		\includegraphics[height=3.8cm, width=5.9cm]{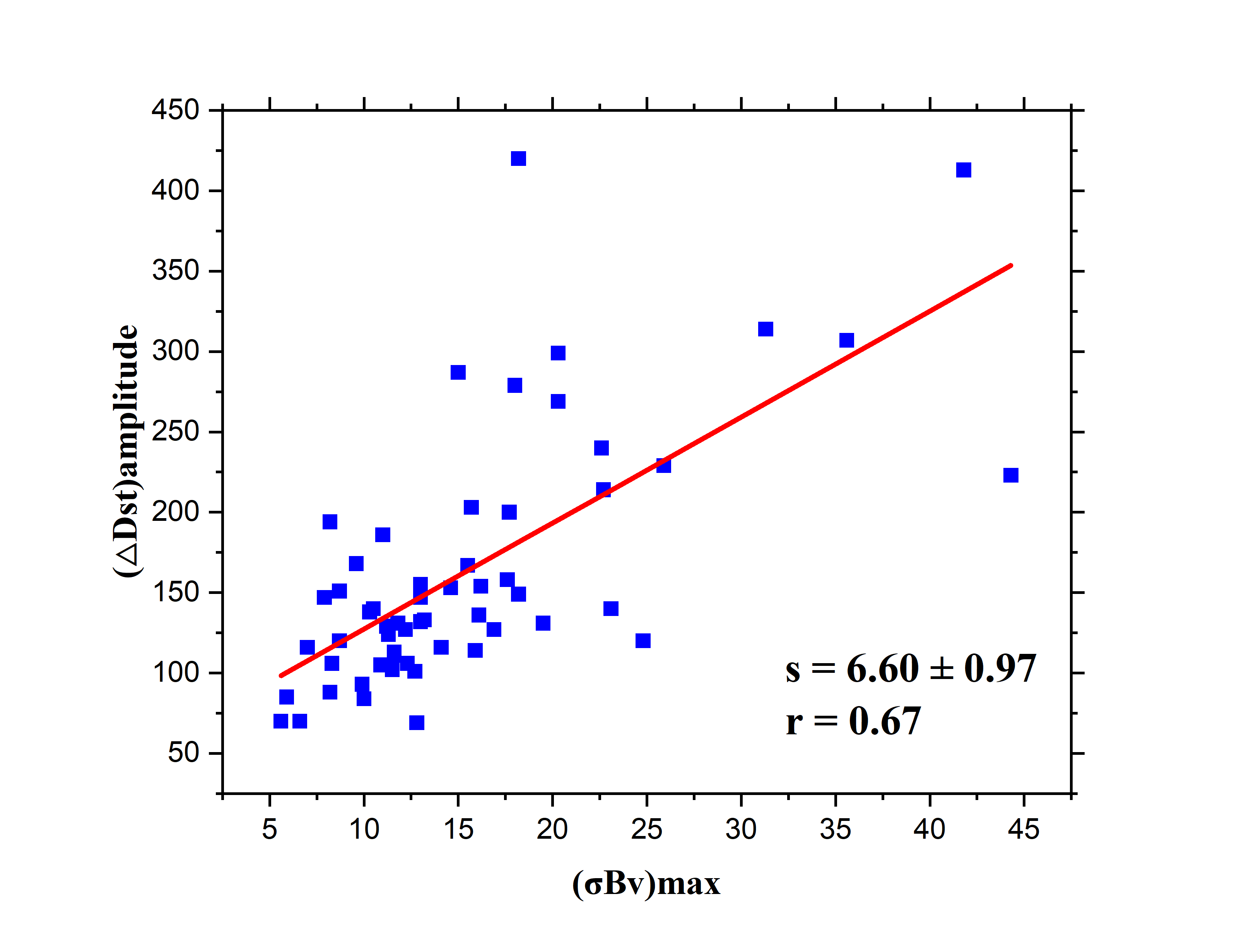}
		\includegraphics[height=3.8cm, width=5.9cm]{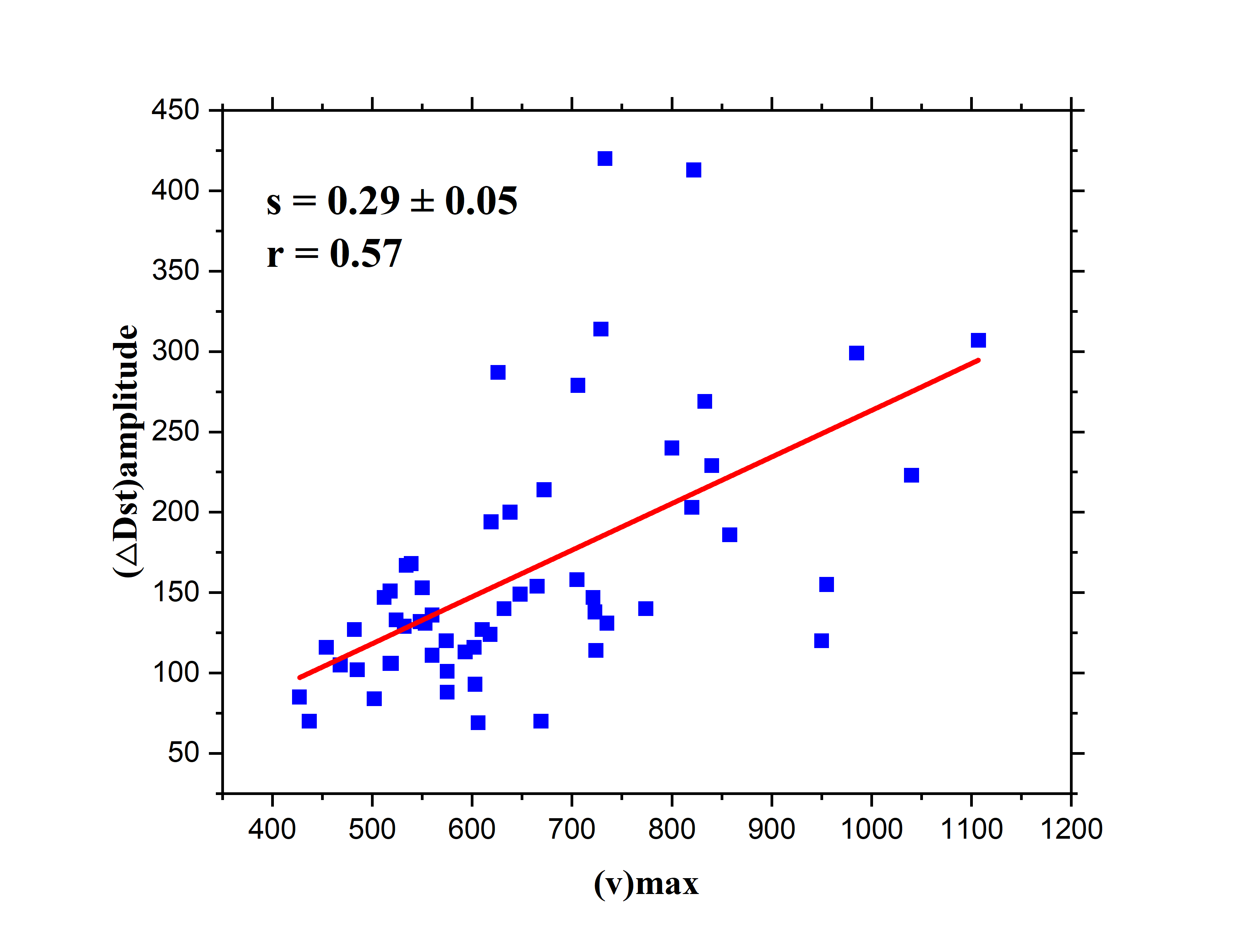}
		\includegraphics[height=3.8cm, width=5.9cm]{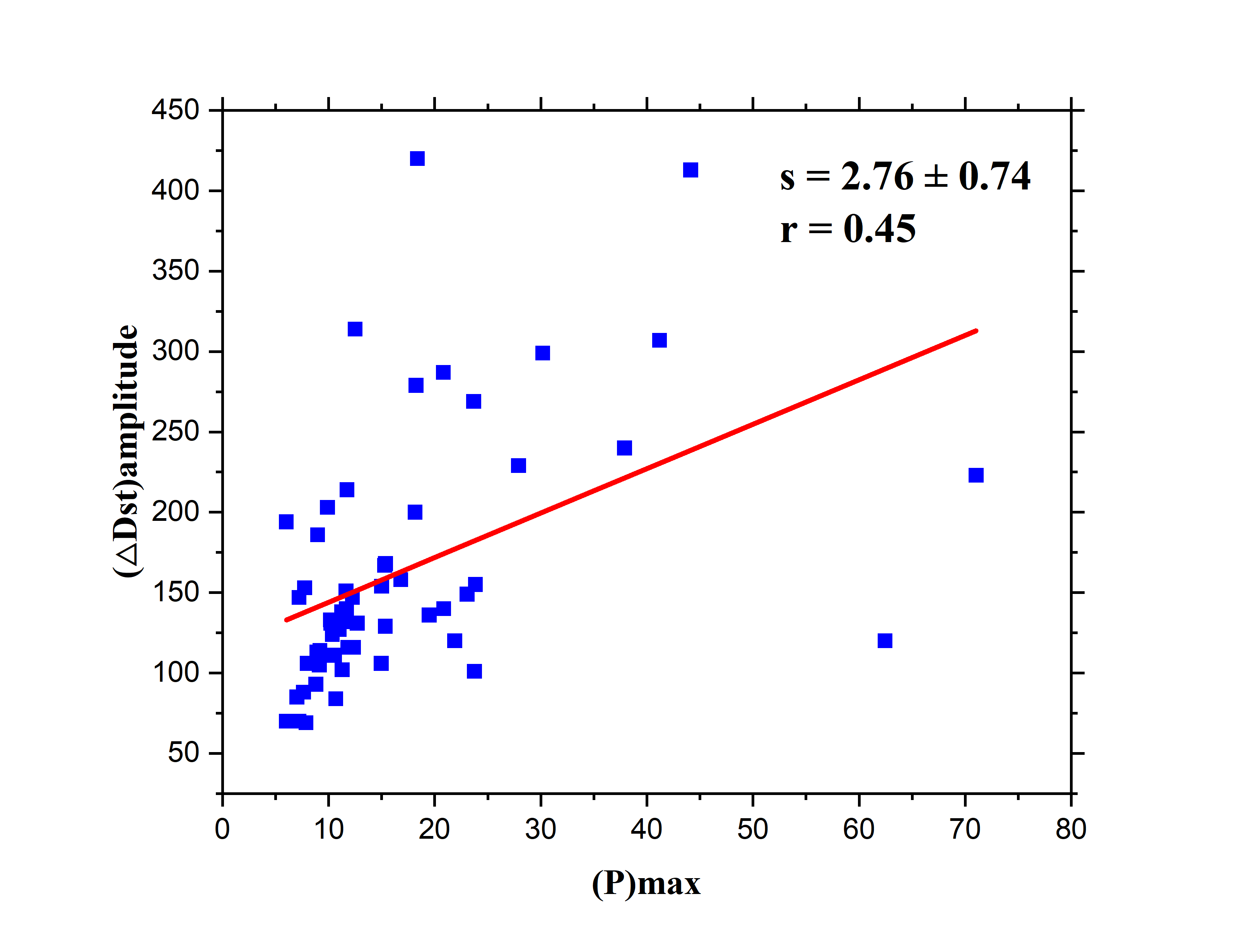}
		\\[\smallskipamount]
		\includegraphics[height=3.8cm, width=5.9cm]{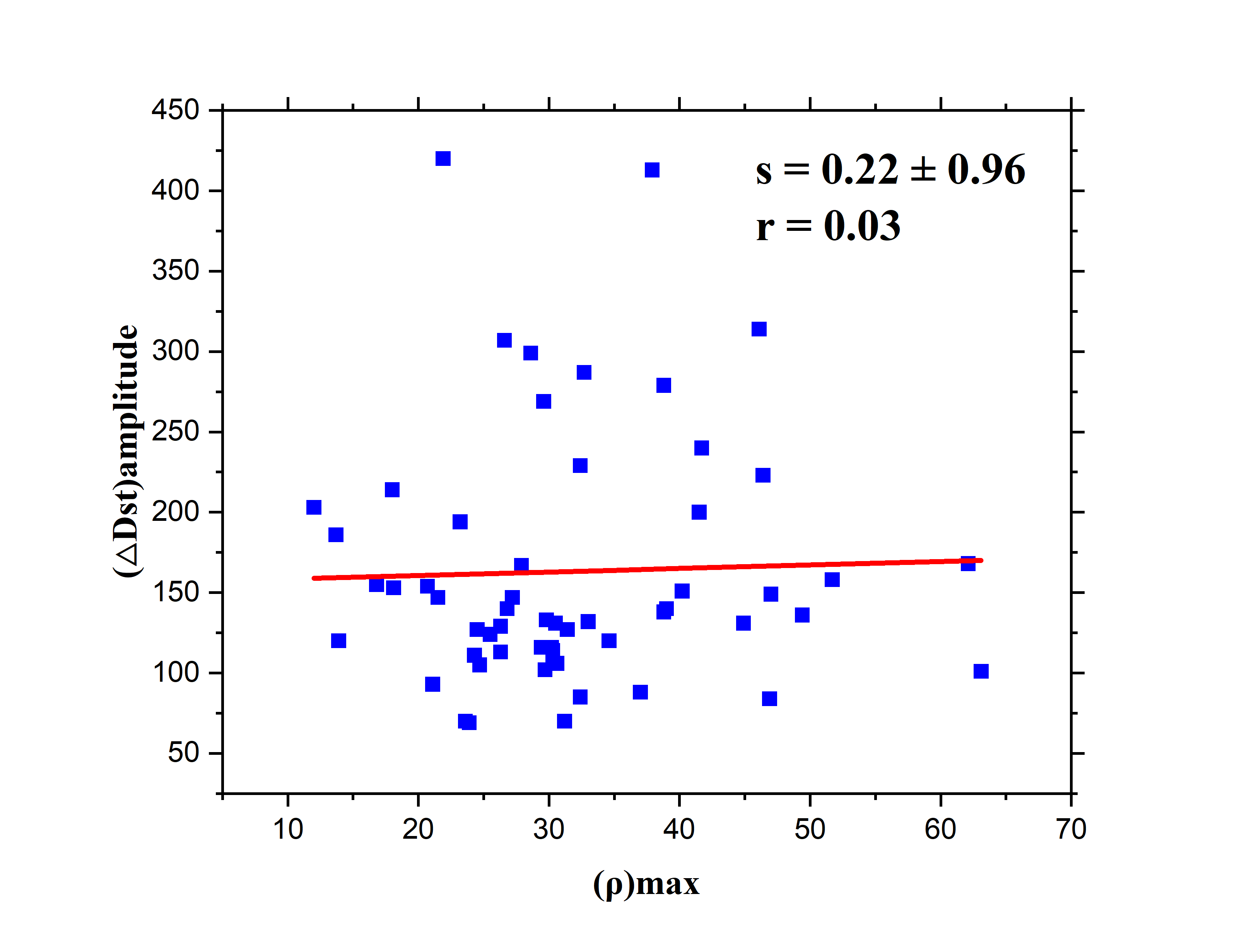}
		\includegraphics[height=3.8cm, width=5.9cm]{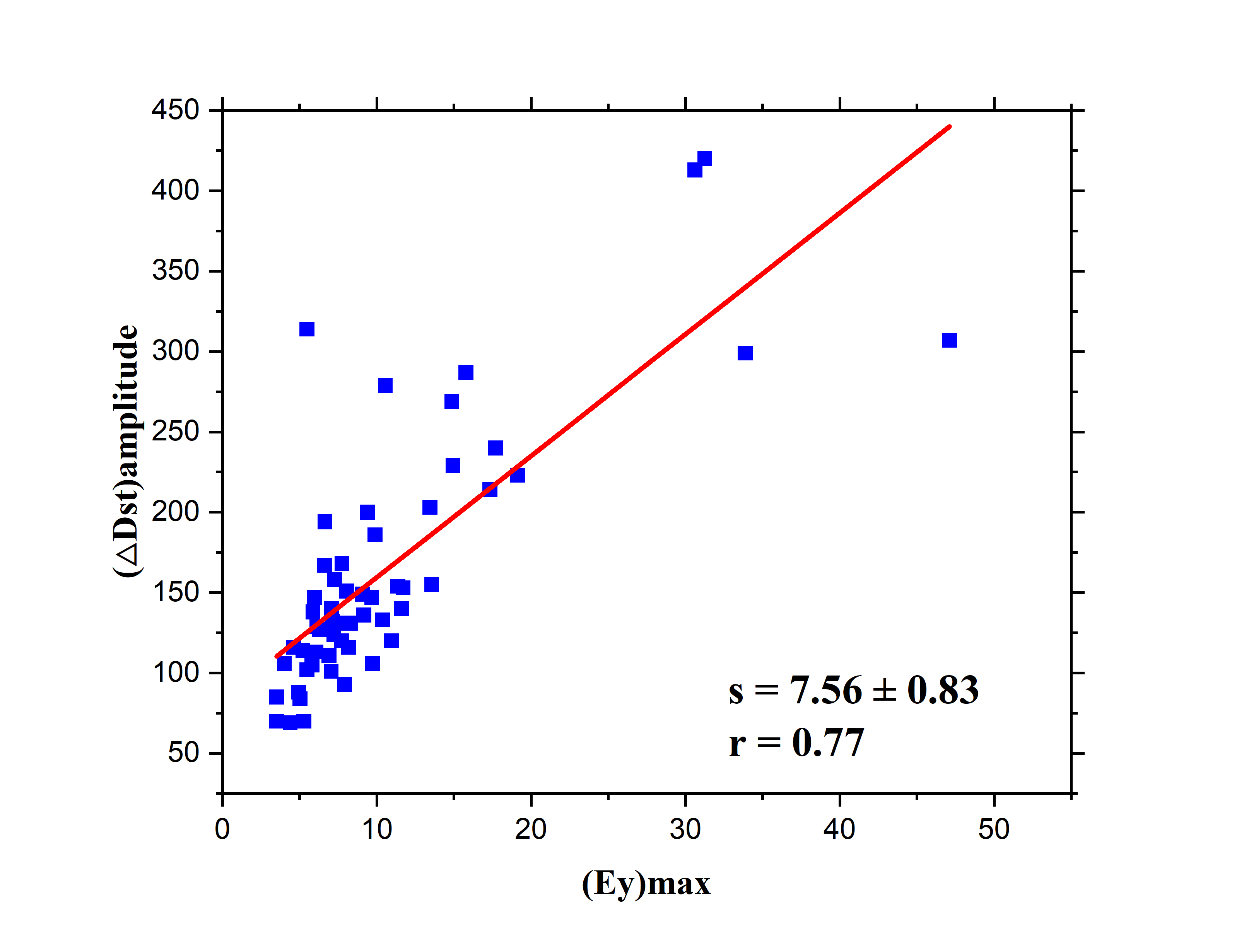}
		\includegraphics[height=3.8cm, width=5.9cm]{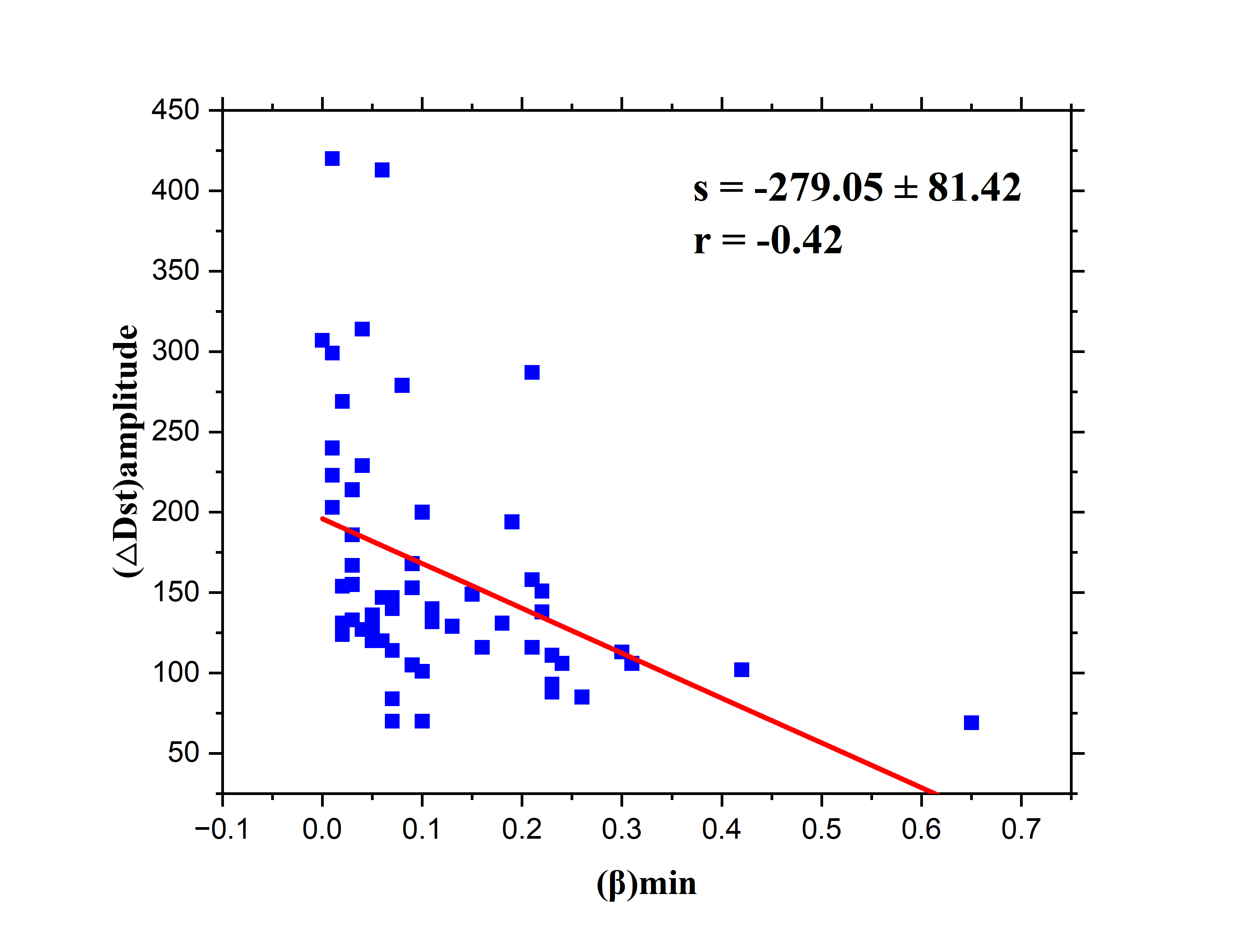}
		\\[\smallskipamount]
		\includegraphics[height=3.8cm, width=5.9cm]{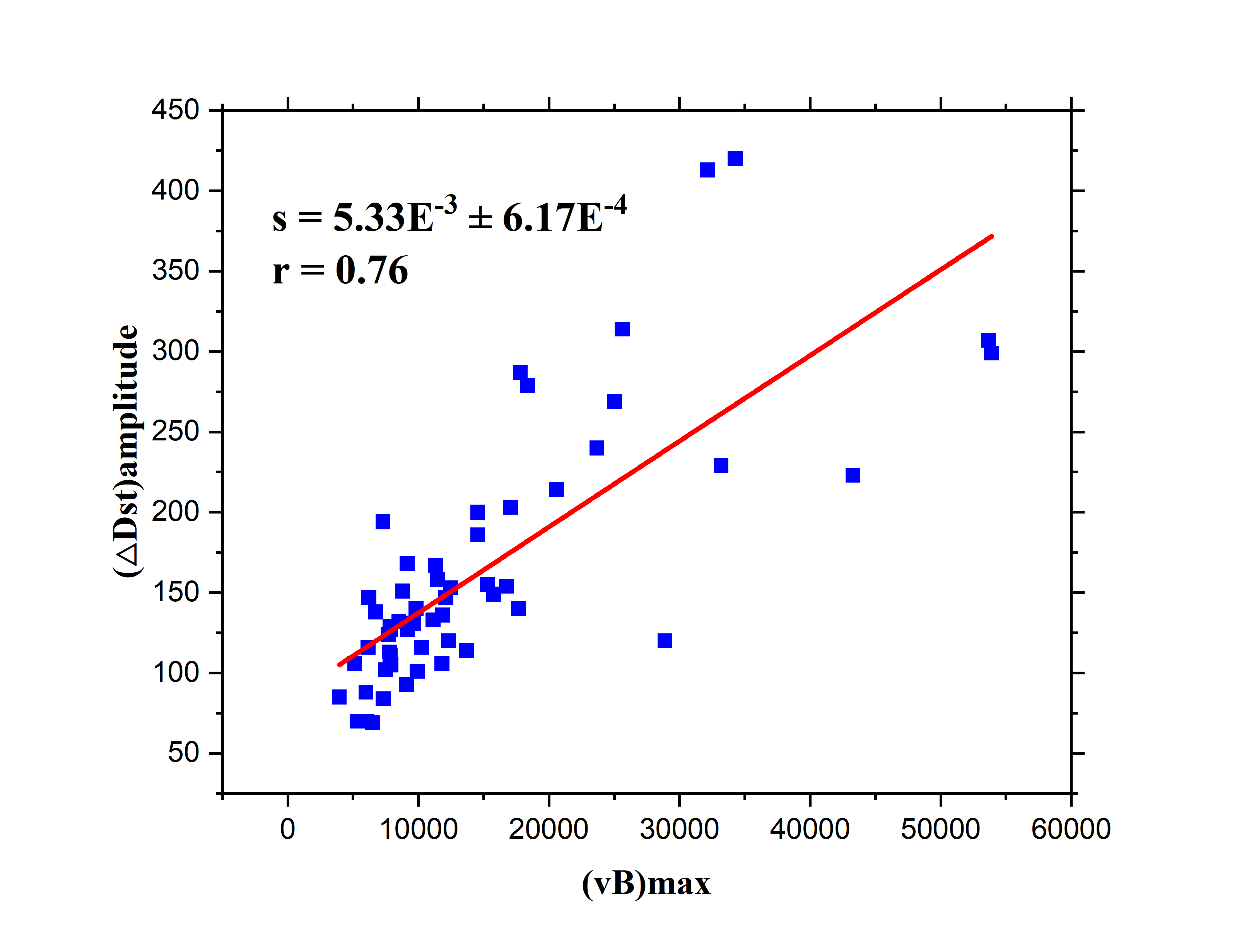}
		\includegraphics[height=3.8cm, width=5.9cm]{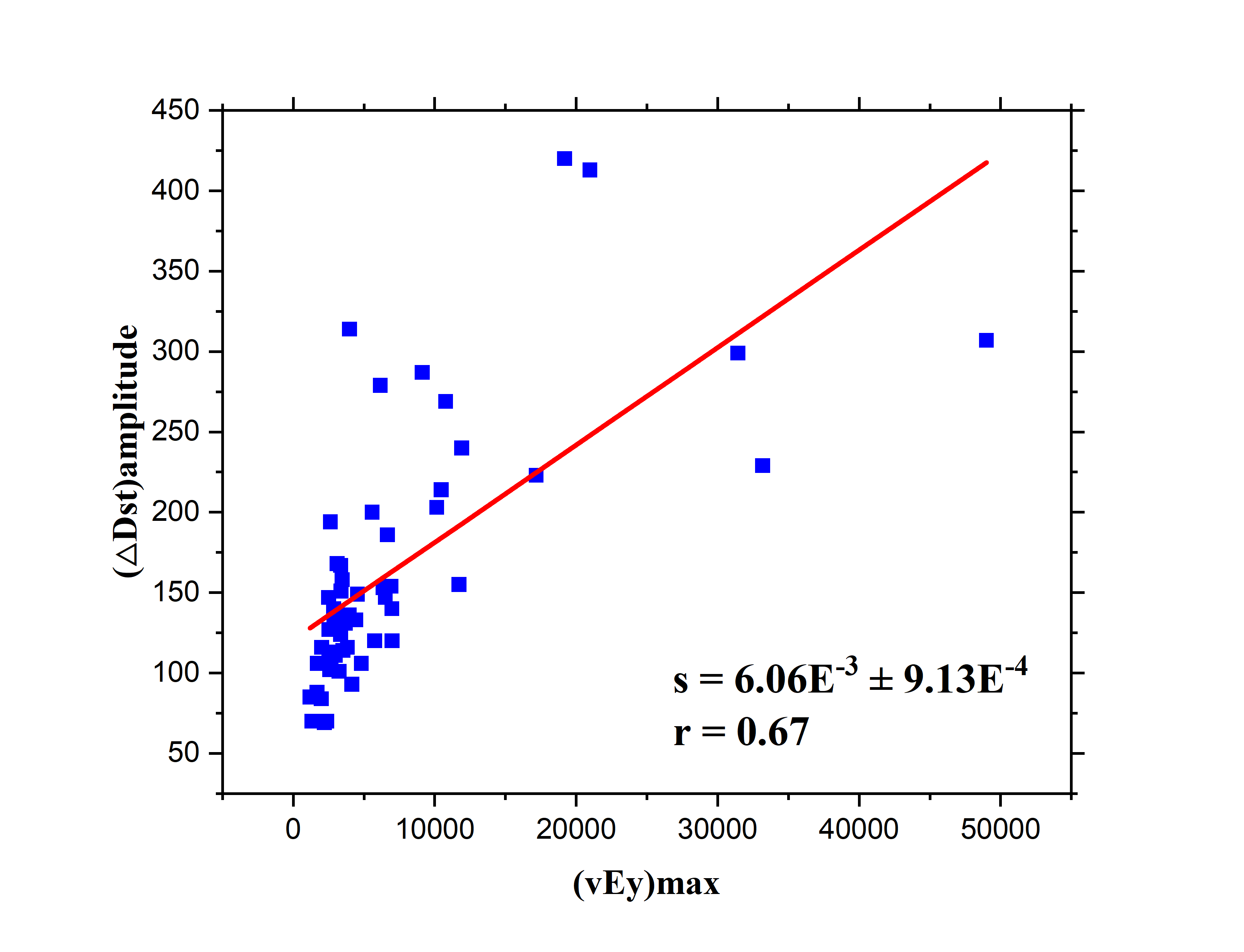}
		\includegraphics[height=3.8cm, width=5.9cm]{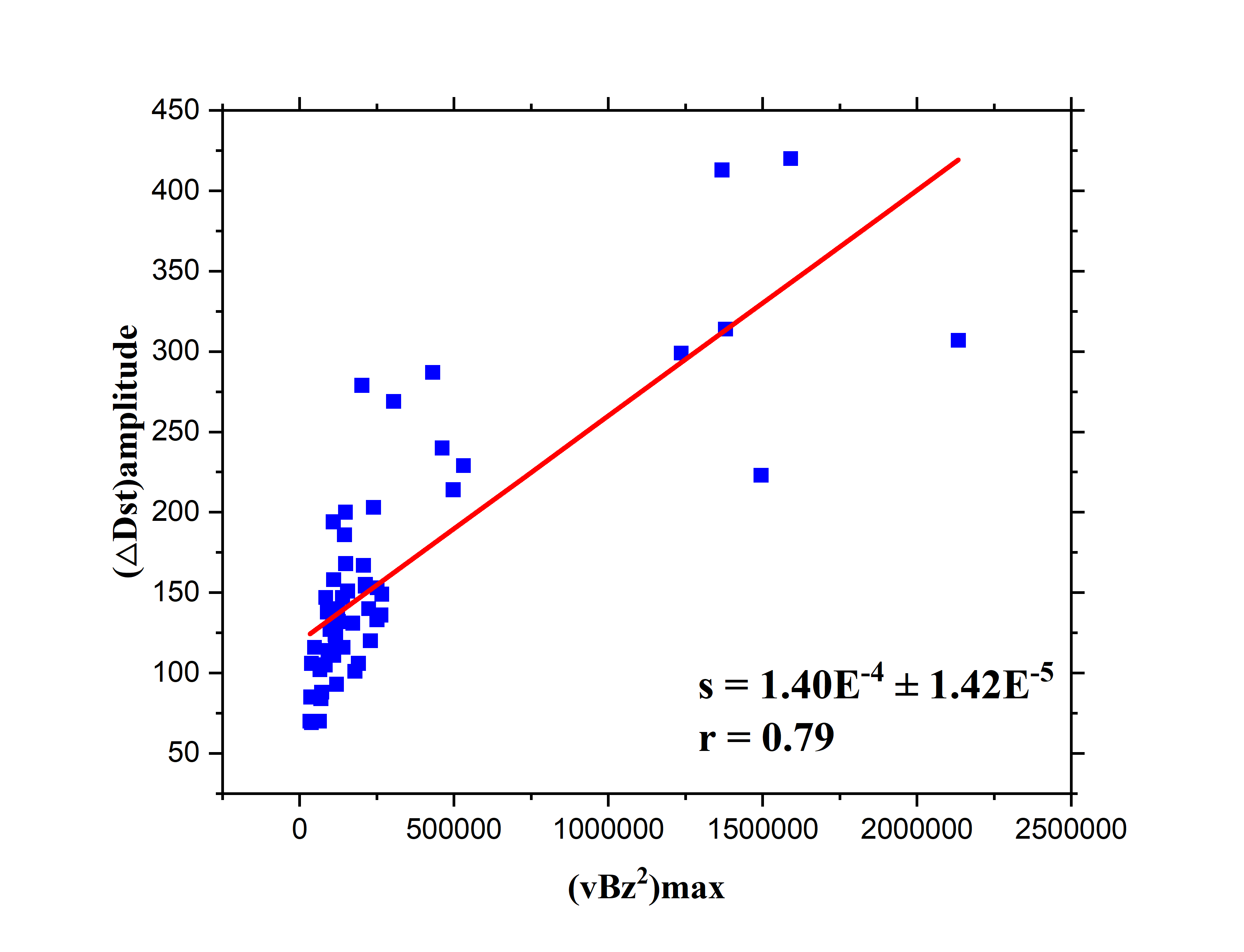}
		\\[\smallskipamount]
		\includegraphics[height=3.8cm, width=5.9cm]{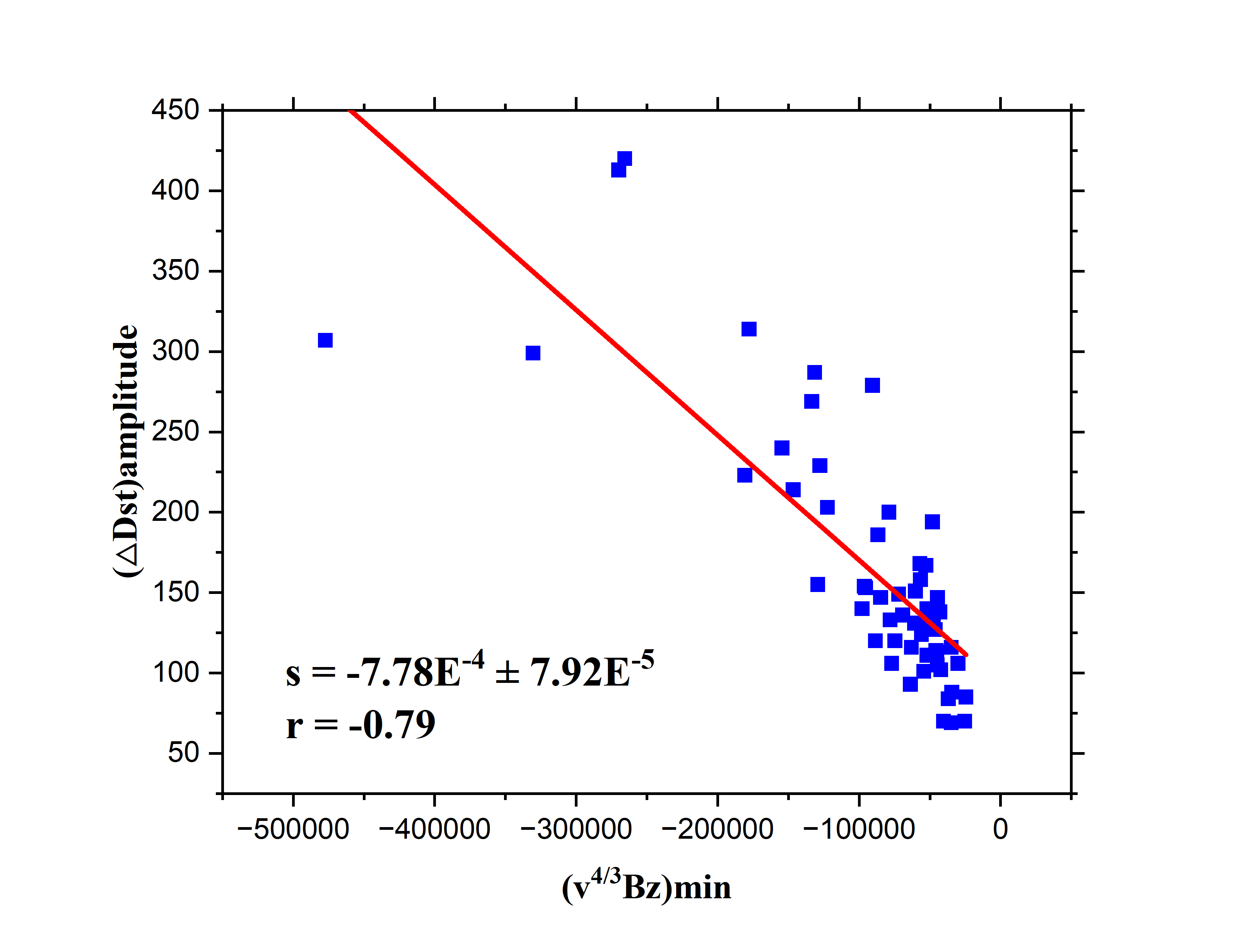}
		\includegraphics[height=3.8cm, width=5.9cm]{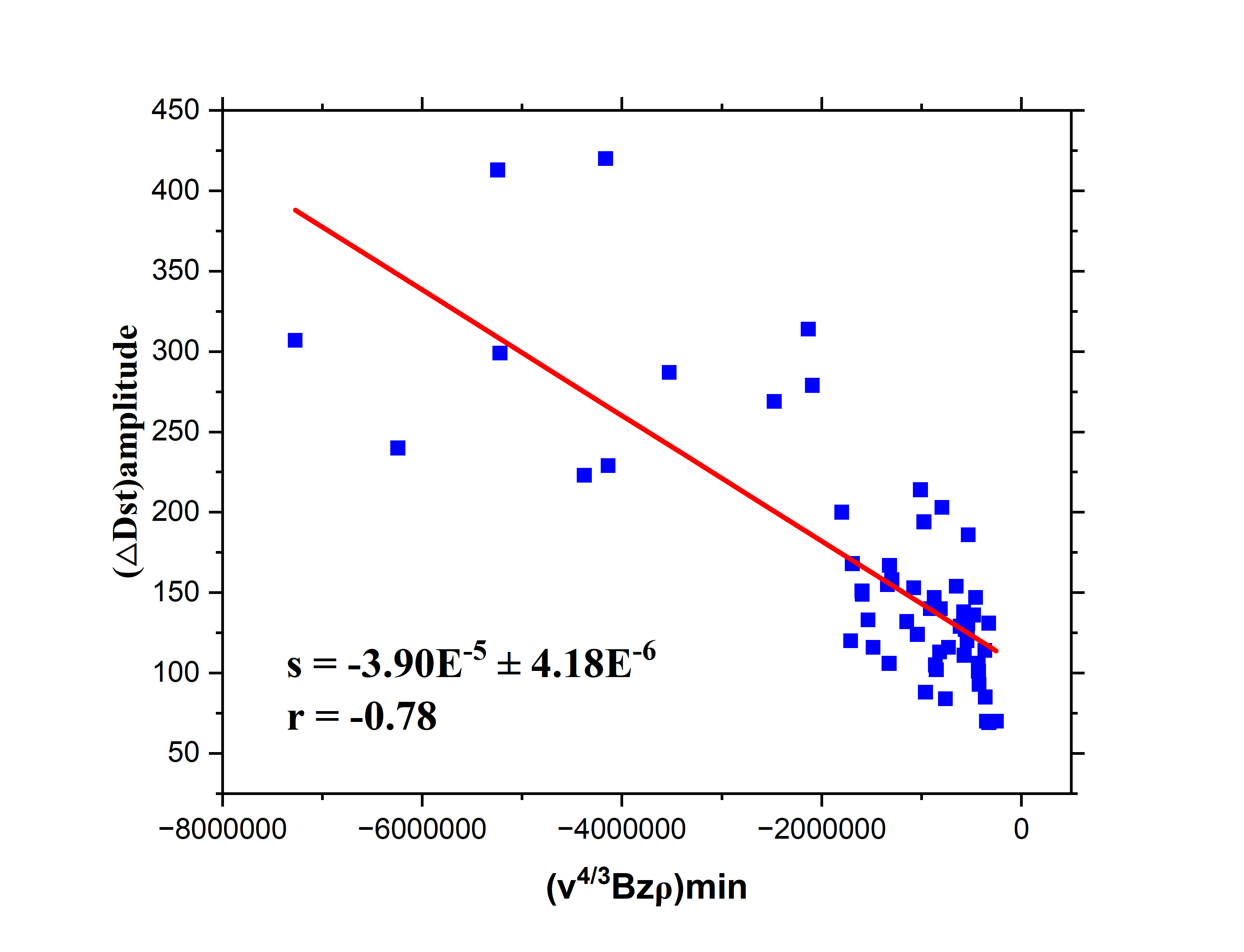}
		\includegraphics[height=3.8cm, width=5.9cm]{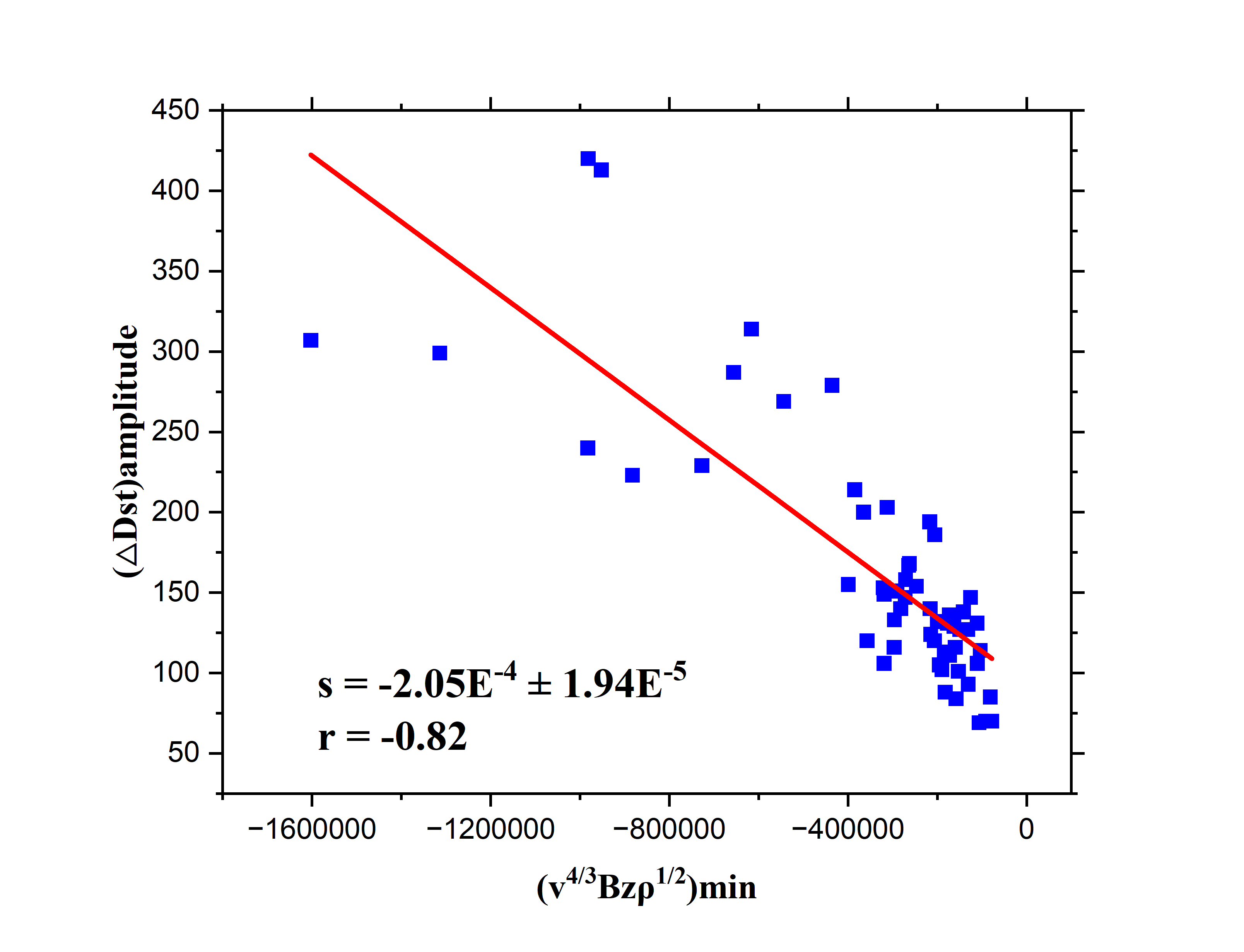}
		\\[\smallskipamount]
		\includegraphics[height=3.8cm, width=5.9cm]{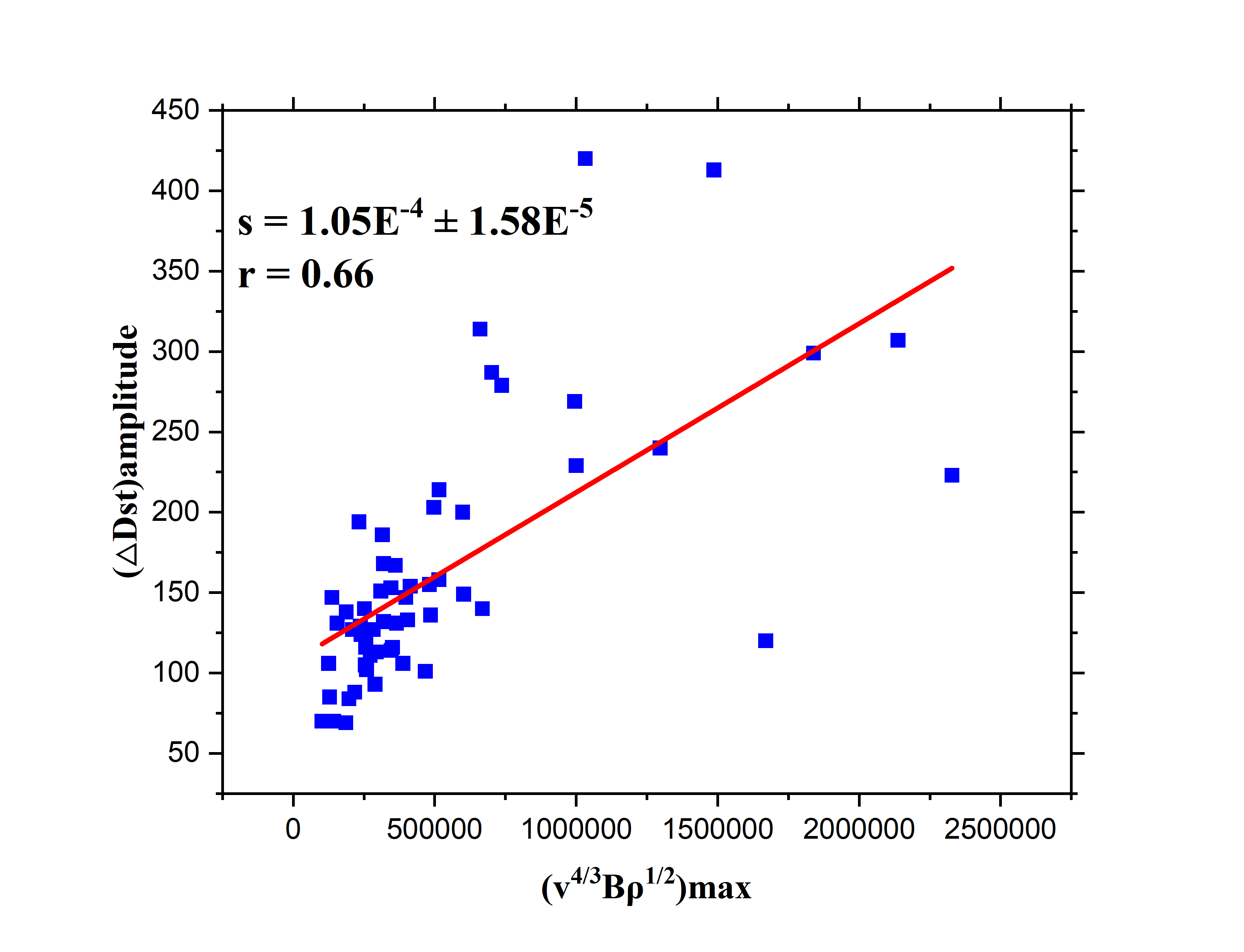}
		\caption{\small The scatter plots show the best linear relationship between the $\Delta$Dst amplitude during GS with the relevant SW plasma and field parameters and their derived functions. }\label{fig:changeDstcorr}
	\end{figure*}

We investigated the linear correlation between geomagnetic indices [Dst (nT), $\Delta$Dst (nT), AE (nT) \& ap (nT)], SW plasma and field parameters and the derived CFs to get best fit parameters. To that end, we took the peak values from all parameters/functions and fitted them linearly. To explain the correlations between geomagnetic indices and SW plasma and field parameters and product CFs, we considered following SW plasma and field parameters and product functions: 
	\begin{enumerate}
\item Single parameters i.e.,(B, Bz, P, $\sigma$B$_\mathrm{m}$, $\sigma$B$_\mathrm{v}$, $\beta$, $\rho$ and v)
\item Dual parameter CFs of IMFs B, Bz and IP plasma parameters v i.e., [Ey (Bzv), vB, vBz$^2$, vEy (Bzv$^{2}$) and v$^{\frac{4}{3}}$Bz] and
\item Triple parameter CFs of IMFs B, Bz and IP plasma parameters v, $\rho$ i.e., (v$^{\frac{4}{3}}$Bz$\rho$, v$^{\frac{4}{3}}$Bz$\rho^{\frac{1}{2}}$ and v$^{\frac{4}{3}}$B$\rho^{\frac{1}{2}}$).
	\end{enumerate}
It may be noted that in these expressions, (Bz) is taken as the southward Bz only. The SW parameters among them may be described as viscous terms (e.g., $ \rho $, v, v$^2$, $\rho^{\frac{1}{2}}$ , v$ ^{\frac{4}{3}}$), single merging term (Bz), electric field related/merging terms (Bz, vB, vBz$ ^2 $, Bzv$ ^2 $, and v$^{\frac{4}{3}}$Bz) and electric field related terms/merging terms coupled with viscous terms \citep[see,][]{1990P&SS...38..627G, 2007JGRA..112.1206N, 2008JGRA..113.4218N}.     \\
For each of the aforementioned categories (1,2,3), we have chosen the two best correlated parameters for discussion from among the scatter plots with best fit linear curve between the intensity of GS (as measured by Dst$ _\mathrm{min}$) and maximum value of SW plasma and field parameters/functions reached during around the peak geomagnetic disturbance.\\
Figure \ref{fig:Dstcorr} shows scatter plot with the linear relationship between the amplitude of geomagnetic index Dst$ _\mathrm{min}$ with maximum value of SW plasma and field parameters and their products or CFs. The intensity of GS (Dst$ _\mathrm{min}$) as obtained from index Dst has the highest correlation with single parameter Bz showing best fit parameter Pearson's correlation coefficient ($0.84$), with the best fit equation [Dst$_\mathrm{min}=[6.48\pm0.55$](Bz$_\mathrm{min}$)+($-26.70\pm12.03$)]. Our result agree with the previous studies \citep[e.g.,][]{1974PhDT........10P, 2008PhyS...78d5902M, Echer2008InterplanetaryCC} as they discussed that compared to the maximums of \sw\ density and \sw\ speed, minimum Dst has a higher correlation with the maximum southward Bz component of the IMF.
The second best correlated single parameter Pearson's correlation coefficient ($-0.82$) with Dst is the IMF strength B.\\
Among the studied dual parameters (see Table \ref{tabl2}), Dst has the highest correlation with CF vBz$^2$ showing best fit Pearson's correlation coefficient ($-0.80$), with the best fit equation [Dst$_\mathrm{min}=[-1.36\mathrm{E}^{-4}\pm1.38\mathrm{E}^{5}$](vBz$^2)_\mathrm{max}+(-109.74\pm7.46)$]. 
The second best correlation dual parameter CF showing Pearson's correlated coefficient ($0.79$) with Dst is v$^{\frac{4}{3}}$Bz.
As regards the CF involving three parameters, the Dst has the highest correlation with triple parameter CF v$^{\frac{4}{3}}$Bz$\rho^{1/2}$ showing best fit parameter Pearson's correlation coefficient ($0.79$), with the best fit equation [Dst$_\mathrm{min}=[1.93\mathrm{E}^{-4}\pm2.00\mathrm{E}^{5}$](v$ ^{\frac{4}{3}}$Bz$\rho^{\frac{1}{2}})_\mathrm{min}+(-109.74\pm7.46)$]. The second best correlated triple parameter CF showing Pearson's correlation coefficient ($0.74$) with Dst is v$ ^{\frac{4}{3}} $Bz$\rho$.\\
	Figure \ref{fig:changeDstcorr} shows scatter plot of the linear relationship between the amplitude geomagnetic index $\Delta$Dst with the peak values of SW plasma and field parameters and their functions. In general, $\Delta$Dst relation with SW parameters is almost similar.
	$\Delta$Dst has the highest correlation with single parameter IMFs Bz \& B (r$=-0.84$ or $0.85$) showing the best fit equation [$\Delta$Dst$_\mathrm{max}=[-6.68\pm0.57$](Bz$_\mathrm{min})+(33.83\pm12.38)$]. 
	$\Delta$Dst has the highest correlation with dual parameter CFs vBz$^2$ and v$^{\frac{4}{3}}$Bz showing best fit parameter Pearson's correlation coefficient ($0.79$ \& $-0.79$), with the best fit equation [$\Delta$Dst$=[1.40\mathrm{E}^{-4}\pm1.42\mathrm{E}^{5}$](vBz$^2)_\mathrm{max}+(119.64\pm7.73)$] and [$\Delta$Dst$=[-7.78\mathrm{E}^{-4}\pm7.92\mathrm{E}^{5}$](v$^{\frac{4}{3}}$Bz)$_\mathrm{min}+(92.22\pm9.61)$] respectively.
 	Second best correlated dual parameter CF showing Pearson's correlation coefficient ($0.77$) with $\Delta$Dst is Ey.
	$\Delta$Dst has the highest correlation with triple parameter CF v$^{\frac{4}{3}}$Bz$\rho^{\frac{1}{2}}$ showing best fit parameter Pearson's correlation coefficient ($-0.82$); with the best fit equation 
[$\Delta$(Dst)$=[2.05\mathrm{E}^{-4}\pm1.94\mathrm{E}^{5}$](v$ ^{\frac{4}{3}} $Bz$\rho^{\frac{1}{2}})_\mathrm{min}+(92.93\pm8.99)$]. 
Thus, in this case, the merging term (v$^{\frac{4}{3}}$Bz) when coupled with viscous term ($\rho^{\frac{1}{2}}$) is a better CF representing $\Delta$Dst \citep[see,][]{2008JGRA..113.4218N}.
The second best correlation triple parameter CF $\Delta$Dst showing Pearson's correlation coefficient $-0.78$ is v$^{\frac{4}{3}}$Bz$\rho$.\\

\begin{figure*}
	\centering
		\includegraphics[height=3.8cm, width=5.9cm]{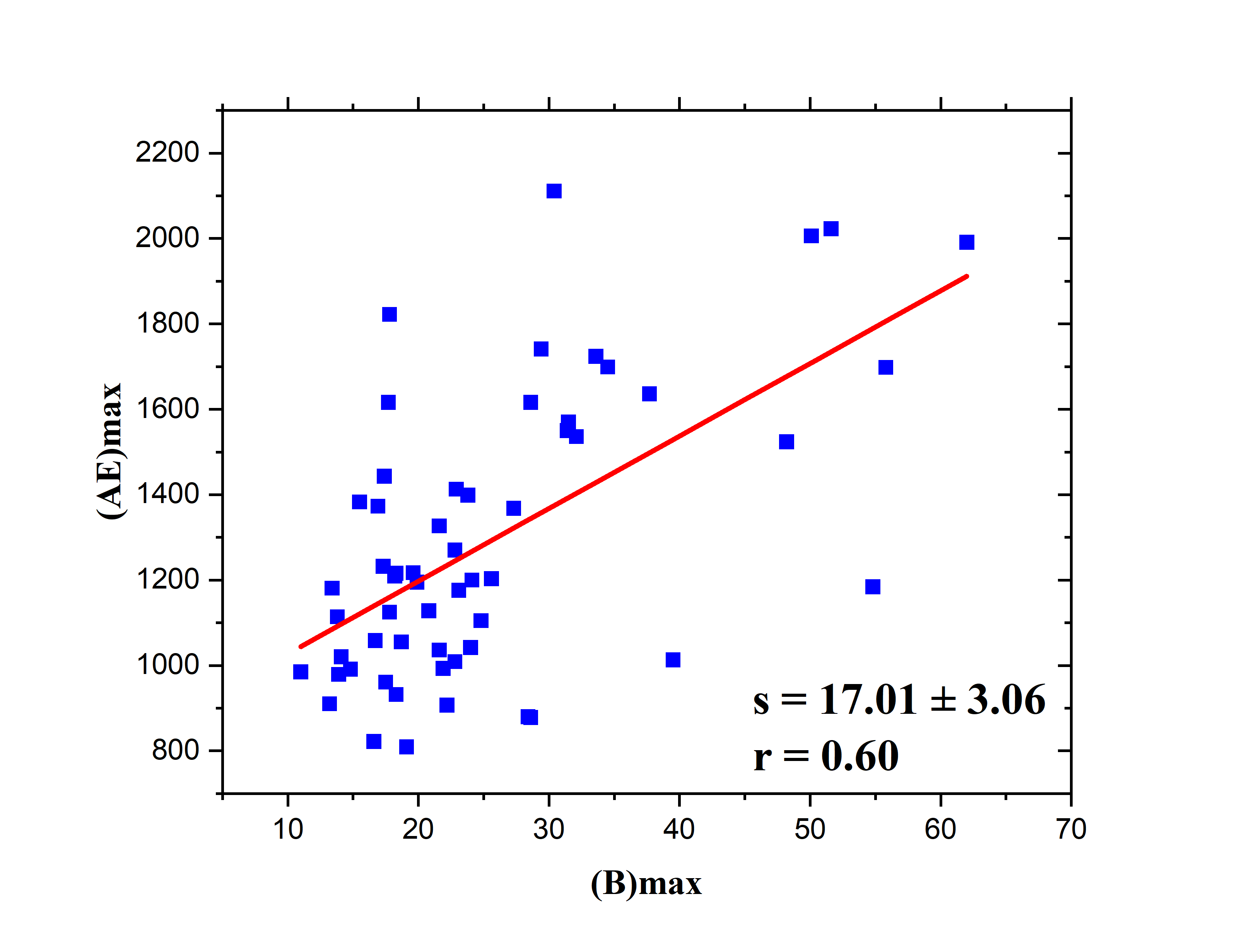}
		\includegraphics[height=3.8cm, width=5.9cm]{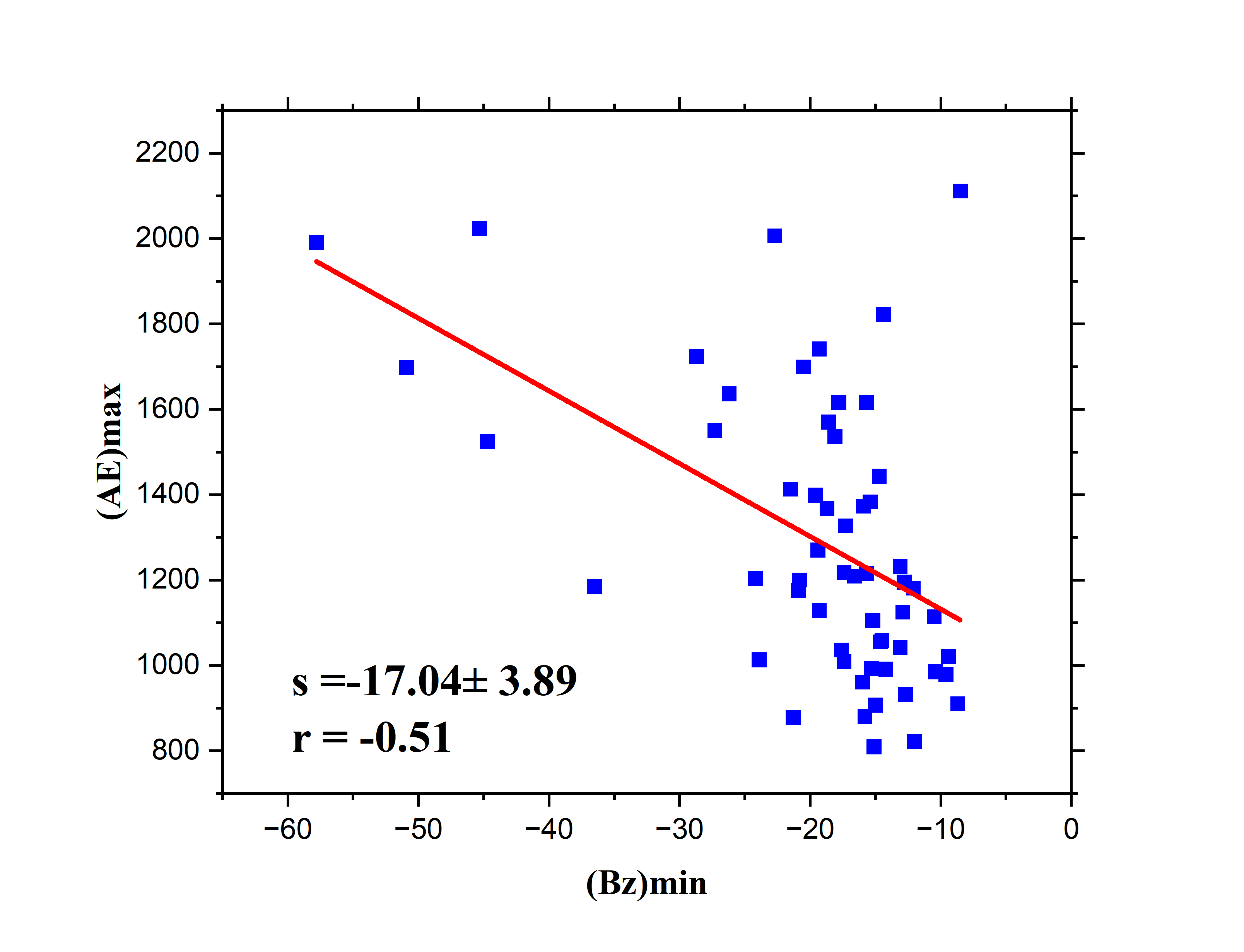}
		\includegraphics[height=3.8cm, width=5.9cm]{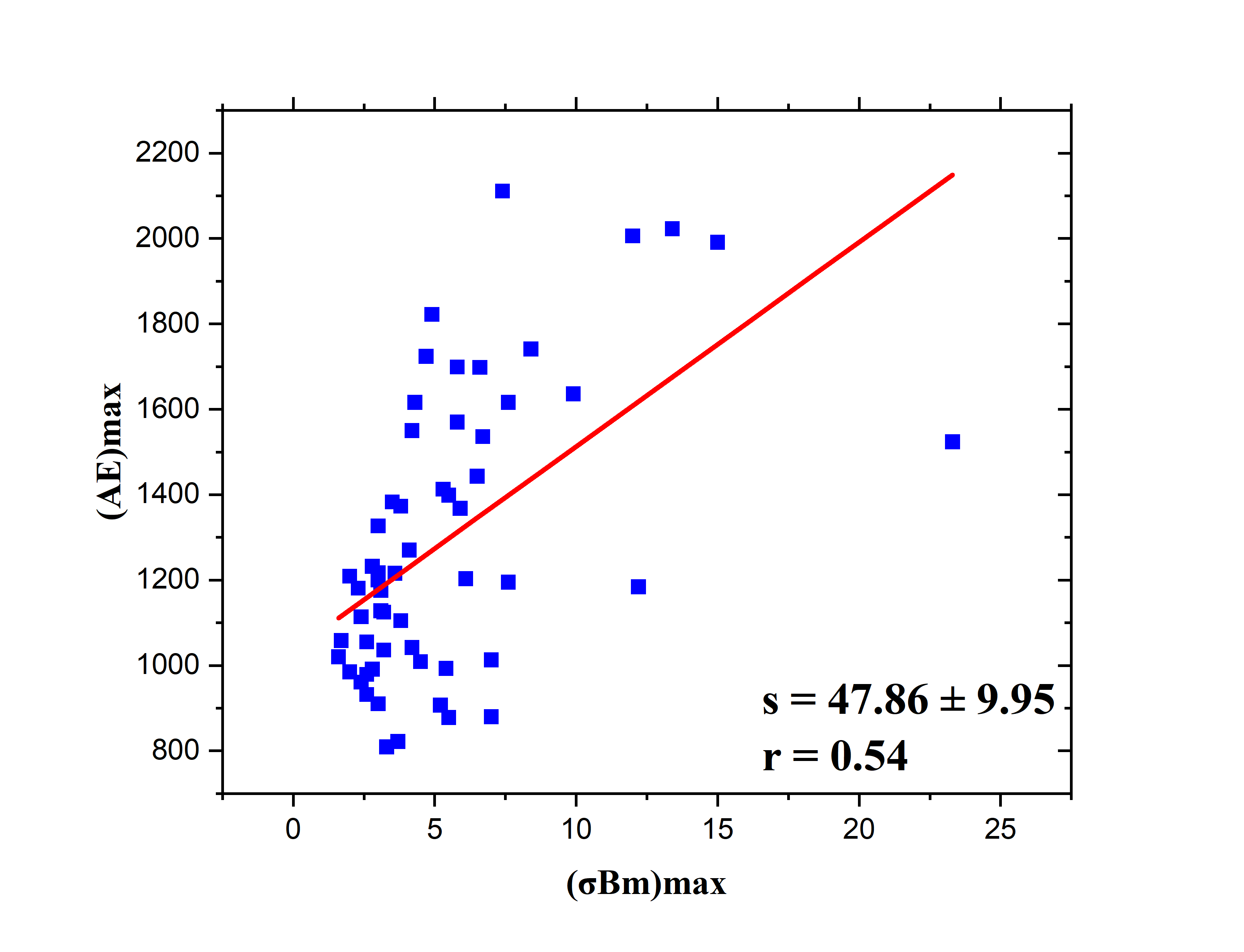}
		\\[\smallskipamount]
		\includegraphics[height=3.8cm, width=5.9cm]{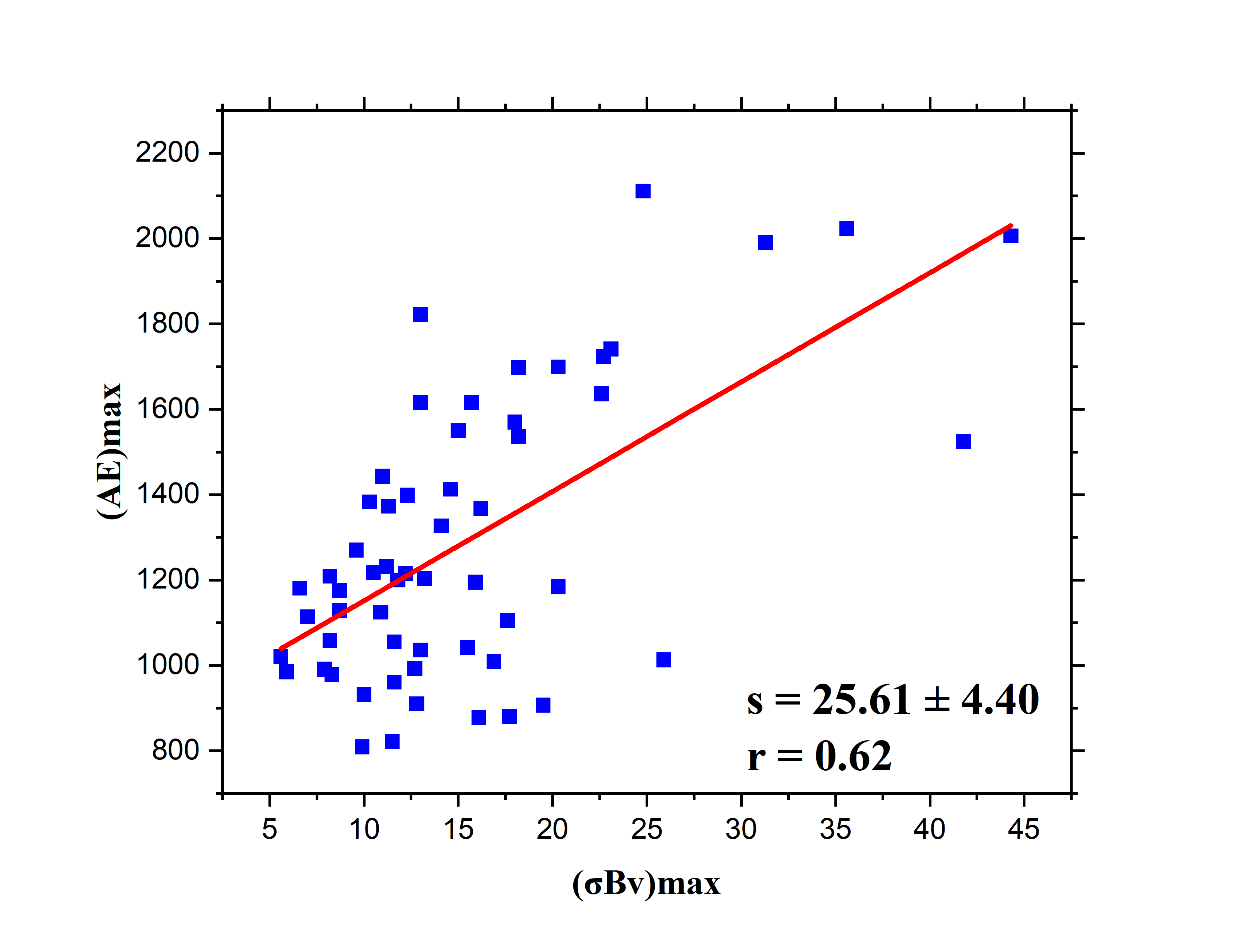}
		\includegraphics[height=3.8cm, width=5.9cm]{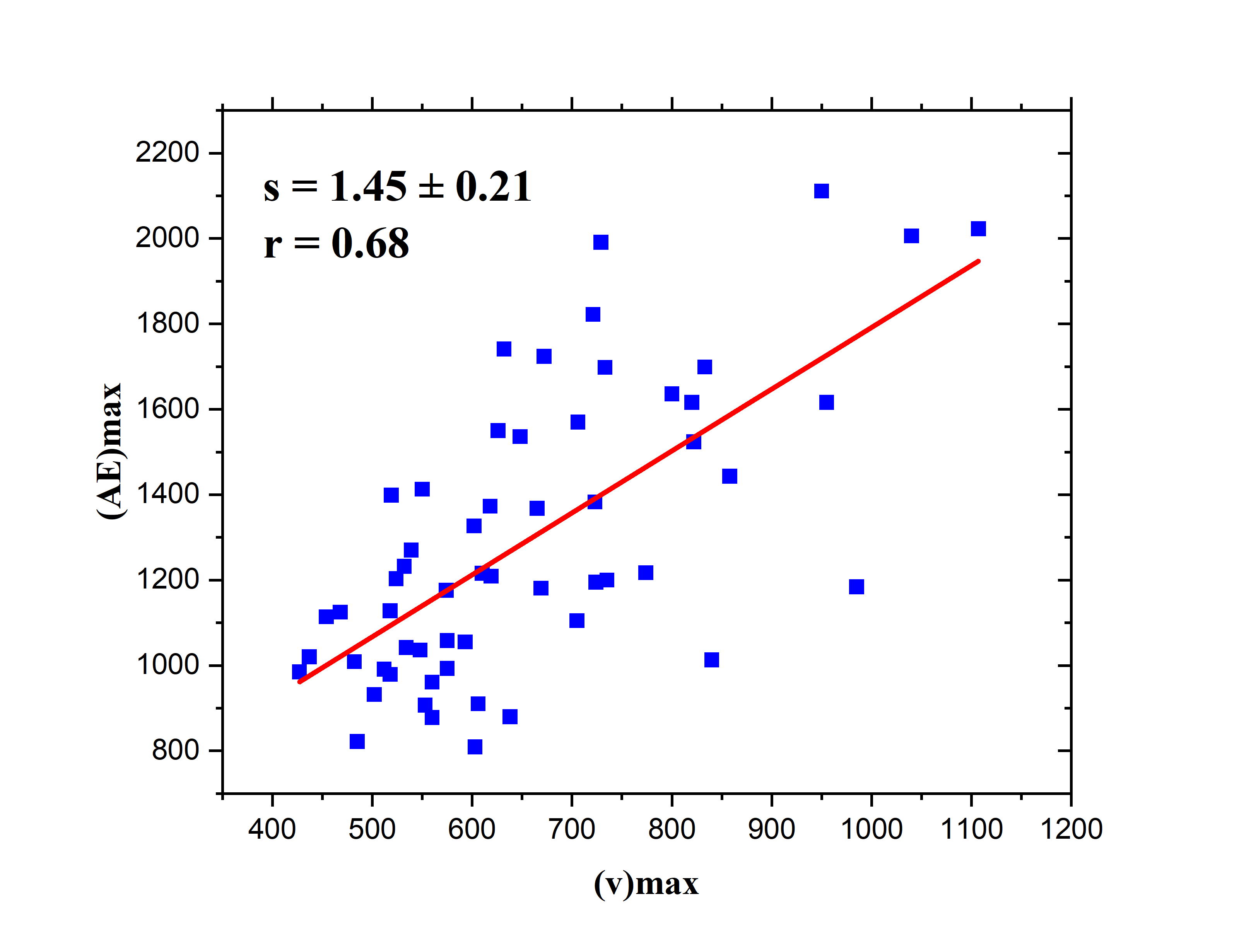}
		\includegraphics[height=3.8cm, width=5.9cm]{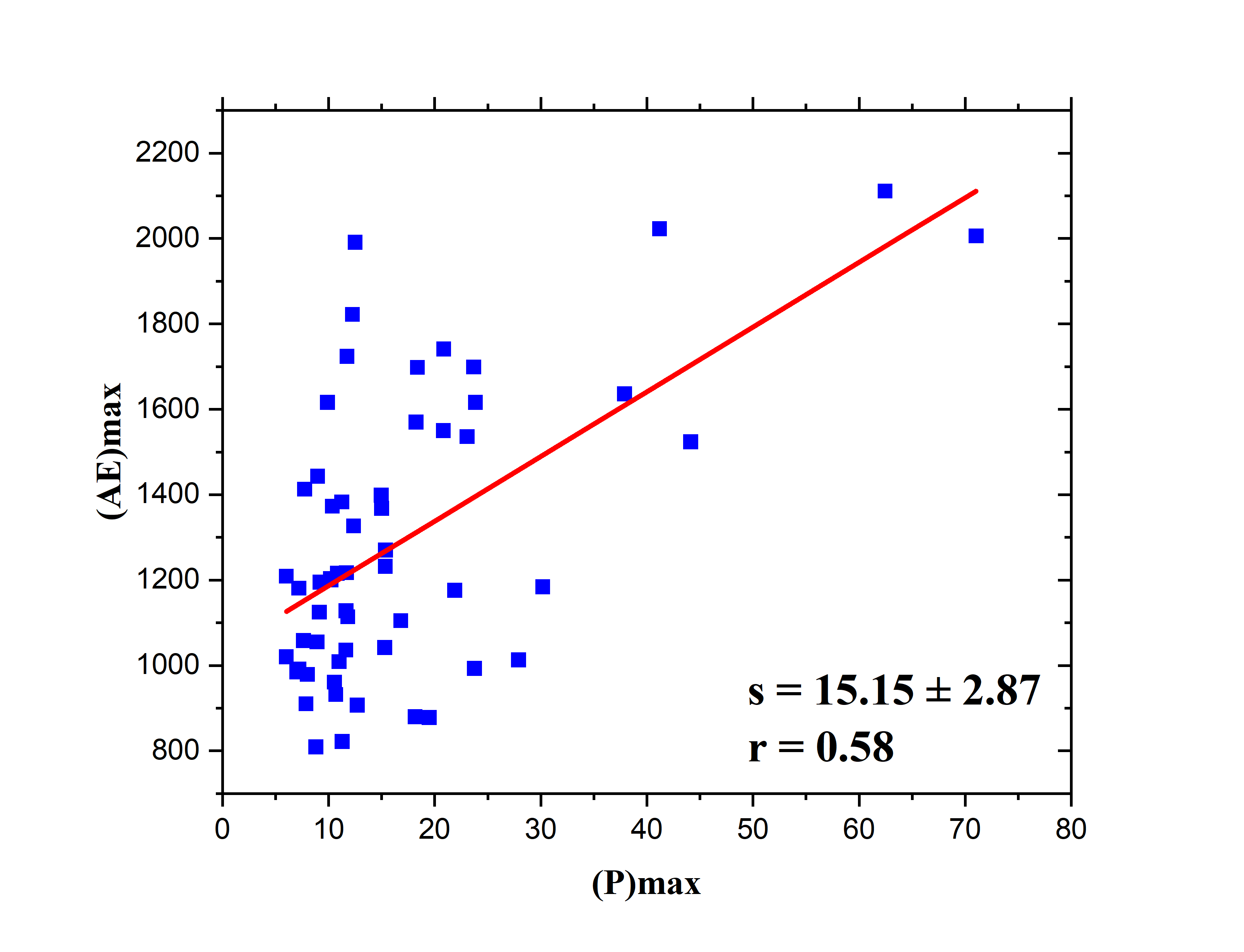}
		\\[\smallskipamount]
		\includegraphics[height=3.8cm, width=5.9cm]{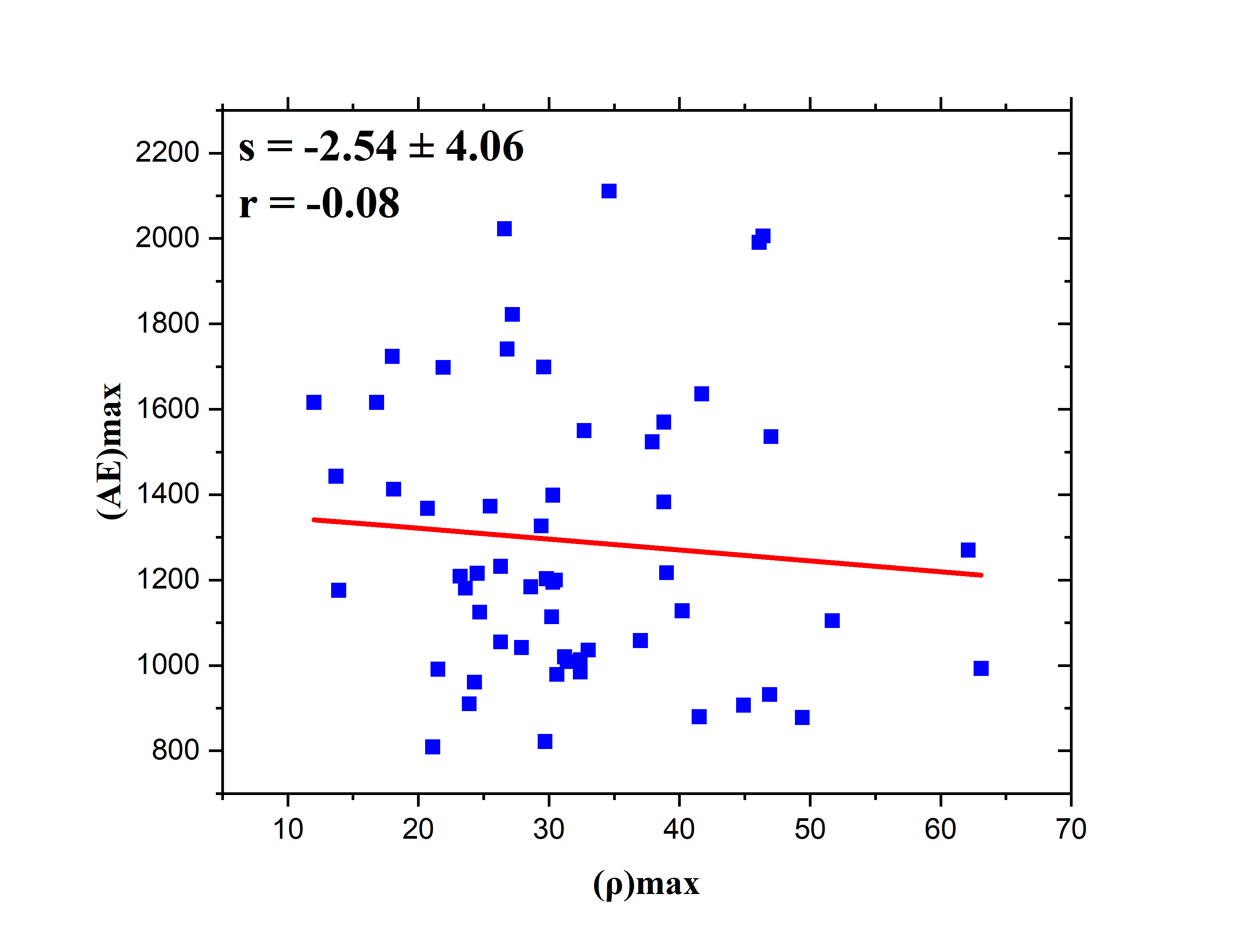}
		\includegraphics[height=3.8cm, width=5.9cm]{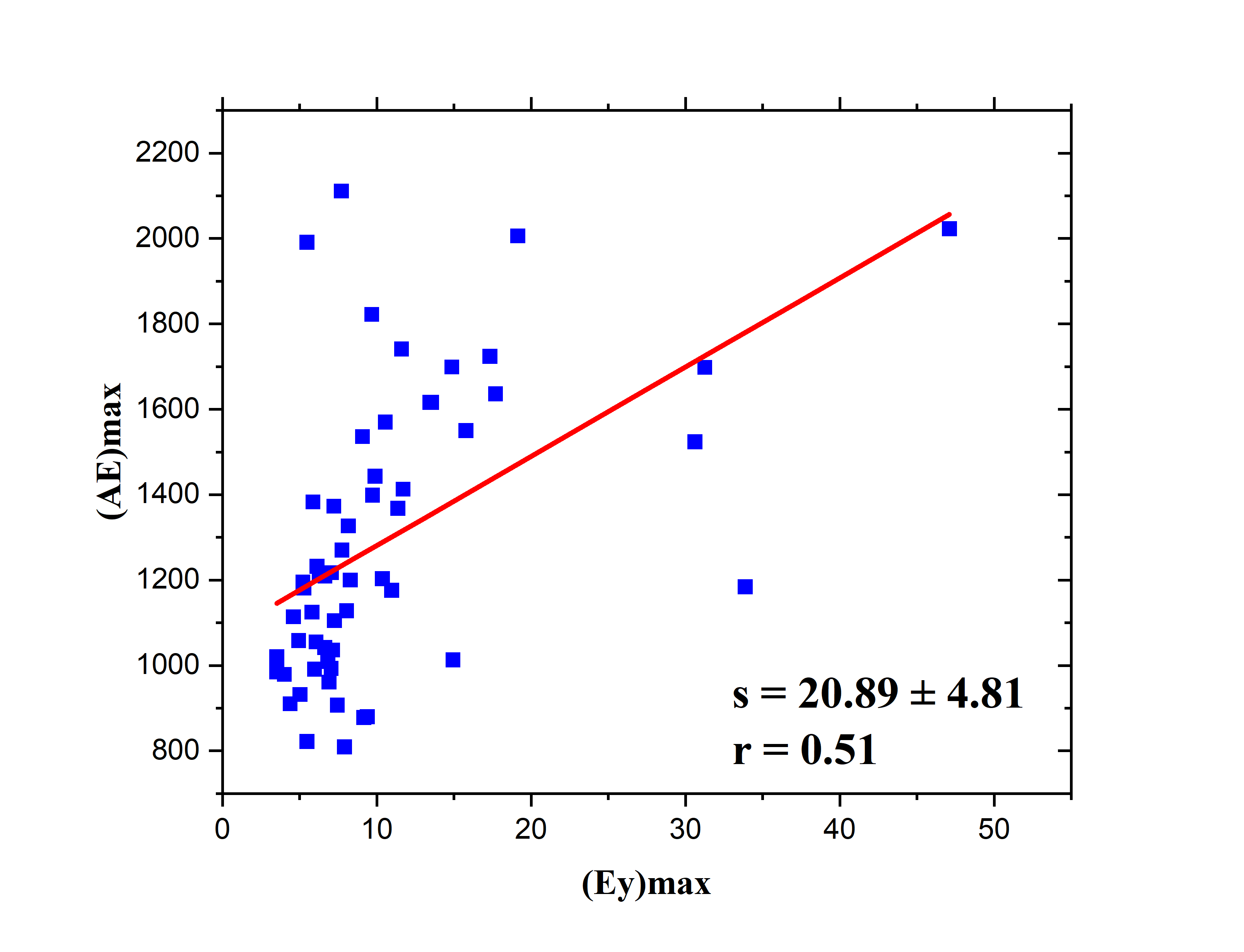}
		\includegraphics[height=3.8cm, width=5.9cm]{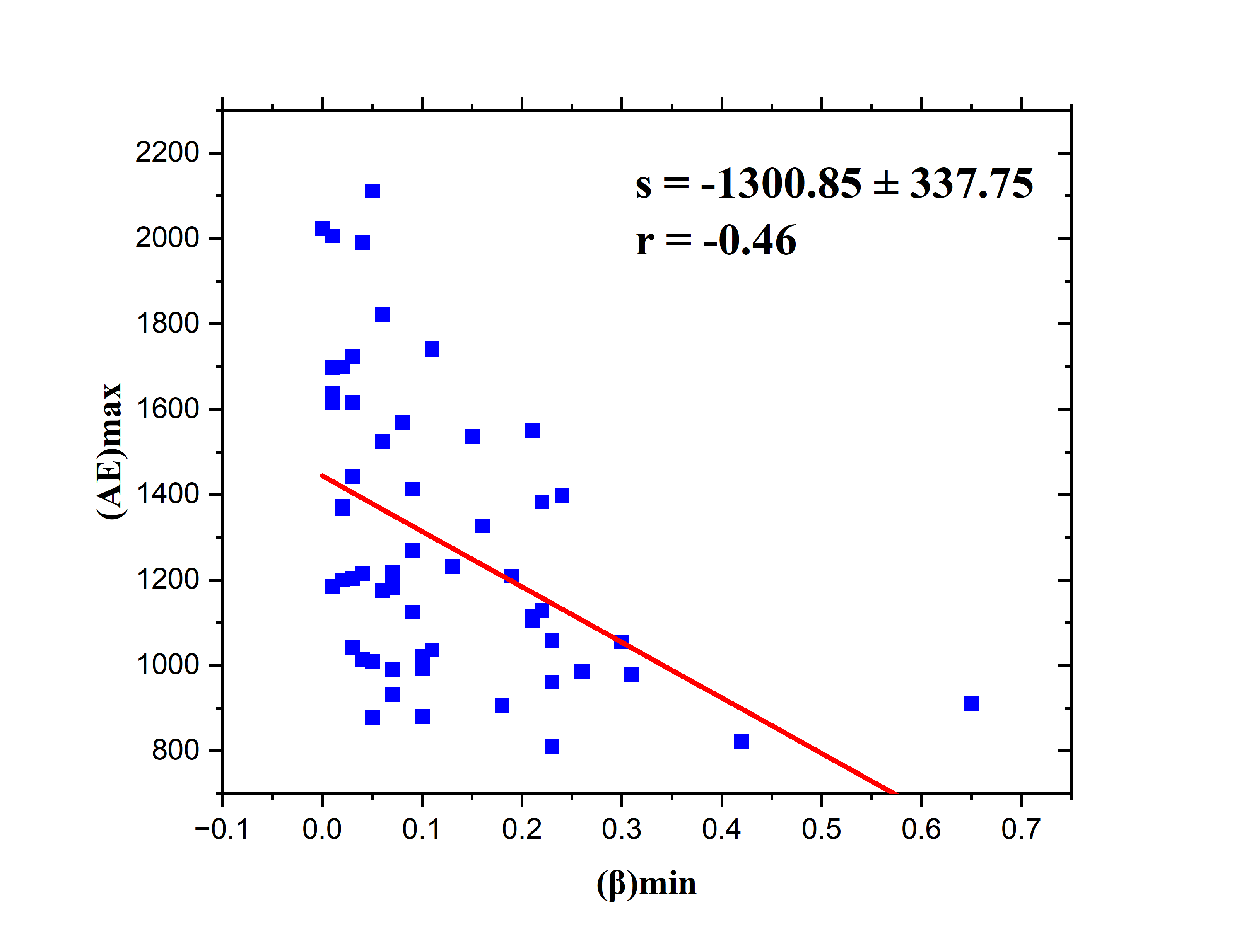}
		\\[\smallskipamount]
		\includegraphics[height=3.8cm, width=5.9cm]{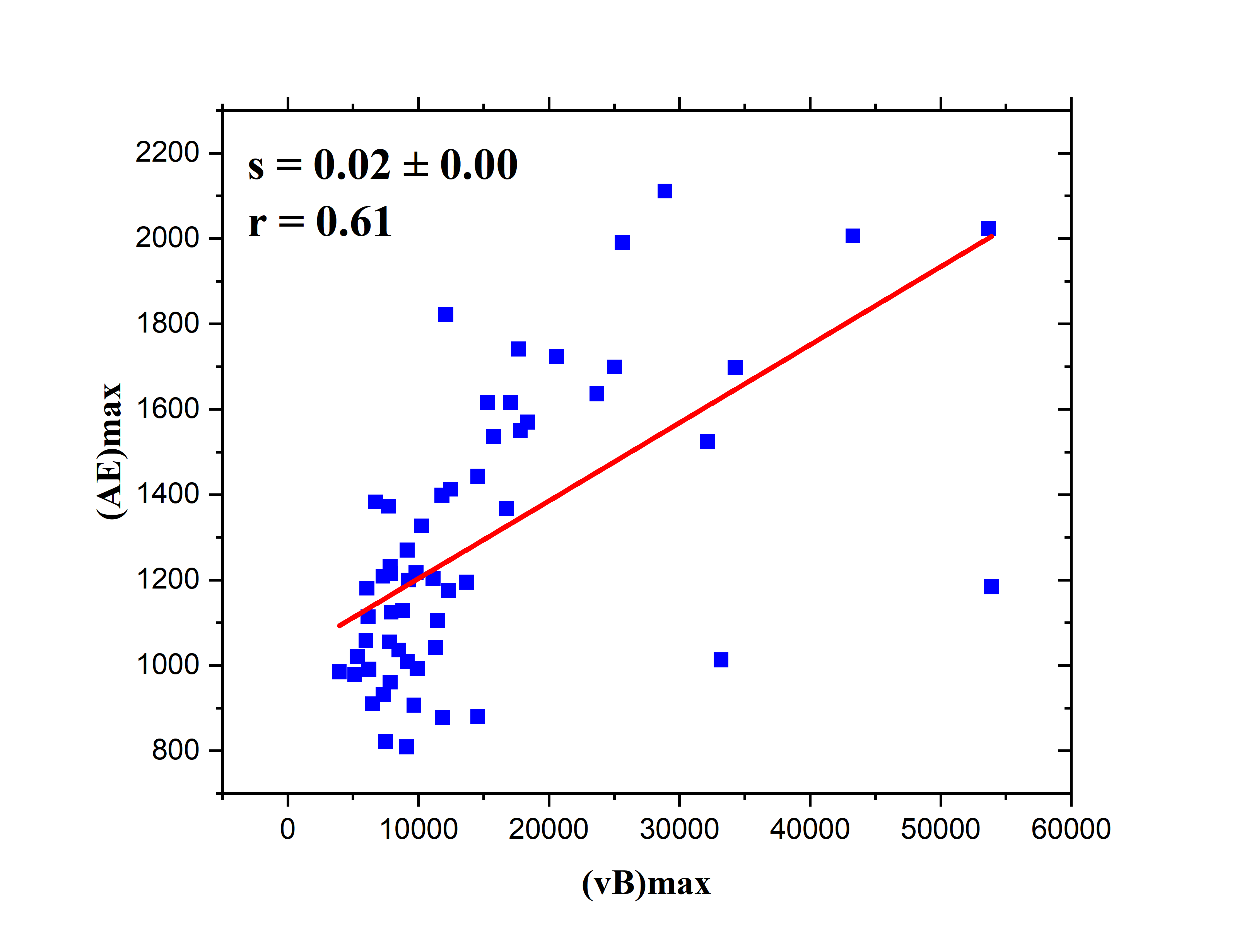}
		\includegraphics[height=3.8cm, width=5.9cm]{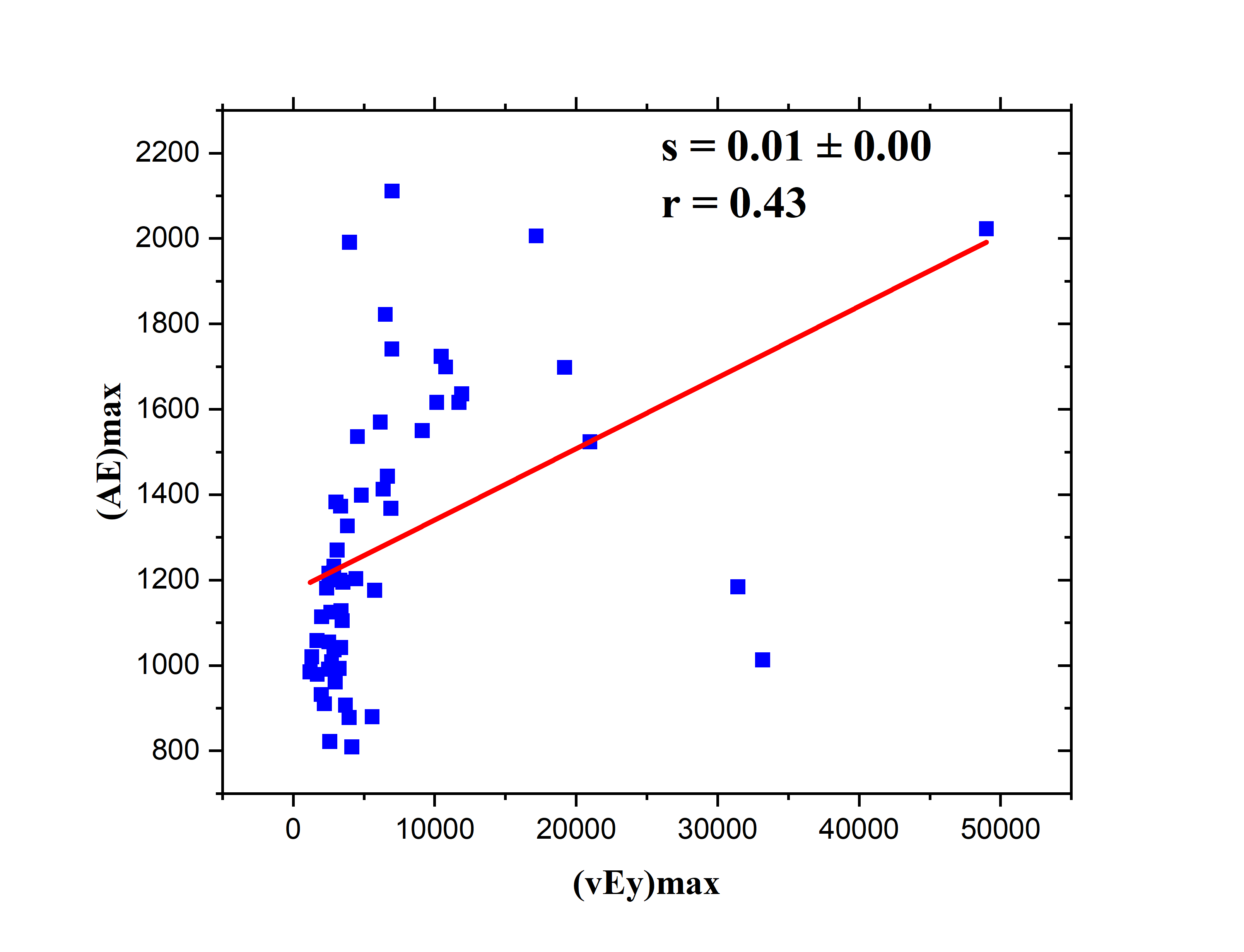}
		\includegraphics[height=3.8cm, width=5.9cm]{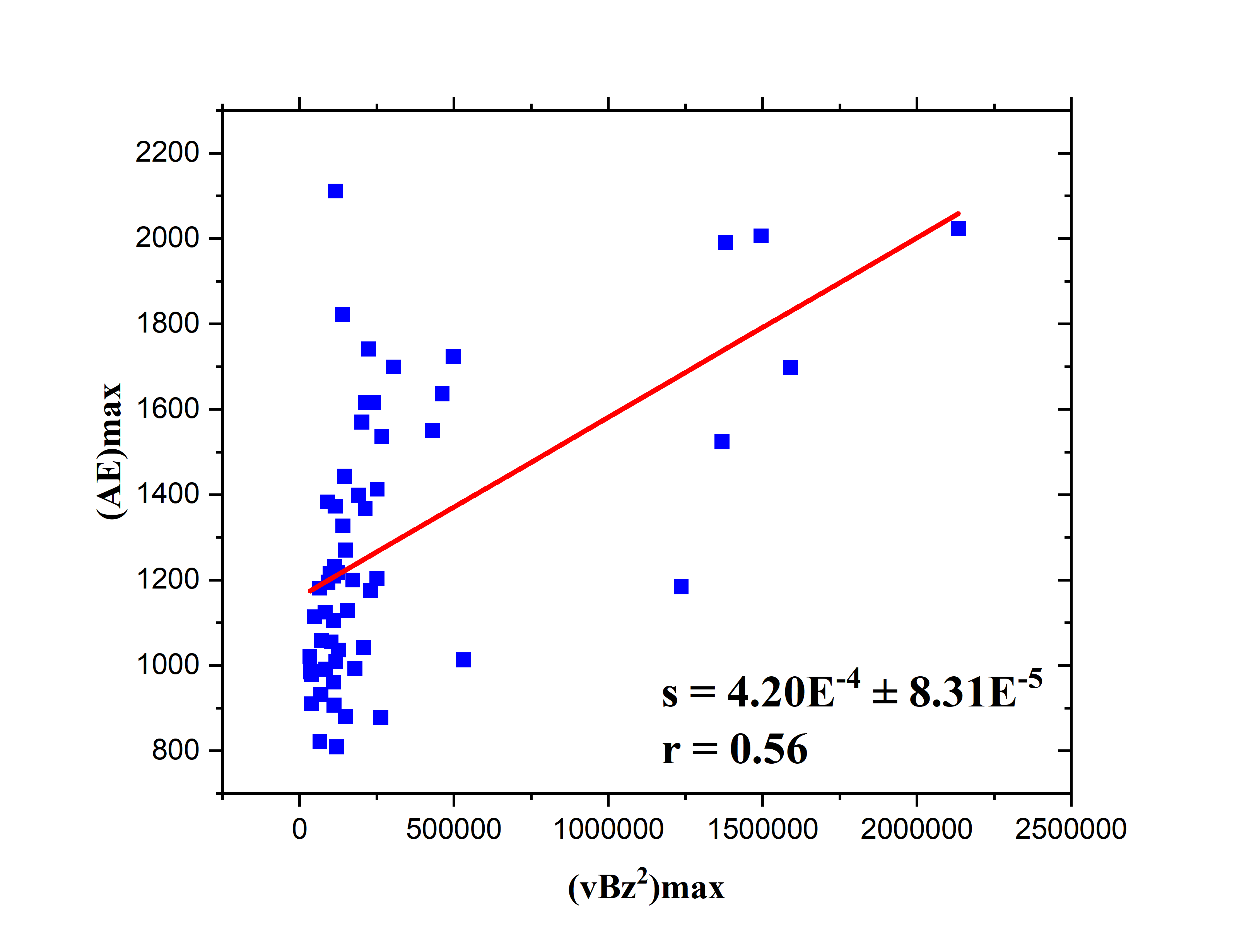}
		\\[\smallskipamount]
		\includegraphics[height=3.8cm, width=5.9cm]{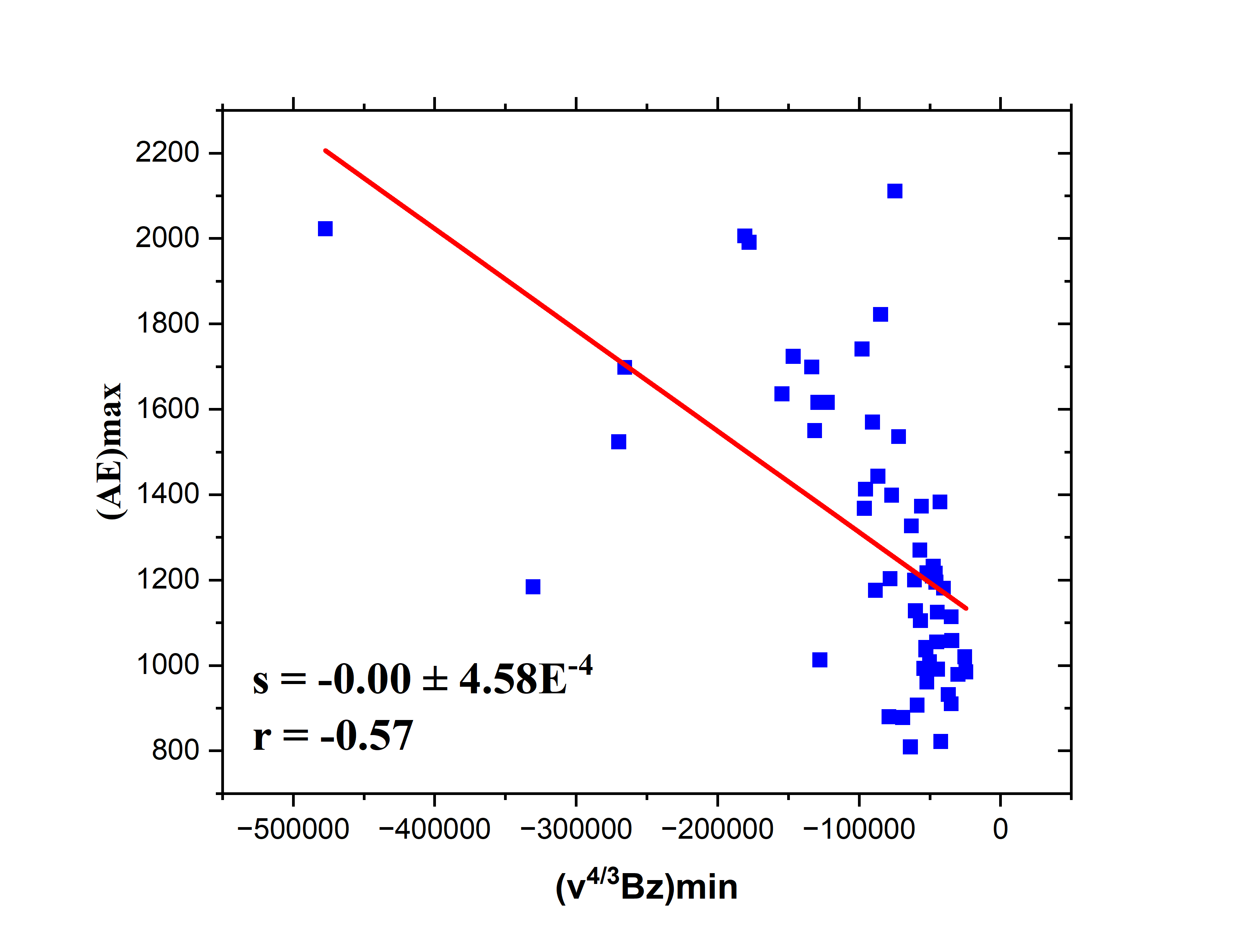}
		\includegraphics[height=3.8cm, width=5.9cm]{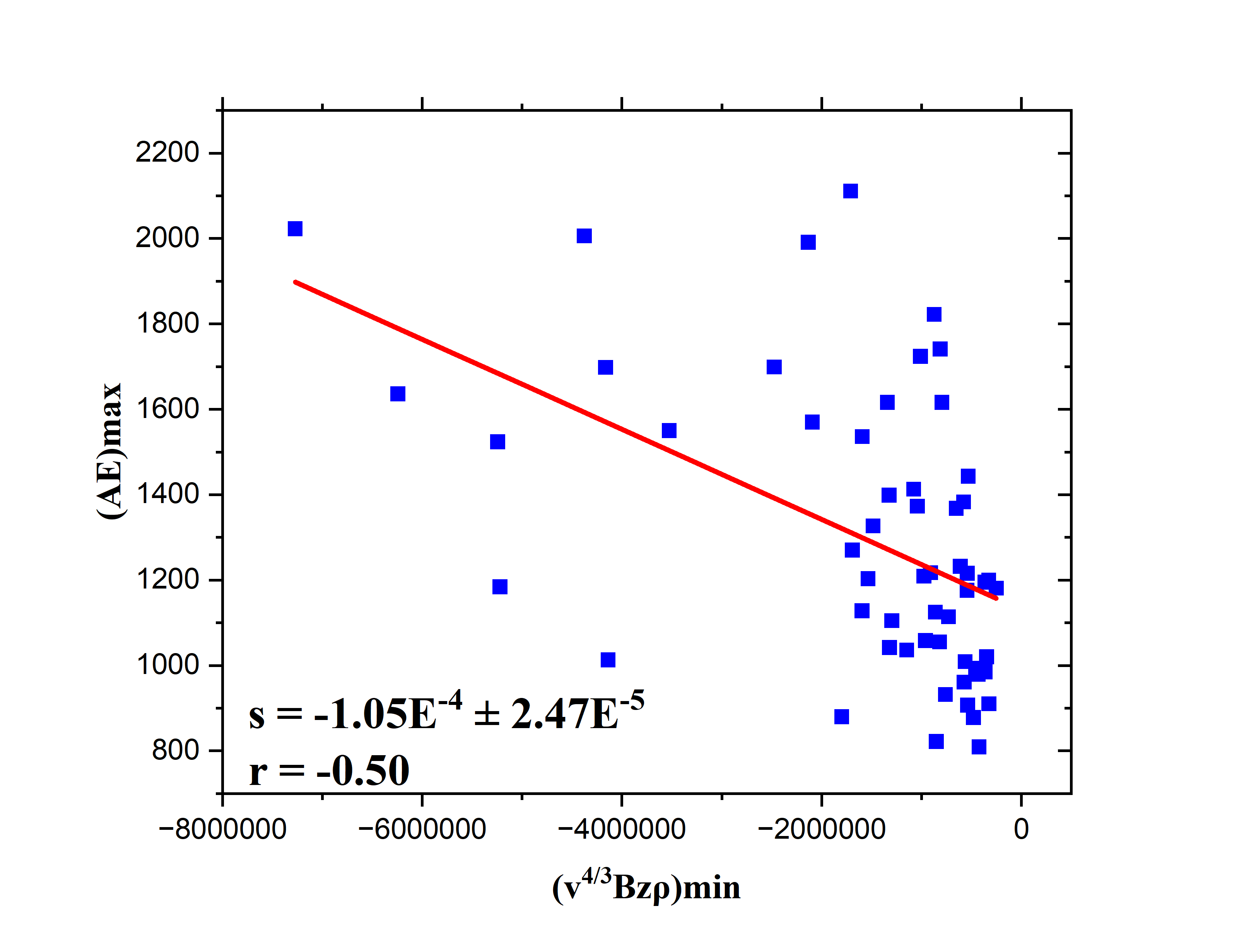}
		\includegraphics[height=3.8cm, width=5.9cm]{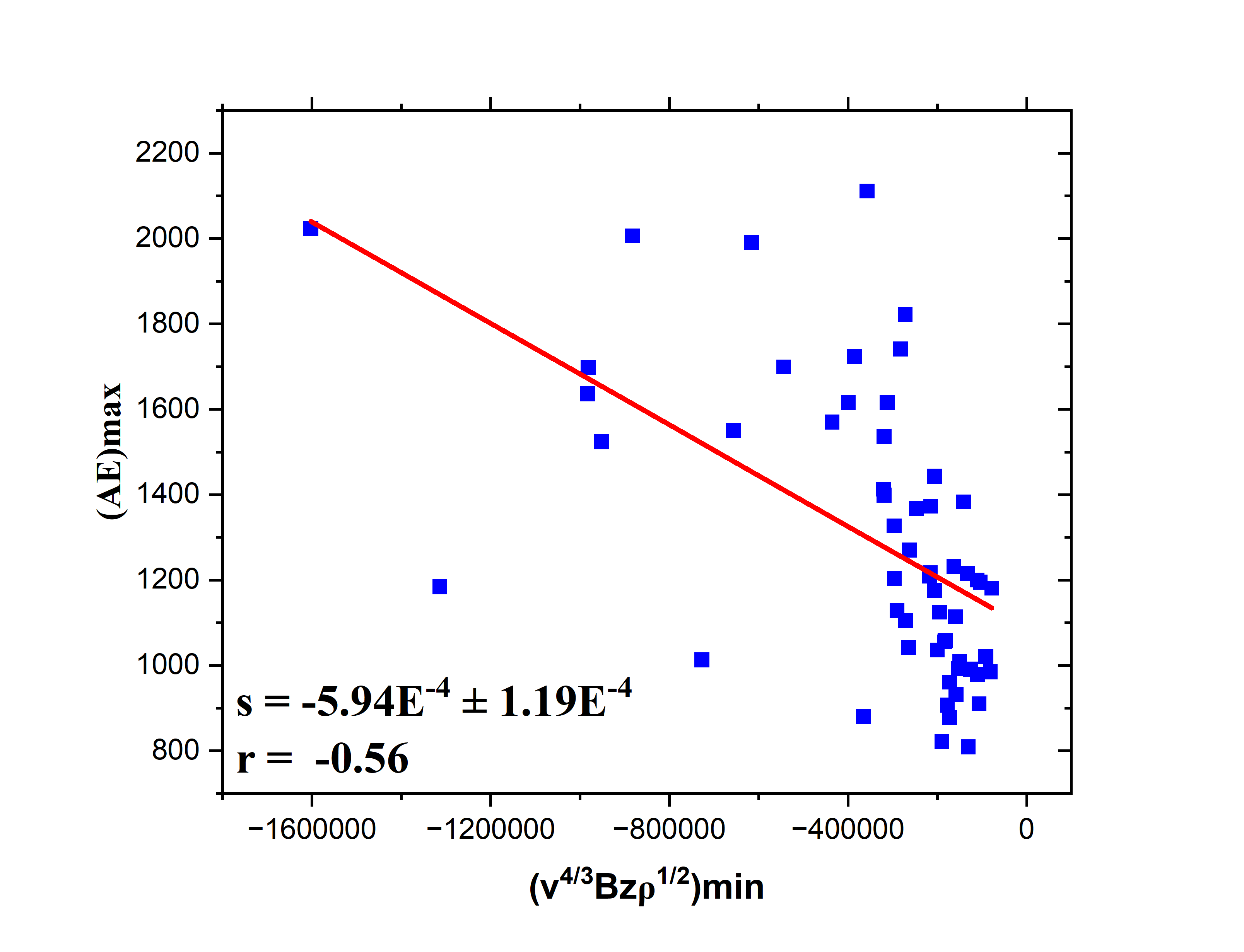}
		\\[\smallskipamount]
		\centering
		\includegraphics[height=3.8cm, width=5.9cm]{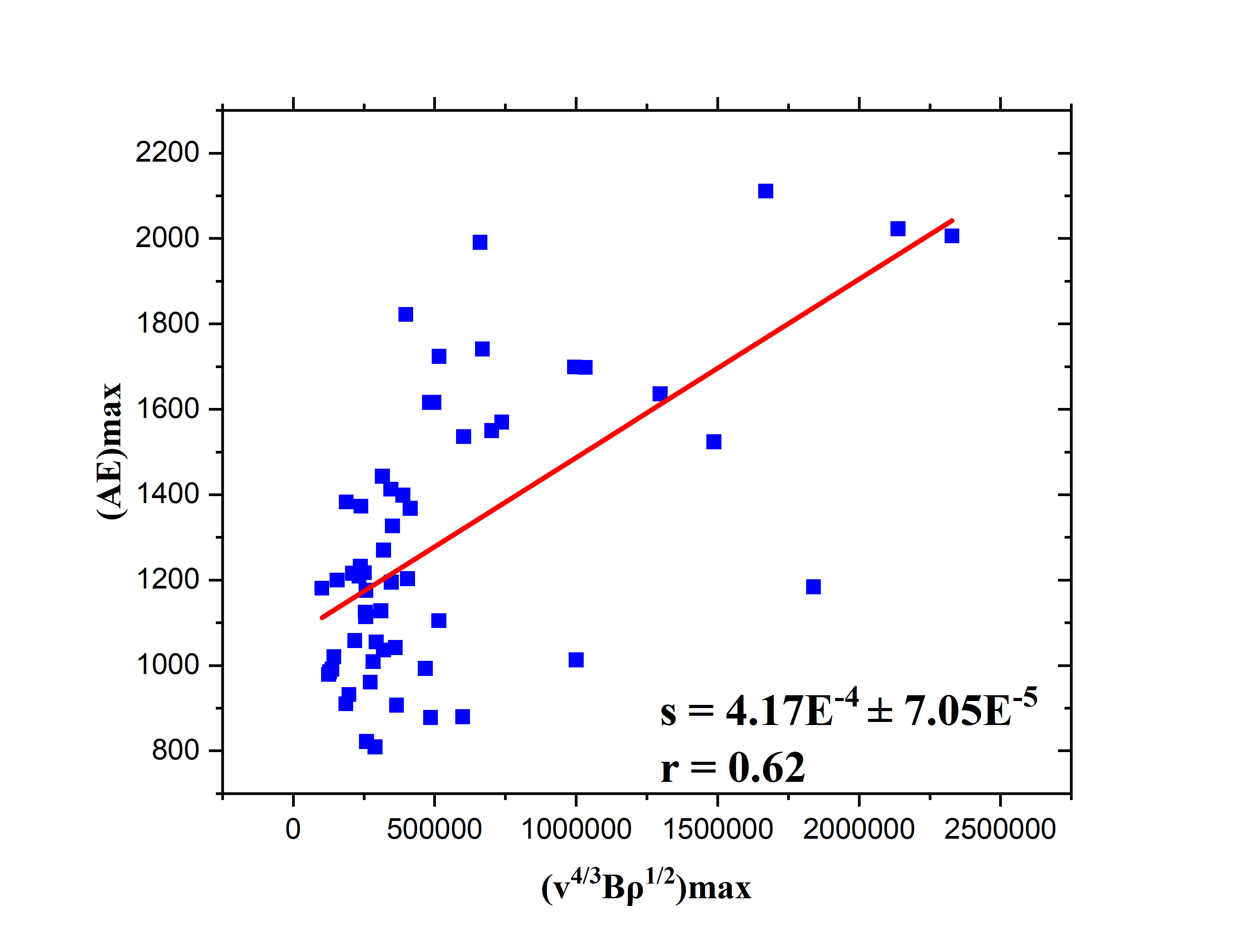}
		\caption{\small The scatter plots represent the best linear fit between the maximum value of AE during geomagnetic disturbances and with the corresponding peak values SW plasma and field parameters and their derive(nT)d functions. }\label{fig:AEcorr}
	\end{figure*}
\begin{figure*}
	\centering
		\includegraphics[height=3.8cm, width=5.9cm]{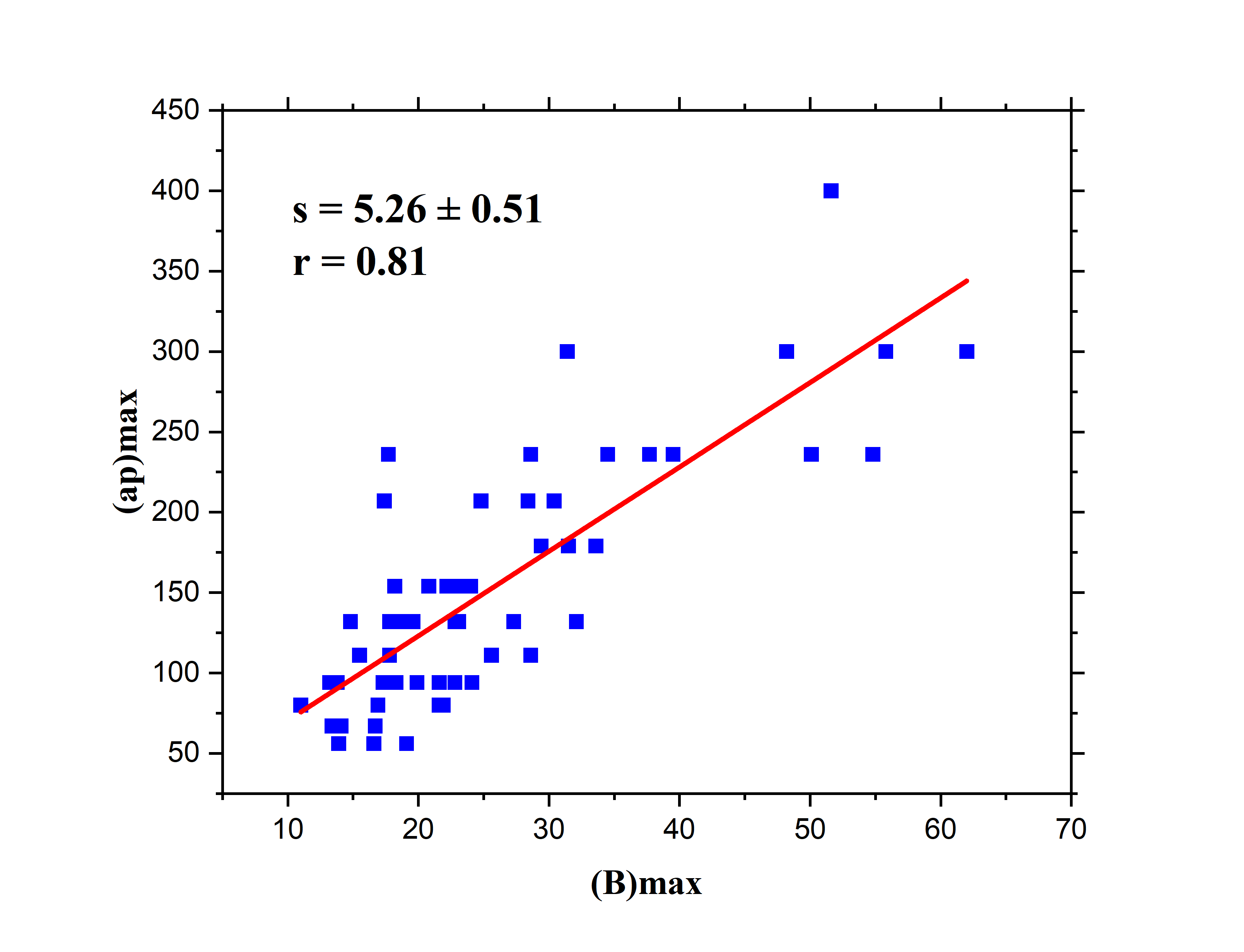}
		\includegraphics[height=3.8cm, width=5.9cm]{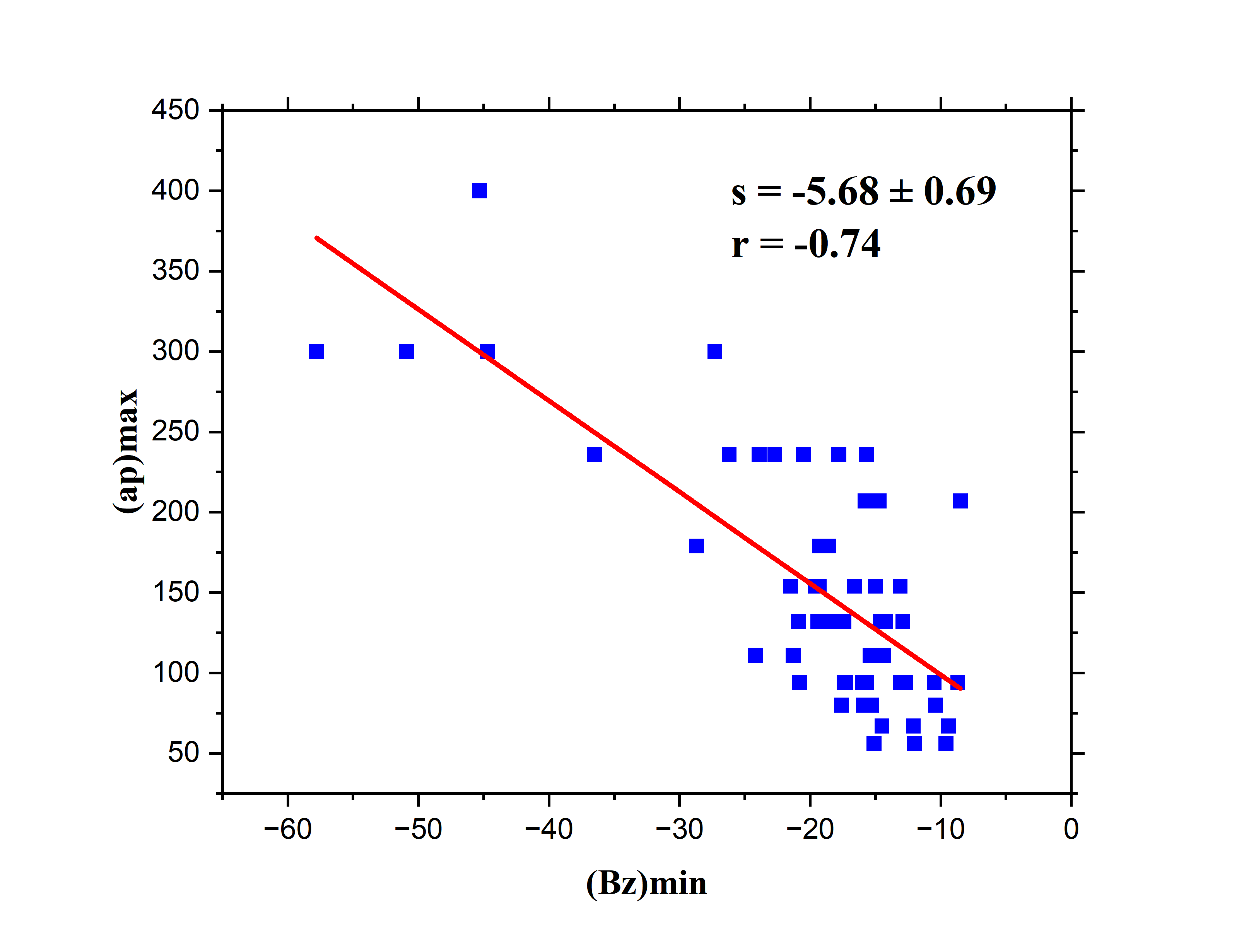}
		\includegraphics[height=3.8cm, width=5.9cm]{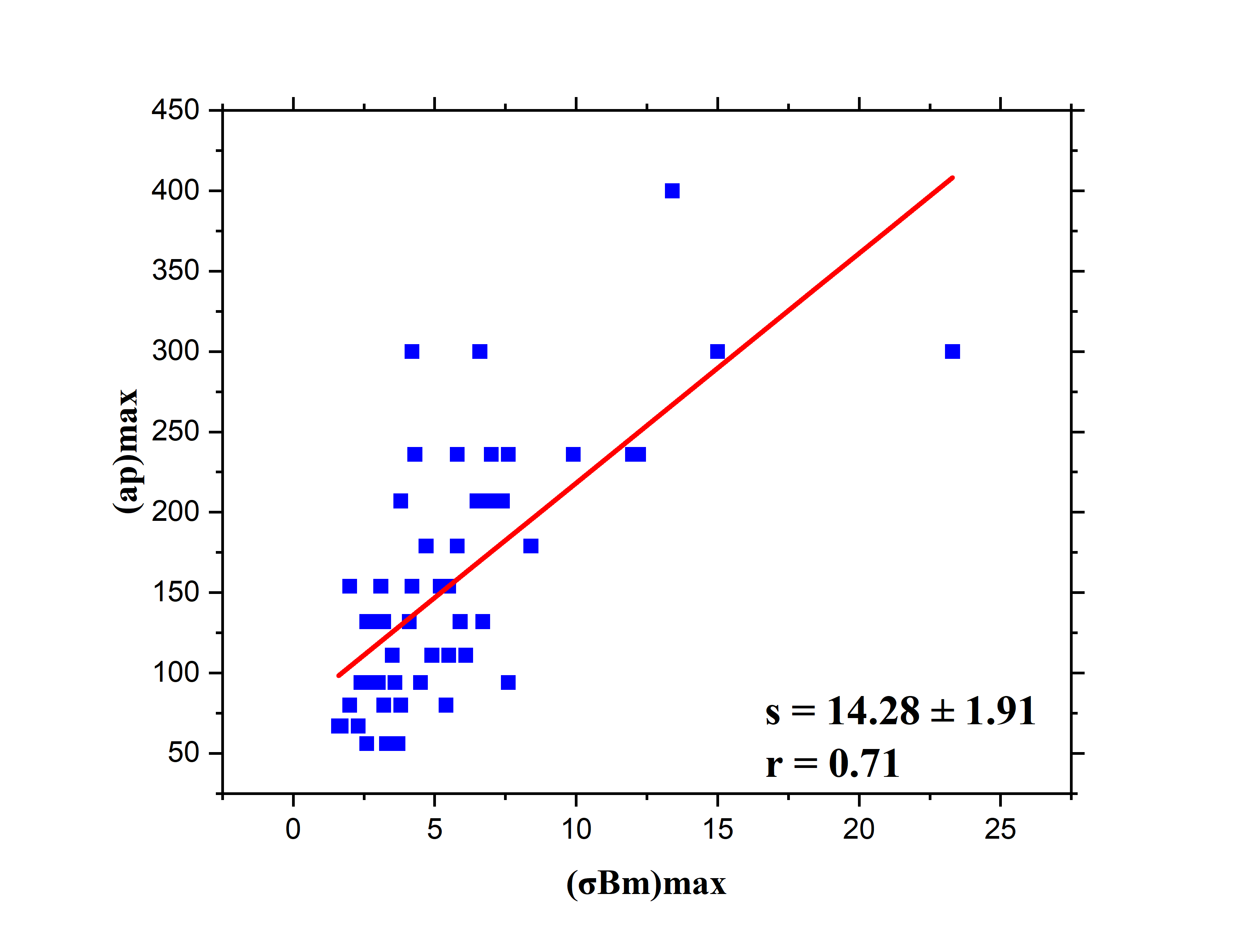}
		\\[\smallskipamount]
		\includegraphics[height=3.8cm, width=5.9cm]{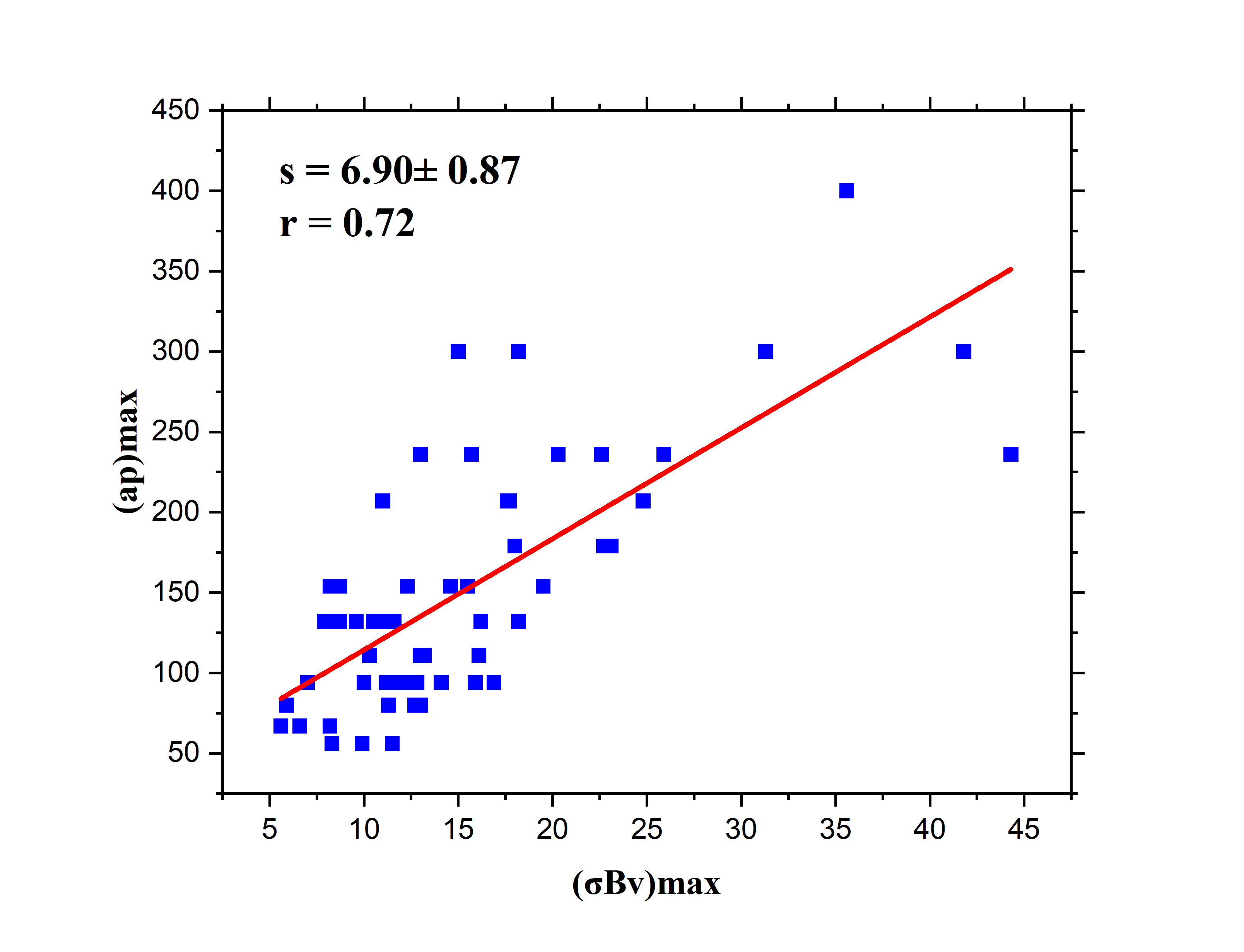}
		\includegraphics[height=3.8cm, width=5.9cm]{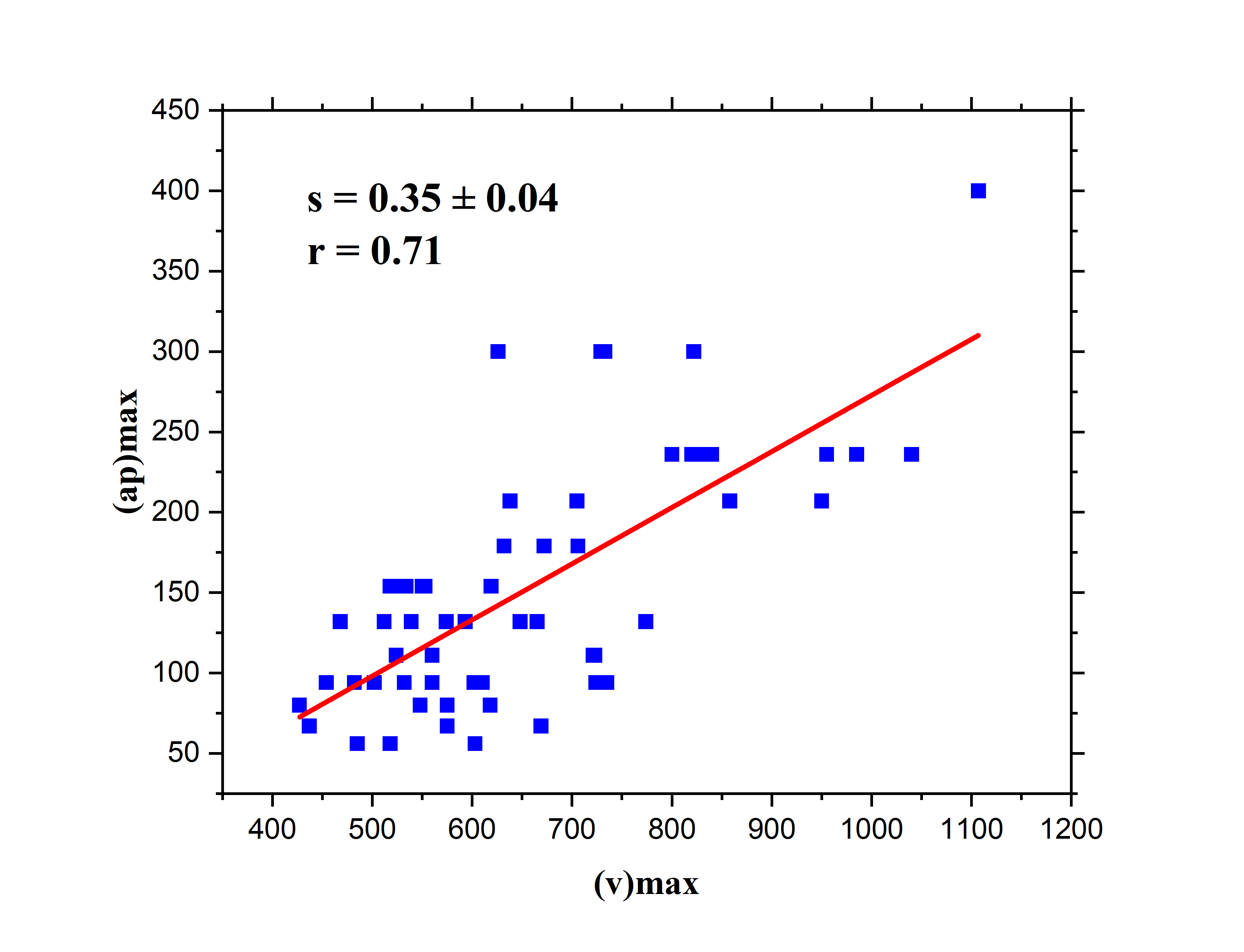}
		\includegraphics[height=3.8cm, width=5.9cm]{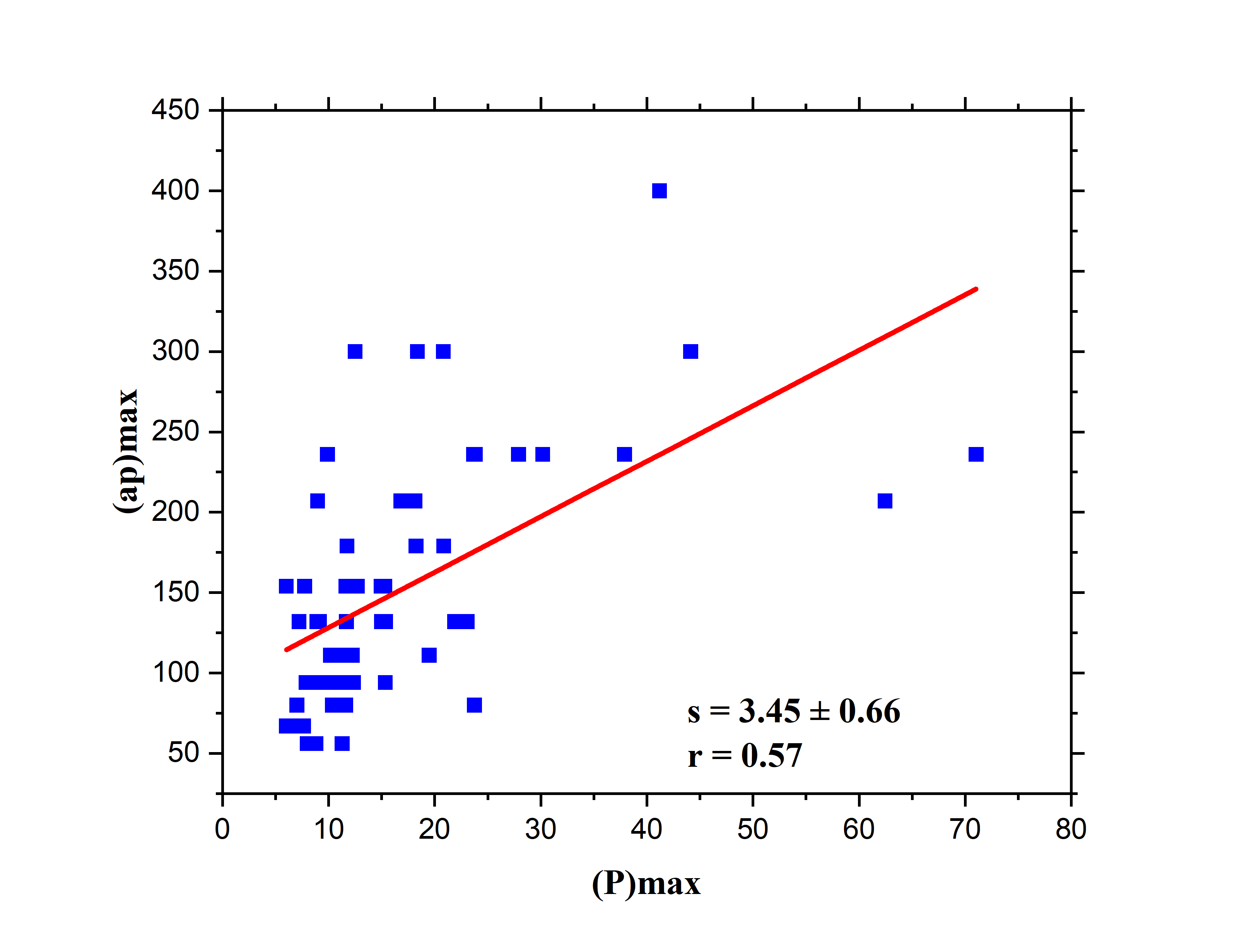}
		\\[\smallskipamount]
		\includegraphics[height=3.8cm, width=5.9cm]{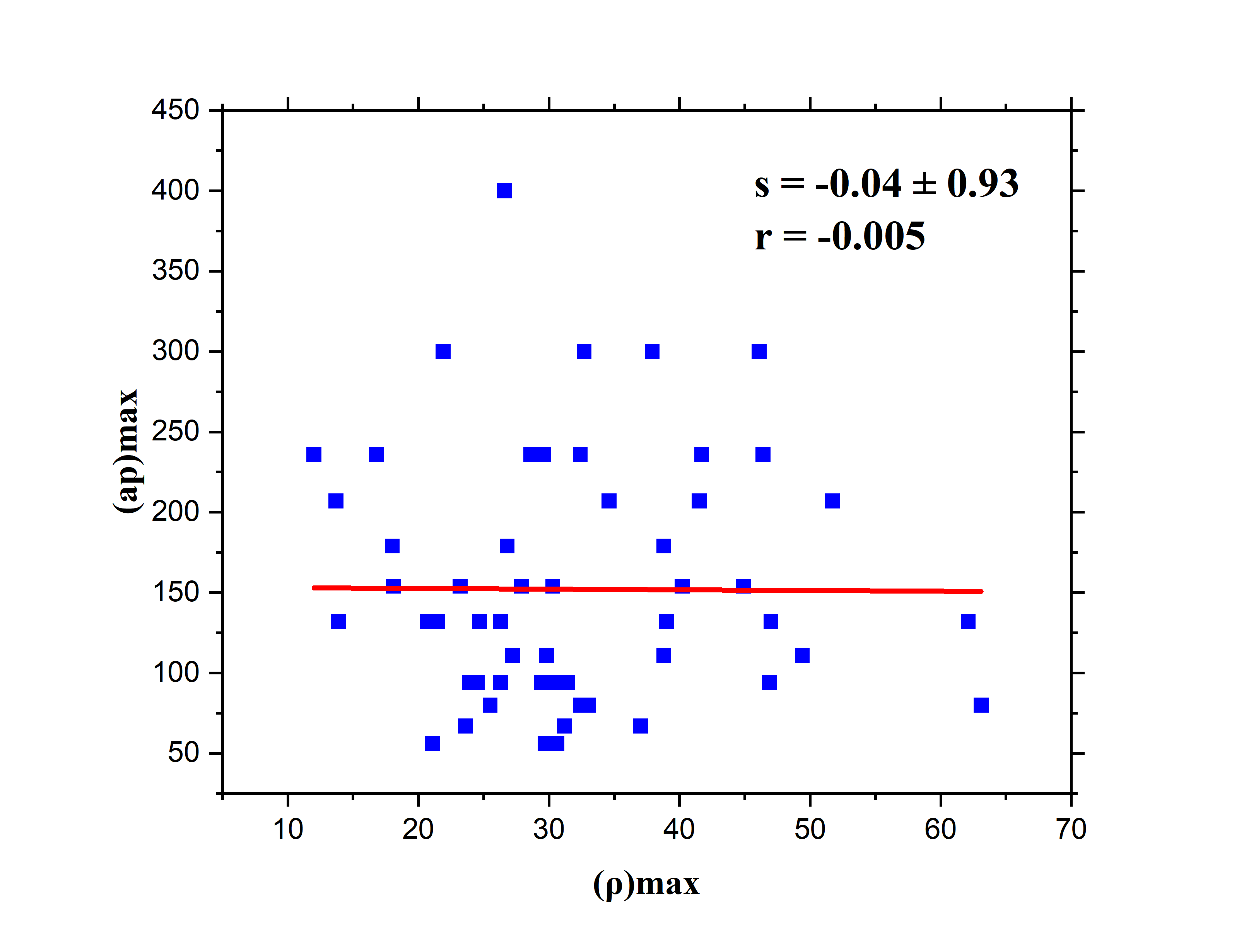}
		\includegraphics[height=3.8cm, width=5.9cm]{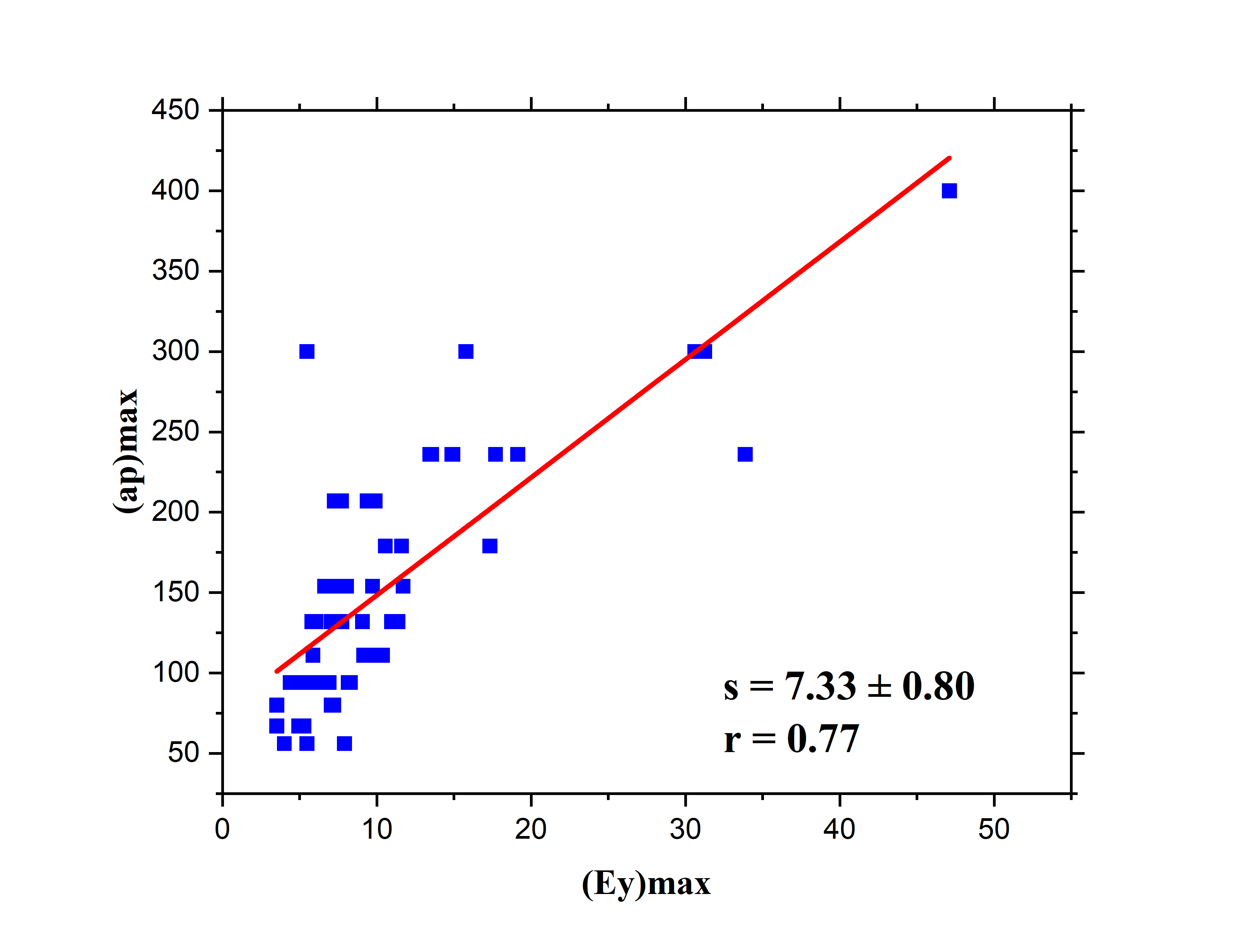}
		\includegraphics[height=3.8cm, width=5.9cm]{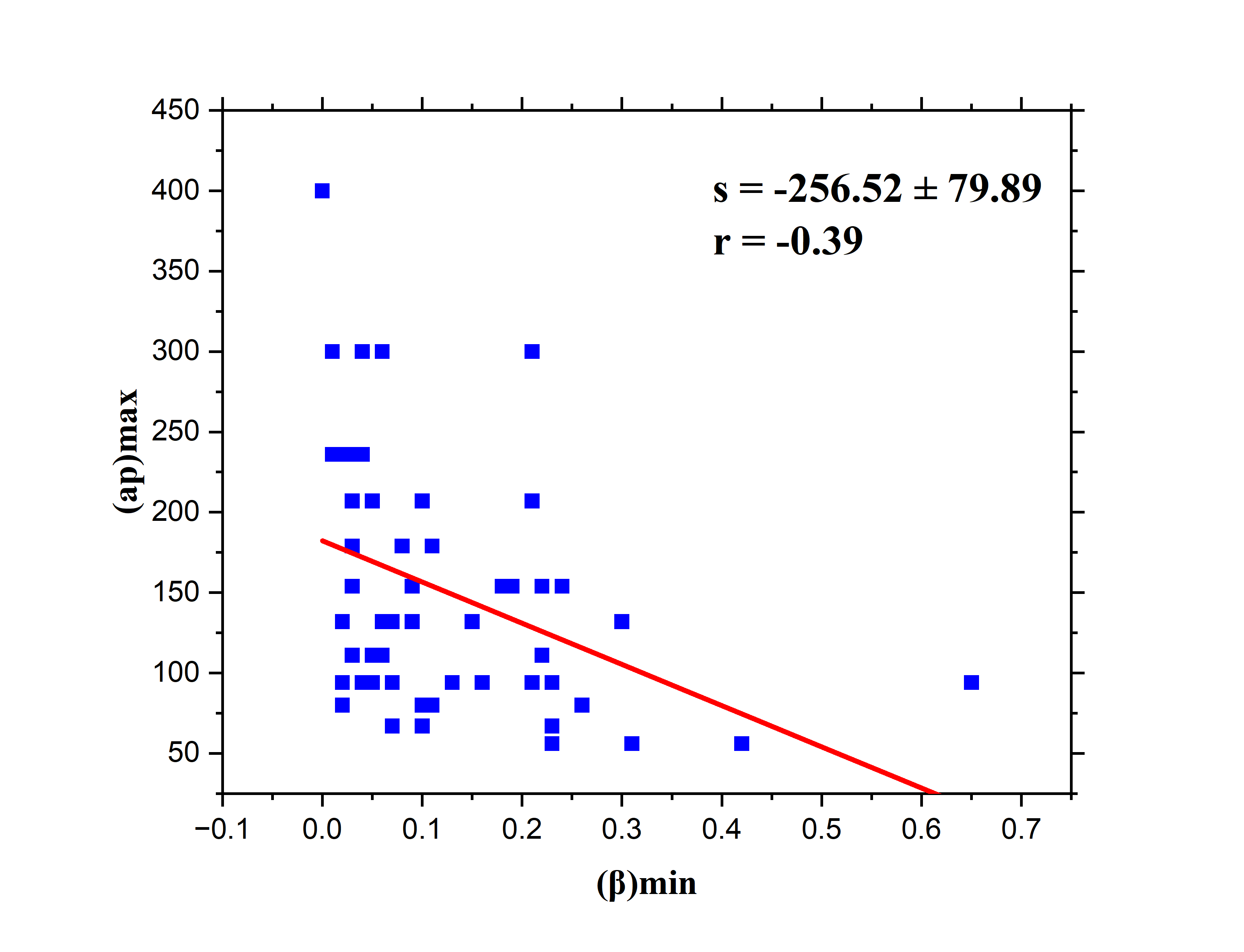}
		\\[\smallskipamount]
		\includegraphics[height=3.8cm, width=5.9cm]{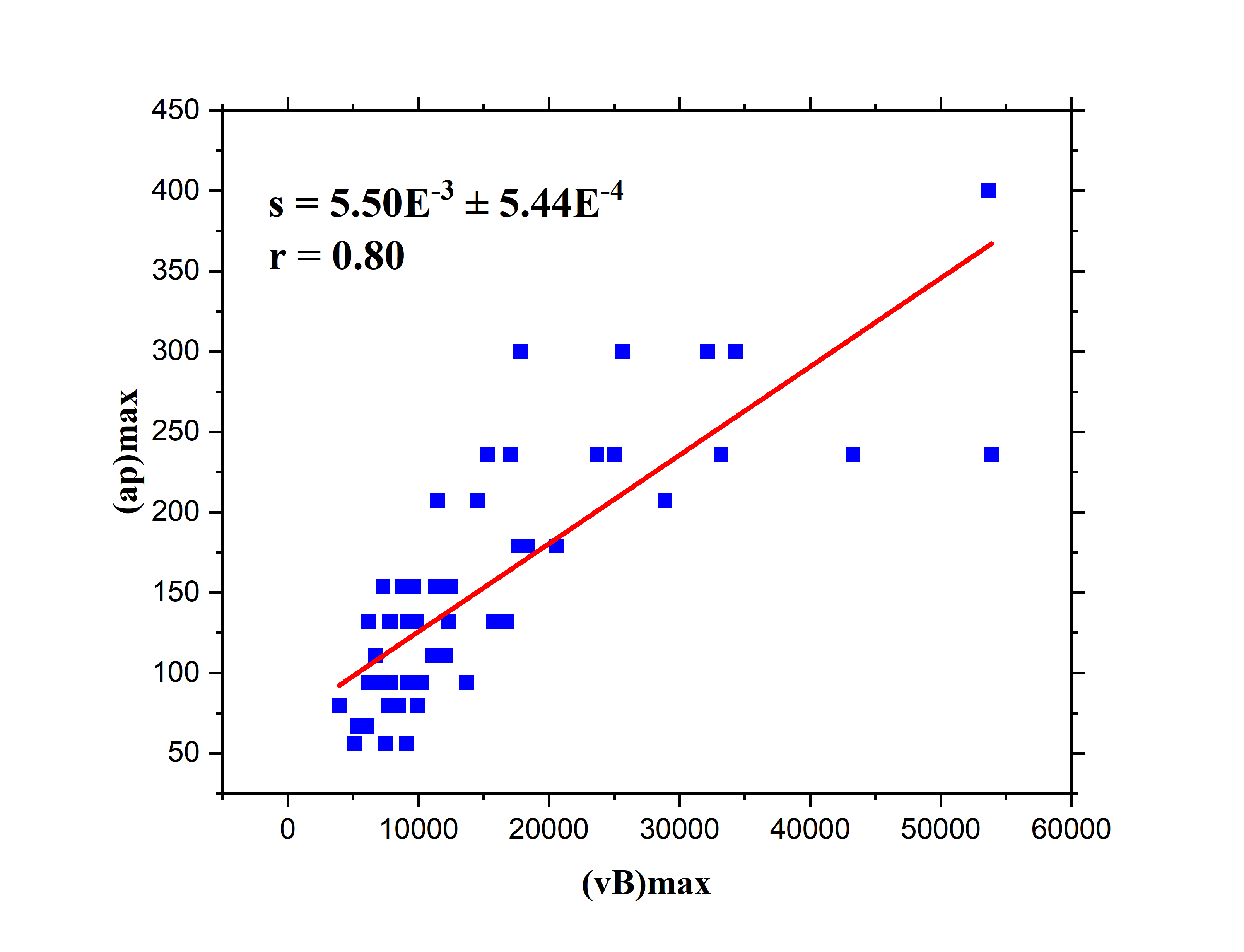}
		\includegraphics[height=3.8cm, width=5.9cm]{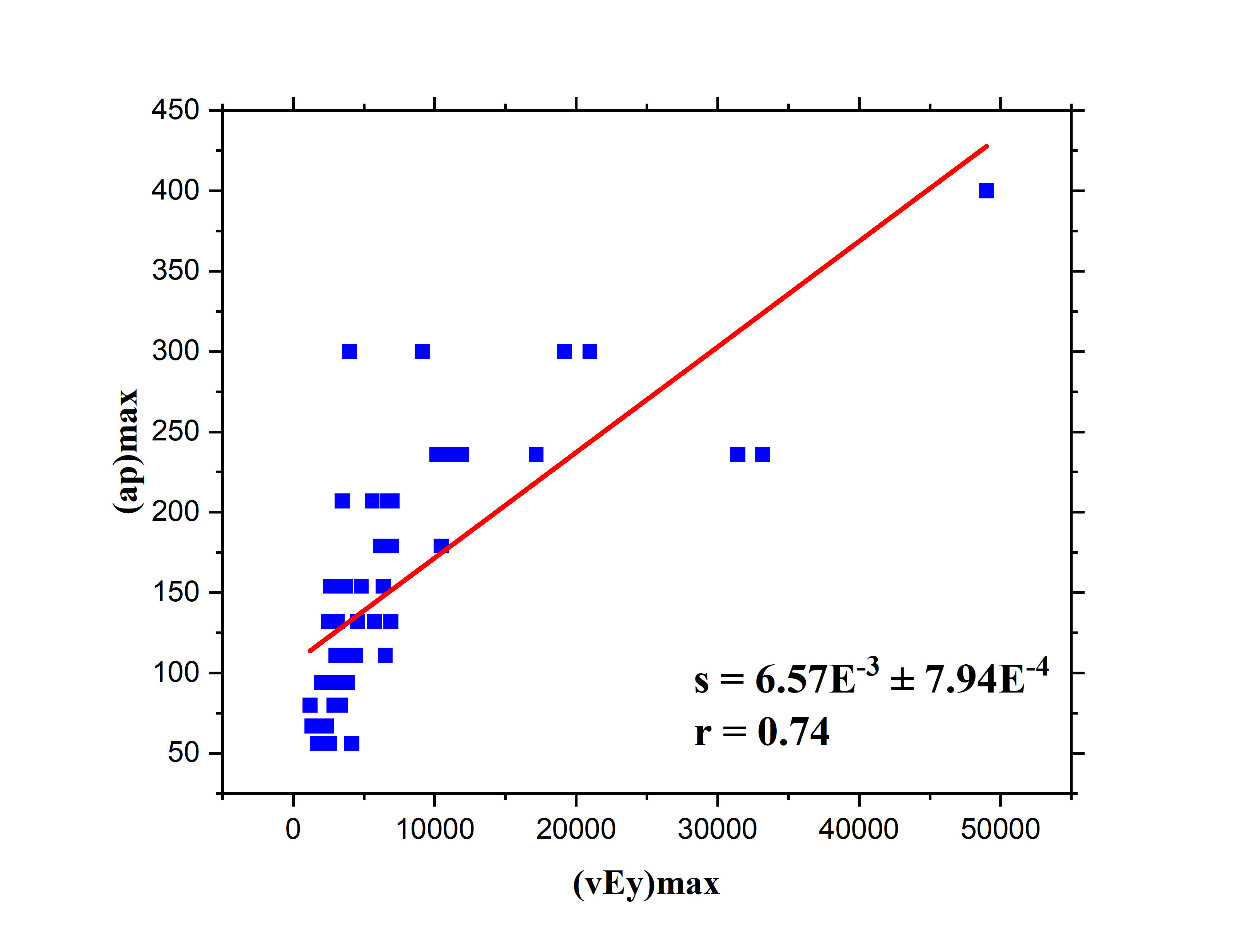}
		\includegraphics[height=3.8cm, width=5.9cm]{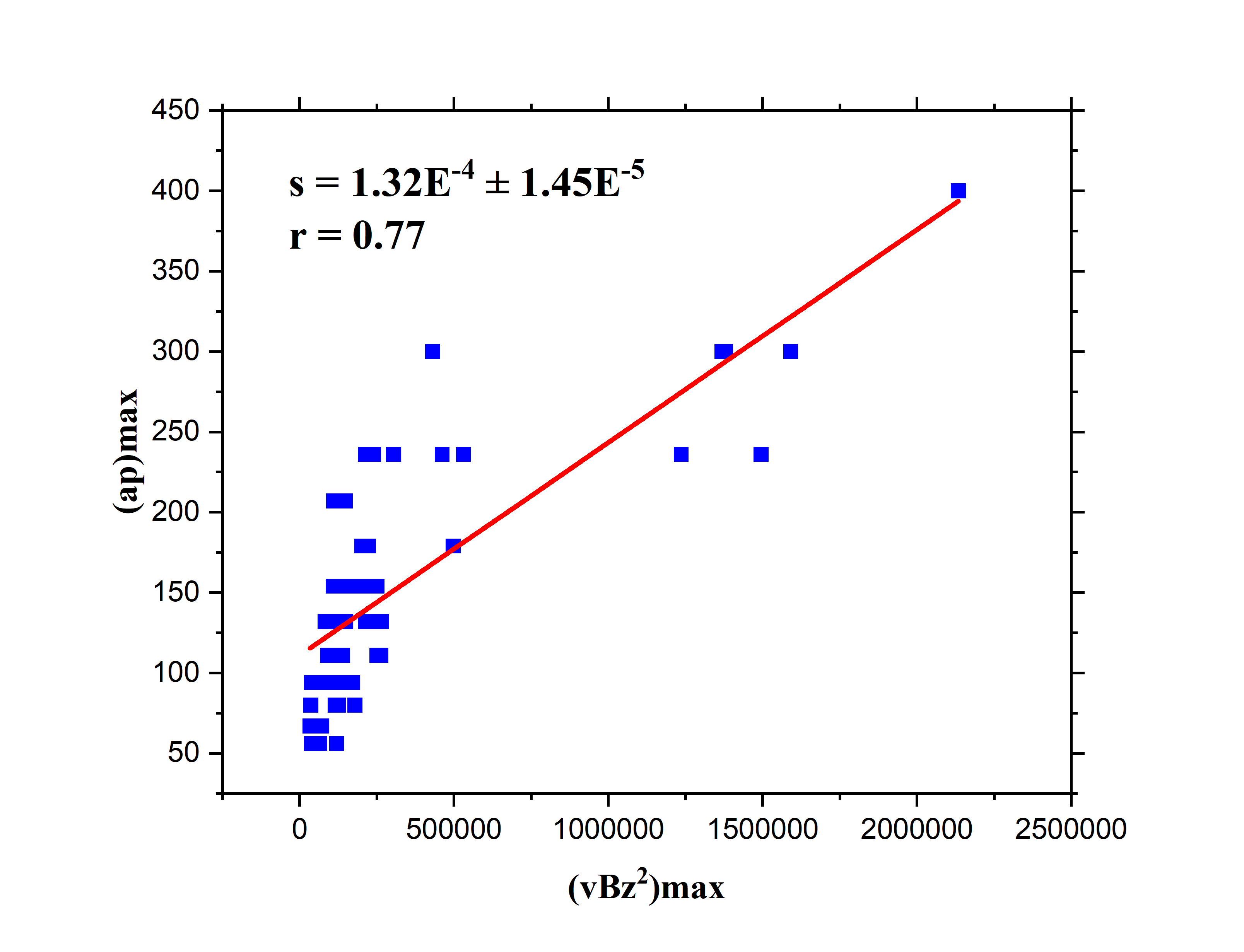}
		\\[\smallskipamount]
		\includegraphics[height=3.8cm, width=5.9cm]{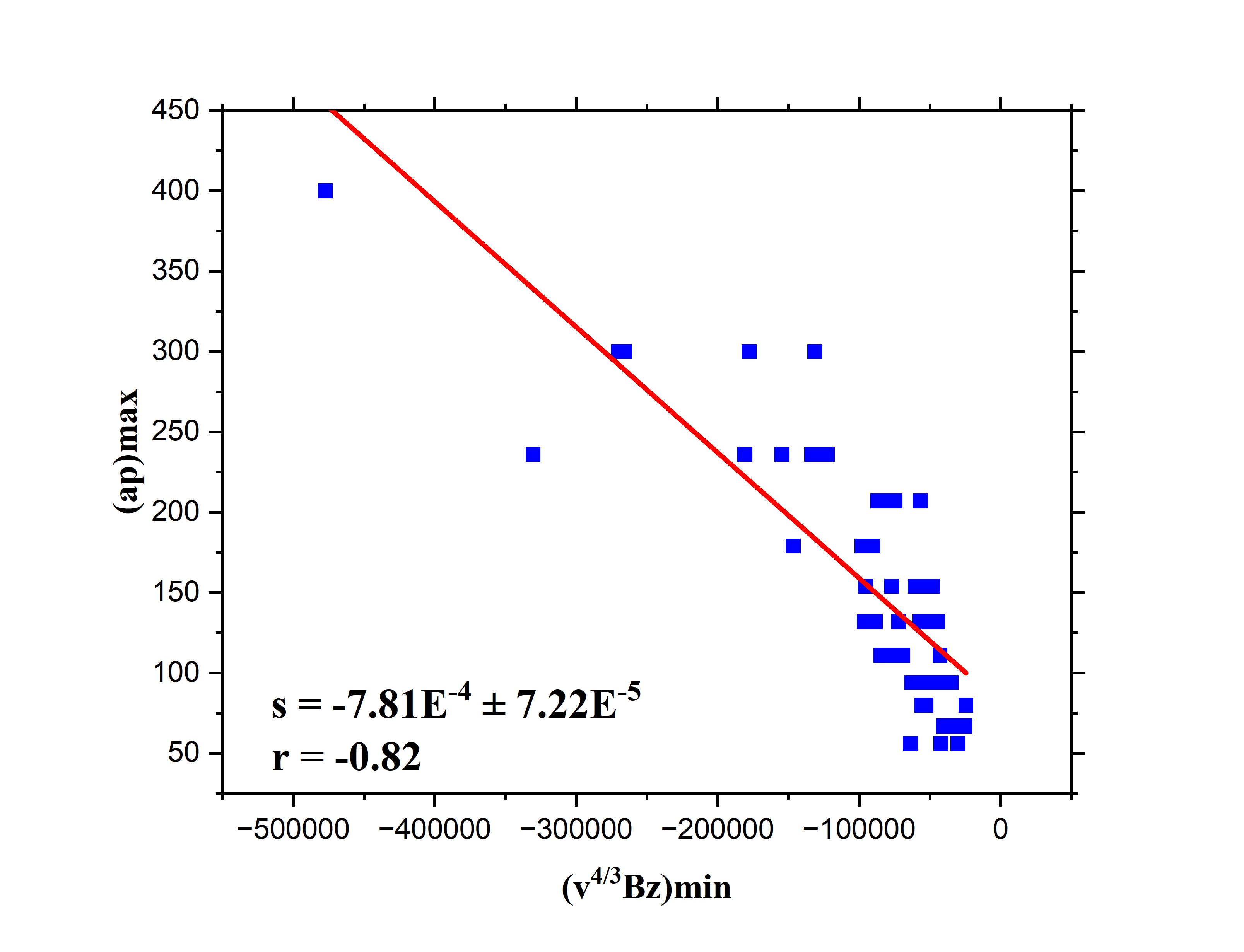}
		\includegraphics[height=3.8cm, width=5.9cm]{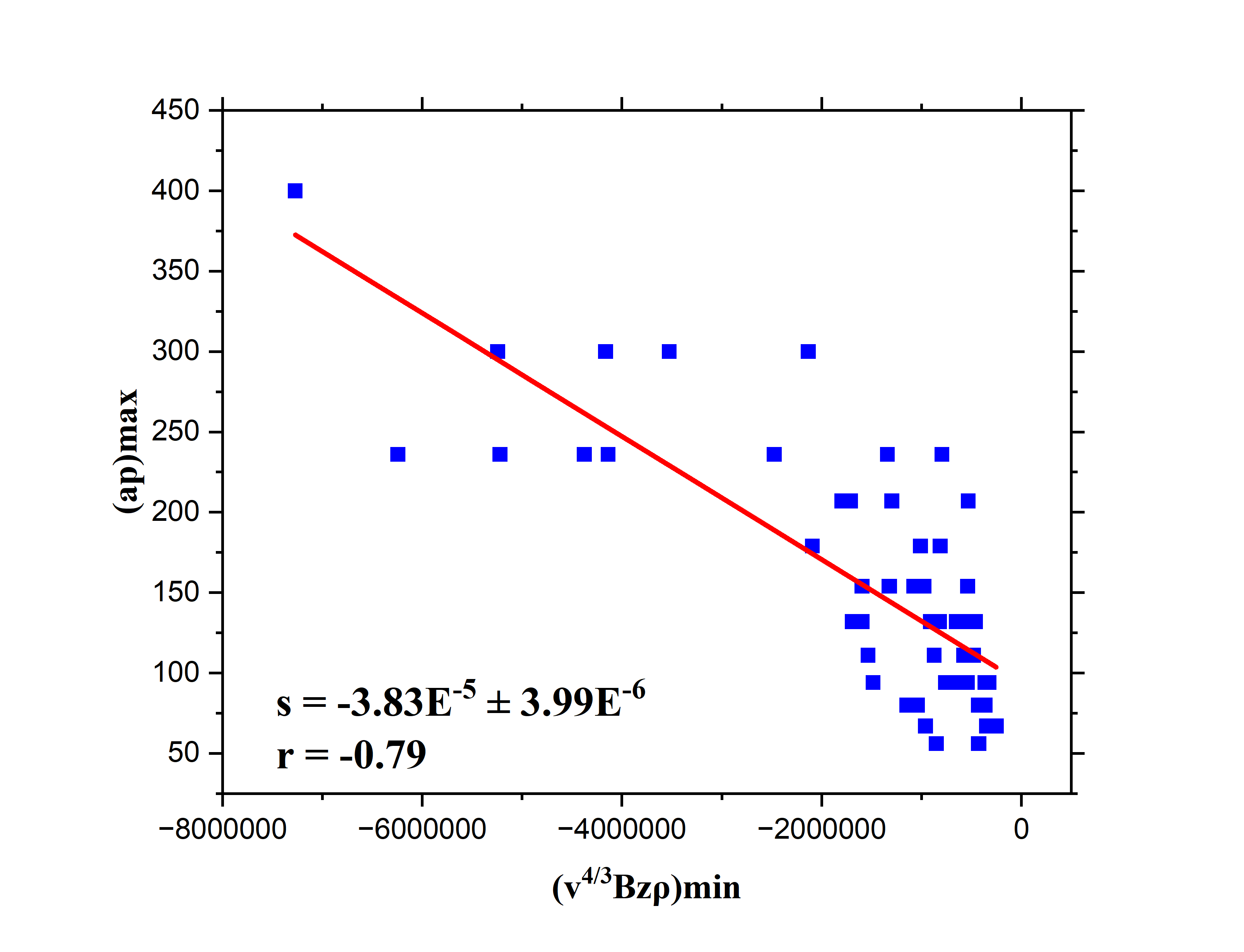}
		\includegraphics[height=3.8cm, width=5.9cm]{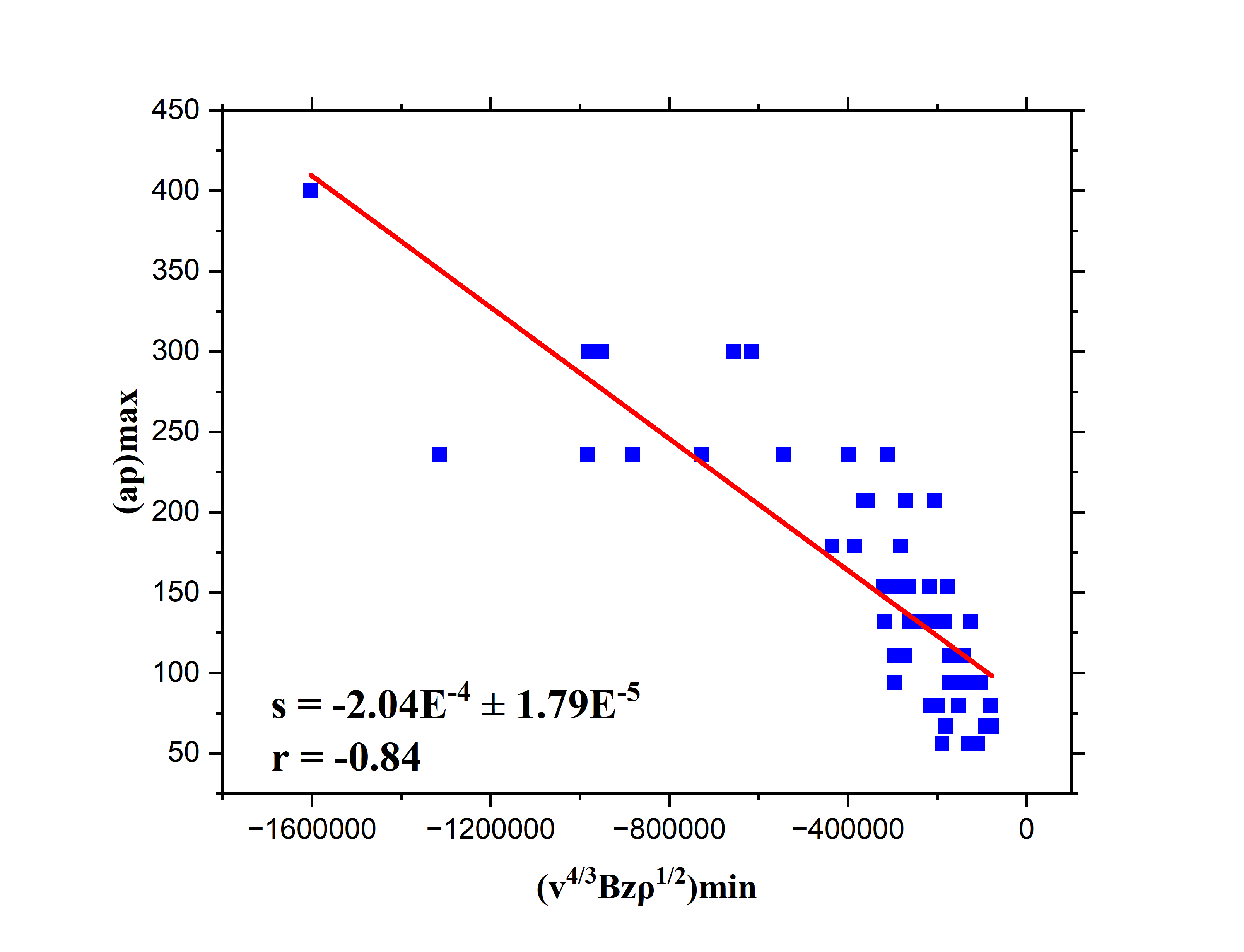}
		\\[\smallskipamount]
		\centering
		\includegraphics[height=3.8cm, width=5.9cm]{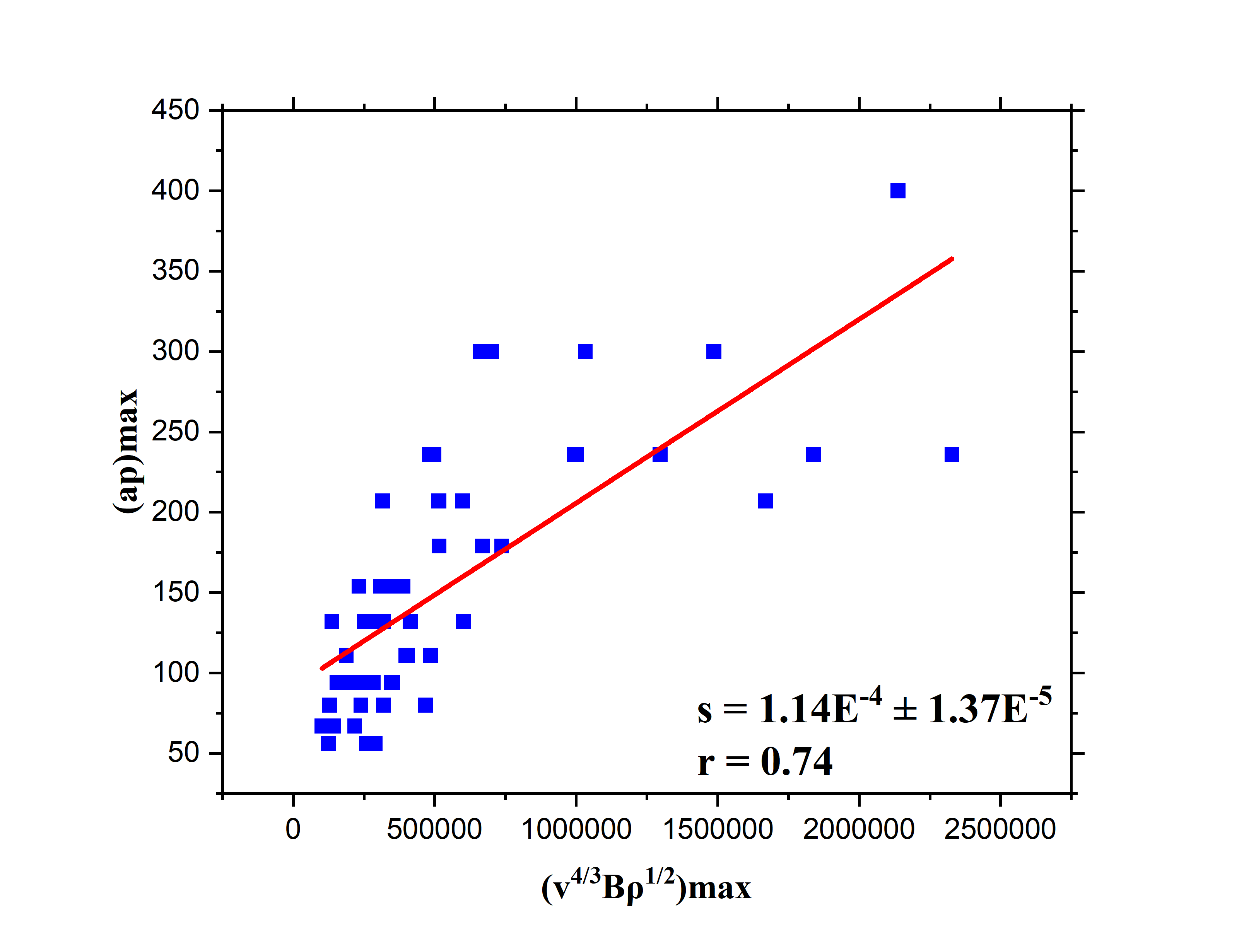}	
		\caption{\small The scatter plots show the best linear fit between the maximum values of ap during geomagnetic disturbances and the corresponding peak values of SW plasma and field parameters and their derived functions. }\label{fig:apcorr}
	\end{figure*}
\begin{figure*}
	\centering
 		\includegraphics[width=0.97\textwidth, height=13.1cm]{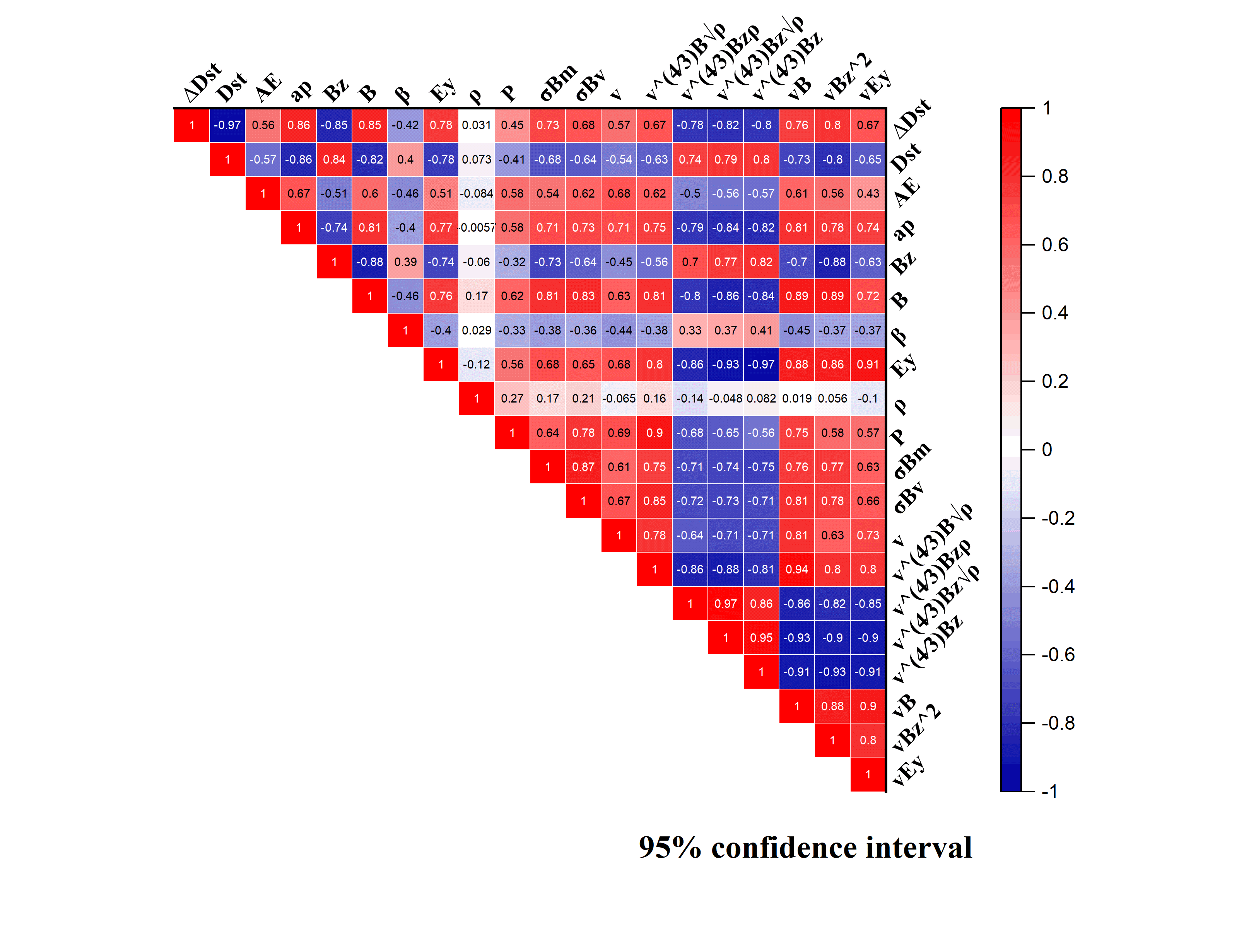}
		\caption{\small Triangular matrix showing the statistical significance of the linear correlation between geomagnetic indices and SW plasma and field parameters and their products. Red and blue represents strong positive and negative correlations respectively, while white shows no relation at 2-$\sigma $ confidence interval. The numbers in the boxes are correlation coefficients.}\label{stat2}
	\end{figure*}

Similar correlation analysis involves peak value reached in mid-latitude geomagnetic index (ap$_\mathrm{max}$) as well as polar geomagnetic index (AE$_\mathrm{max}$) with maximum value of field-plasma parameter (functions) during/around the time of peak geomagnetic disturbances. These correlations are plotted in Figures \ref{fig:AEcorr} and \ref{fig:apcorr} and results tabulated in Table \ref{tabl2}.
Figure \ref{fig:AEcorr} shows scatter plot of the linear relationship between the peak values of geomagnetic index AE during the GS with SW plasma and field parameters and their derivatives/functions. 
AE has the highest correlation with single parameter SW velocity v showing best fit parameter Pearson's correlation coefficient ($0.68$), with the best fit equation [AE$_\mathrm{max}=(1.47 \pm 0.24)\mathrm{v_\mathrm{max}}+(305.18 \pm 166.87)$]. Our finding confirms that the high latitude geomagnetic index AE has strong relation with plasma velocity which is in agreement with the previous results of \citep{Joshi2010BehaviorOP}.
The second best correlated single parameter showing best fit Pearson's correlation coefficient ($0.62$) with AE is the IMF vector $\sigma$B$_\mathrm{v}$ which is an indicator of field fluctuations. 
AE has the highest correlation with dual parameter CF of IP electric field vB showing best fit parameter Pearson's correlation coefficient ($0.61$), with the best fit equation [AE$_\mathrm{max}=[0.02\pm3.57\mathrm{E}^{-3}$](vB)$_\mathrm{max}+(984.63\pm66.08)$]. The second best correlated dual parameter CF showing best fit Pearson's correlation coefficient ($-0.57$) is v$ ^{\frac{4}{3}}$ Bz.
AE has the highest correlation with triple parameter CF v$ ^{\frac{4}{3}} $B$\rho^{\frac{1}{2}}$ showing best fit parameter Pearson's correlation coefficient ($0.62$), with the best fit equation [AE$_\mathrm{max}=[4.40\mathrm{E}^{-4}\pm7.95\mathrm{E}^{5}$](v$ ^{\frac{4}{3}}$B$\rho^{\frac{1}{2}})_\mathrm{max}+(1038.41\pm57.70)$]. 
The second best correlation triple parameter CF showing best fit Pearson's correlation coefficient ($-0.56$) is v$^{\frac{4}{3}}$Bz$\rho^{\frac{1}{2}}$.\\
	Figure \ref{fig:apcorr} shows scatter plot of the linear relationship between the peak values of geomagnetic index ap and other SW plasma and field parameters and derivatives. 
The ap has the highest correlation with single parameter IMF B showing best fit parameter Pearson's correlation coefficient ($0.81$), with the best fit equation 
[ap$_{\mathrm{max}}=[5.26\pm0.51]$(B$_{\mathrm{max}})+(17.83\pm14.41)$]. 
The second best correlation single parameter showing best fit Pearson's correlation coefficient ($-0.74$) with ap is the IMF Bz. 
The ap has the highest correlation with dual parameter CFs v$^{\frac{4}{3}}$Bz showing best fit parameter Pearson's correlation coefficient ($-0.82$), with the best fit equation [ap$_{\mathrm{max}}=[-7.81\mathrm{E}^{-4}\pm7.22\mathrm{E}^{-5}]$(v$^{\frac{4}{3}}$Bz)$_{\mathrm{min}}+(80.91\pm8.76)$]. The second best correlation dual parameter CF showing best fit Pearson's correlation coefficient $0.80$ is IP electric field (vB). These results support \citep{2018AdSpR..61..348B, 2020JASTP.20405290B} who used Kp.
The ap has the highest correlation with triple parameter CF v$ ^{\frac{4}{3}} $Bz$\rho^{\frac{1}{2}}$ showing best fit parameter Pearson's correlation coefficient ($-0.84$), with the best fit equation [ap$_\mathrm{max}=(-2.04\mathrm{E}^{-4} \pm 1.79\mathrm{E}^{-5})(\mathrm{v}^{\frac{4}{3}}$Bz$\rho^{\frac{1}{2}})_\mathrm{min}+(82.18\pm 8.26)$].
The second best correlation triple parameter CF showing best fit Pearson's correlation coefficient $-0.79$ is v$^{\frac{4}{3}}$Bz$\rho$. 
Thus, for (ap$_\mathrm{max}$) also, as in case of ($\Delta$Dst), a merging term when coupled with a viscous term improved the correlation \citep[e.g.,][]{2008JGRA..113.4218N}.\\
	\begin{table*}
		\small{
			\caption{\small List of the Pearson's linear correlation coefficient and the slope as a result of a linear fitting between geomagnetic indices (Dst, AE \& ap) and SW plasma and field parameters with their products.  \label{tabl2}}
			\centering{	
				\begin{tabular}{l cccccccc}
					\hline\
					\textbf{Parameters}  & \multicolumn{2}{c}{\underline{\textbf{Dst (nT)}}}  &\multicolumn{2}{c}{\underline{\textbf{$\Delta$Dst (nT)}}} &\multicolumn{2}{c}{\underline{\textbf{AE (nT)}}}&\multicolumn{2}{c}{\underline{\textbf{ap (nT)}}}\\
					&\textbf{r}&\textbf{s}&\textbf{r}&\textbf{s}&\textbf{r}&\textbf{s}&\textbf{r}&\textbf{s}\\
					\hline
B &-0.82 &$-5.33\pm0.50$ &0.85 &$5.71\pm0.47$ &0.60 &$17.01\pm3.06$ &0.81&$5.26\pm0.51$ \\
Bz & 0.84 &$6.48\pm0.55$&-0.84 &$-6.68\pm0.57$ &-0.51 &$-17.04\pm4.89$ & -0.74&$-5.68\pm0.69$ \\ 
$\sigma$B$_{\mathrm{m}}$ & -0.67 &$-13.61\pm1.99$&0.73 &$15.16\pm1.90$ &0.54 &$47.86\pm9.95$ & 0.71&$14.28\pm1.91$ \\ 
$\sigma$B$_{\mathrm{v}}$ & -0.64 &$-6.10\pm0.98$&0.67 &$6.60\pm0.97$ &0.62 &$25.61\pm4.40$ & 0.72&$6.90\pm0.87$ \\ 
v &-0.53 &$-0.26\pm0.05$ &0.57 &$0.29\pm0.05$ &0.68 &$1.45\pm0.21$ & 0.71&$0.35\pm0.04$ \\ 
P &-0.41 &$-2.48\pm0.73$ &0.45 &$2.76\pm0.74$ &0.58 &$15.15\pm2.87$ & 0.57&$3.45\pm0.66$ \\
$\rho$ &0.07 &$0.50\pm0.93$ &0.03 &$0.22\pm0.96$ &-0.08 &$-2.54\pm4.06$ & -0.005&$-0.04\pm0.93$ \\ 
Ey &-0.77 &$-7.35\pm0.80$ &0.77 &$7.56\pm0.83$ &0.51 &$20.89\pm4.81$ & 0.77&$7.33\pm0.80$ \\ 
$\beta$ &0.40 &$258.72\pm79.73$ &-0.42 &$-279.05\pm81.42$ &-0.46 &$-1300.85\pm337.75$ & -0.39&$-256.52\pm79.89$ \\ 
vB  &-0.73 &-4.97E$^{-3}\pm$6.30E$^{-4}$ &0.76 &5.33E$^{-3}\pm$6.17E$^{-4}$ &0.61 &$0.02\pm0.00$ & 0.80&5.50E$^{-3}\pm$5.44E$^{-4}$ \\ 
vEy &-0.65 &-5.77E$^{-3}\pm$8.99E$^{-4}$ &0.67 &6.06E$^{-3}\pm$9.13E$^{-4}$ &0.43 &$0.01\pm$0.00 & 0.74&6.57E$^{-3}\pm$7.94E$^{-4}$ \\ 
vBz$^2$ &-0.80 &-1.36E$^{-4}\pm$1.38E$^{-5}$ &0.79 &1.40E$^{-4}\pm$1.42E$^{-5}$ &0.56 &4.20E$^{-4}\pm$8.31E$^{-5}$ & 0.77&1.32E$^{-4}\pm$1.45E$^{-5}$ \\ 
v$^{\frac{4}{3}}$Bz &0.79 &7.52E$^{-4}\pm$7.74E$^{-5}$ &-0.79 &-7.78E$^{-4}\pm$7.92E$^{-5}$ &-0.57 &$-0.00\pm$4.58E$^{-4}$ & -0.82&-7.81E$^{-4}\pm$7.22E$^{-5}$\\ 
v$^{\frac{4}{3}}$Bz$\rho$ &0.74 &3.59E$^{-5}\pm$4.36E$^{-6}$ &-0.78 &-3.90E$^{-5}\pm$4.18E$^{-6}$ &-0.50 &-1.05E$^{-4}\pm$2.47E$^{-5}$ & -0.79&-3.83E$^{-5}\pm$3.99E$^{-6}$\\
v$^{\frac{4}{3}}$Bz$\rho^{\frac{1}{2}}$ &0.79 &1.93E$^{-4}\pm$2.00E$^{-5}$ &-0.82 &-2.05E$^{-4}\pm$1.94E$^{-5}$ &-0.56 &-5.94E$^{-4}\pm$1.19E$^{-4}$ & -0.84&-2.04E$^{-4}\pm$1.79E$^{-5}$\\ 
v$^{\frac{4}{3}}$B$\rho^{\frac{1}{2}}$ &-0.62 &-9.58E$^{-5}\pm$1.60E$^{-5}$ &0.66 &1.05E$^{-4}\pm$1.58E$^{-5}$ &0.62 &4.17E$^{-4}\pm$7.05E$^{-5}$ & 0.74&1.14E$^{-4}\pm$1.37E$^{-5}$\\
\hline
\multicolumn{4}{l}{\footnotesize \textbf{r}, represents Pearson's correlation coefficient.}\\
\multicolumn{4}{l}{\footnotesize \textbf{s}, stands for the slope of the linear fit characteristic.}\\
\end{tabular}}}
	\end{table*}
Figure \ref{stat2} represents the statistical hypothesis test of Pearson's correlation coefficient for the geomagnetic indices with the SW plasma and field parameters and derivatives with 95\% confidence level. The Pearson coefficients are shown in the square of upper triangular matrix as recorded in the Table \ref{tabl2} and Figures \ref{fig:Dstcorr} and \ref{fig:changeDstcorr}. The red and blue values are highly significant parameters while the white colored boxes show statistically insignificant or poorly significant. For those highly significant correlations the null hypothesis (H$_0=0$) is rejected. In this regard the CF v$^{\frac{4}{3}}$Bz$ \rho^{\frac{1}{2}}$ is the strongly correlated with storm indices, southward Bz magnitude is strongly related with Dst in particular.
	
	\begin{table*}
	\centering
		\small 
		\caption{\small Statistical significance test of Pearson's correlation coefficient with 95\% confidence interval for correlation between geomagnetic indices and SW plasma and field parameters and derivatives for Figures \ref{fig:Dstcorr}, \ref{fig:changeDstcorr}, \ref{fig:AEcorr} \& \ref{fig:apcorr}.  The calculated t-score
			and p-value are recorded. T-score ($t=\frac{\bar{x}-\mu}{s_d}\sqrt{n}$) where, $ \bar{x} $ is the sample mean, 	  
			$ \mu $ is the population mean,
			$s_d$ is the standard deviation of the sample, and
			n is the sample size. \label{tabl5}}
		\centering
		\begin{tabular}{l cccccccc}
			\hline
			\textbf{Parameters}  & \multicolumn{2}{c}{\underline{\textbf{Dst (nT)}}}  &\multicolumn{2}{c}{\underline{\textbf{$\Delta$Dst (nT)}}} &\multicolumn{2}{c}{\underline{\textbf{AE (nT)}}}&\multicolumn{2}{c}{\underline{\textbf{ap (nT)}}}\\
			&\textbf{t-score}&\textbf{p-value}&\textbf{t-score}&\textbf{p-value}&\textbf{t-score}&\textbf{p-value}&\textbf{t-score}&\textbf{p-value}\\
			\hline
			B &-15.55 &5.52E$ ^{-22} $ &-15.09 &2.20E$ ^{-21} $ &-26.11 &4.93E$ ^{-33} $ &-14.22 & 3.05E$ ^{-20} $\\
			Bz & -14.71 &6.76E$ ^{-21} $&-15.80 &2.68E$ ^{-22} $ &-26.21&4.01E$ ^{-33} $ &-15.41 &8.36E$ ^{-22} $ \\ 
			$\sigma$B$_{\mathrm{m}}$ &15.06 &2.35E$ ^{-21} $ &-15.69 &3.67E$ ^{-22} $ &26.19 &4.16E$ ^{-33} $ &-15.03 & 2.62E$ ^{-21} $\\ 
			$\sigma$B$_{\mathrm{v}}$ &15.48 &6.84E$ ^{-22} $ &-15.18 &1.65E$ ^{-21} $ &-26.19 &4.22E$ ^{-33} $ &-14.59 & 9.93E$ ^{-21} $\\ 	
			v &29.44 &9.17E$ ^{-36} $ &29.02 &1.98E$ ^{-35} $ &-15.77 &2.96E$ ^{-22} $ &33.09 & 1.86E$ ^{-38} $\\ 
			P &-15.48 &6.80E$^{-22} $ &-14.96 &3.21E$^{-21} $ &-26.32 &3.22E$^{-33} $ &14.63 & 8.85E$^{-21} $\\	
$\rho$ &-18.19 &3.58E$ ^{-25} $ &-12.55 &6.49E$ ^{-18} $ &25.46 &1.78E$ ^{-32} $ &11.77 & 8.91E$ ^{-17} $\\
			Ey &-14.84 &4.66E$ ^{-21} $ &-15.88 &2.09E$ ^{-22} $ &-26.21 &3.97E$ ^{-33} $ &-15.21 & 1.52E$ ^{-21} $\\ 
			$\beta$ &15.07 &2.31E$ ^{-21} $ &15.64 &4.25E$ ^{-22} $ &-26.15 &4.51E$ ^{-33} $ &-15.02 & 2.70E$ ^{-21}$\\
			vB &10.05 &3.71E$ ^{-14} $ &9.95 &5.53E$ ^{-14} $ &9.32 &5.43E$ ^{-13} $ &9.95& 5.34E$ ^{-14} $\\
			vEy &6.22 &6.56E$ ^{-08} $ &6.02 &1.40E$ ^{-07} $ &5.10 &4.12E$ ^{-06} $ &6.03 & 1.34E$ ^{-07} $\\
			vBz$^2$ &5.24 &2.51E$ ^{-06} $ &5.24 &2.54E$ ^{-06} $ &5.22 &2.70E$ ^{-06} $ &5.23 & 2.54E$ ^{-06} $\\ 			
			v$^{\frac{4}{3}}$Bz &-8.53 &1.04E$ ^{-11} $ &-8.55 &9.83E$ ^{-12} $ &-8.63 &7.04E$ ^{-12} $ &8.54 & 9.87E$ ^{-12} $\\ 	
			v$^{\frac{4}{3}}$Bz$\rho$ &-7.27 &1.23E$ ^{-09} $ &-7.27 &1.22E$ ^{-09} $ &-7.28 &1.20E$ ^{-09} $ &-7.27 & 1.23E$ ^{-09} $\\				
			v$^{\frac{4}{3}}$Bz$\rho^{\frac{1}{2}}$ &-8.24 &3.11E$ ^{-11} $ &-8.24 &3.06E$ ^{-11} $ &-8.27 &2.80E$ ^{-11} $ &-8.24& 3.07E$ ^{-11} $\\ 			
			v$^{\frac{4}{3}}$B$\rho^{\frac{1}{2}}$ &8.04 &6.67E$^{-11}$ &8.03 &6.75E$^{-11}$ &8.02 &7.12E$^{-11}$ &8.03 & 6.75E$^{-11}$\\
			\hline
			\multicolumn{9}{l}{\footnotesize All values of t-score and p-value are calculated from 2-$ \sigma $ confidence level.}\\
		\end{tabular}
	\end{table*}	
As discussed earlier, We tested some functional forms involving single terms (e.g., B, Bz, v, $\rho$), two terms (merging/electric field-related/simple relations) (vBz, vB, v$^2$Bz, v Bz$^2$, v$^{\frac{4}{3}}$Bz), and three terms (combined merging and viscus terms) (v Bz$^2$, v$^{\frac{4}{3}}$Bz$\rho$, v$^{\frac{4}{3}}$Bz$\rho^{\frac{1}{2}}$, v$^{\frac{4}{3}}$B$\rho^{\frac{1}{2}}$). However, our list of potential SW-CFs not exhaustive. \citep[see e.g.,][]{1981SSRv...28..121A, 1990P&SS...38..627G, 2007JGRA..112.1206N, 2008JGRA..113.4218N} for a more extensive list of potential CFs. 
However, our selection of SW parameters/functions is guided by some of the successfully employed ones and their variants. Our interest in this work is to find a suitable parameter/function whose magnitude best relates with the magnitude of GS during the MP and the amplitudes of other geomagnetic indices considered here.  \\
Among single SW parameters (B, v, $\rho$) including single-merging (Bz) terms, magnitude of (Bz) is best correlated with intensity of GSs (r $=0.84$) with Dst$_\mathrm{min}$, however magnitude of B is also very closely related (r $=0.82$) almost similar in the case with $\Delta$Dst. 
During big GSs, the peak ap intensity, (ap$_\mathrm{max}$), best correlates with the magnitude of B (r $=0.81$), followed by the magnitude of (Bz) (r $=-0.74$). When compared to Dst and ap, AE index correlations are weaker; peak AE during big GSs (among those included here) best correlates with v (r $=0.62$), followed by B (r $=0.57$). 
Among the functions containing two SW terms (merging /electric field-related/simple relations), the magnitude of vBz$^2$ is most closely associated to the intensity of GS (Dst$_\mathrm{min}$) (r $=0.80$), however the magnitude of (v$^{\frac{4}{3}}$Bz) is also very closely related (r $=0.79$). The term (v$^{\frac{4}{3}}$Bz) is most strongly connected (r $=0.82$) to the peak of ap during large GSs, followed by the term vB (r $=0.80$). Among the parameters considered, the function vB (r $=0.59$) correlates best with AE maximum during GS, followed by (v$^{\frac{4}{3}}$Bz) (r $=0.54$).  
These two parameter electric field related functions form somewhat better than duskward electric field (Ey=-vBz) with peak in all three respective geomagnetic indices during MP of GSs.\\
\cite{2008JGRA..113.4218N} proposed that a merging term combined with any viscous term performs reasonably well for the SW-magnetosphere interaction. We found that a CF incorporating v, Bz, and $\rho$ (i.e. v$^{\frac{4}{3}}$Bz$\rho^{\frac{1}{2}}$) highly correlates with GS intensity (Dst$_\mathrm{min}$) (r $=0.79$) as well as peak ap$_\mathrm{max}$ (r $=0.84$). The correlation of peak AE during GSs improved with v$^{\frac{4}{3}}$B$\rho^{\frac{1}{2}}$ (r $=-0.60$) when compared to other SW parameters (and their combinations) considered in this study. However, (v$^{\frac{4}{3}}$Bz$\rho^{\frac{1}{2}}$) is also a reasonably well associated function with AE$_\mathrm{max}$ (r $=0.53$).
Thus, in conclusion we can treat, with reasonable accuracy, that the parameter (v$^{\frac{4}{3}}$Bz$\rho^{\frac{1}{2}}$) is a consistent parameter to represent not only the intensity of GSs (as represented by Dst$_\mathrm{min}$) but also the peaks expected in mid-latitude index (ap) and polar index (AE). $\Delta$Dst and Dst$_\mathrm{min}$ correlations almost agree with respective parameters.
In conclusion, the magnitude of the function, involving  $\rho$, v and Bz or B (v$^{\frac{4}{3}}$Bz$\rho^{\frac{1}{2}}$) is suggested to be a reasonably good predictor of the intensity of GS (Dst$_\mathrm{min}$ and $\Delta$Dst) as well as the amplitude of mid-latitude geomagnetic index ap (ap$_\mathrm{max}$) during the geomagnetic disturbances.

	\subsection{Inter-relationship between geomagnetic indices}\label{sub:correndex}
\begin{figure*}
		\centering
		\includegraphics[height=4.2cm, width=5.9cm]{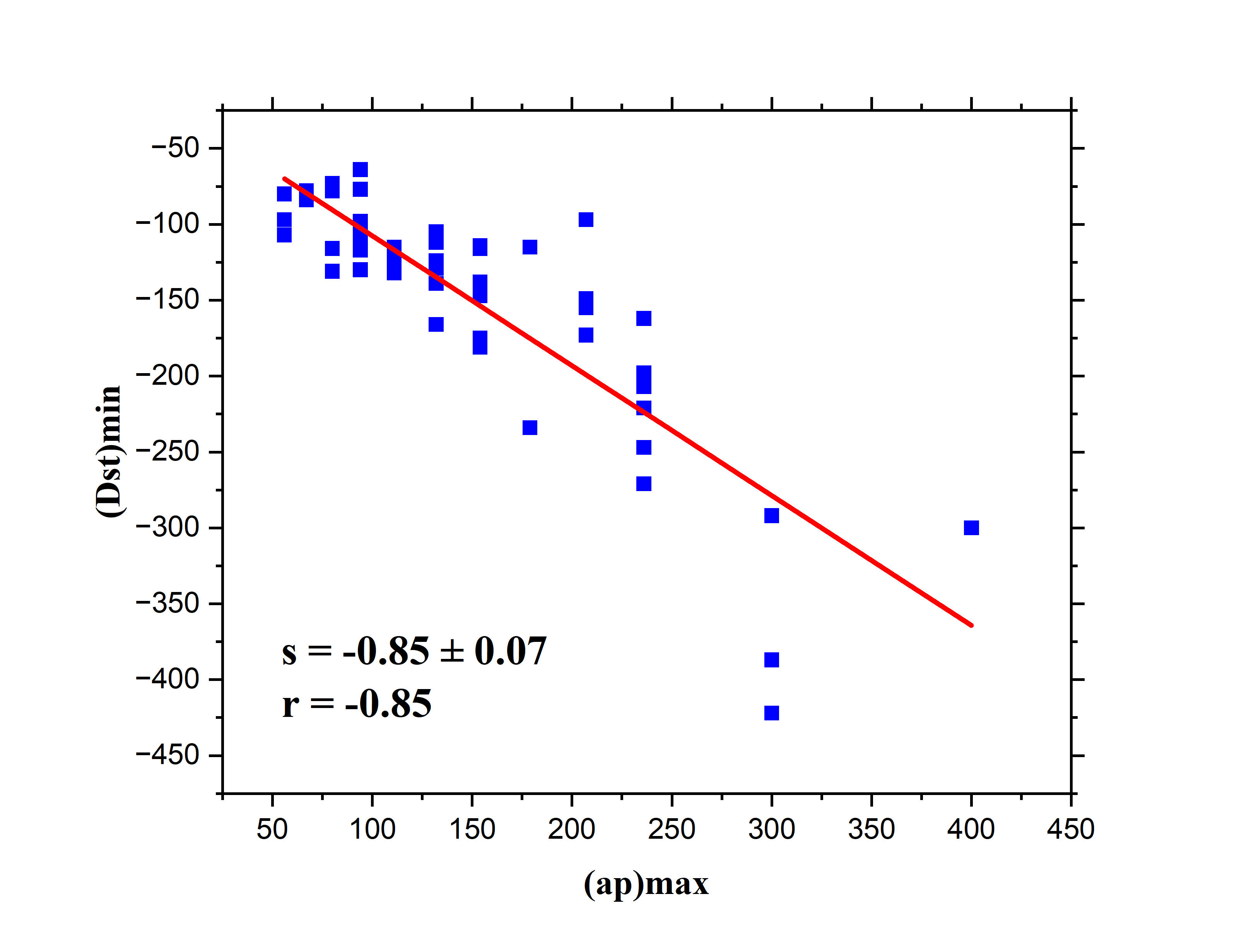}	
		\includegraphics[height=4.2cm, width=5.9cm]{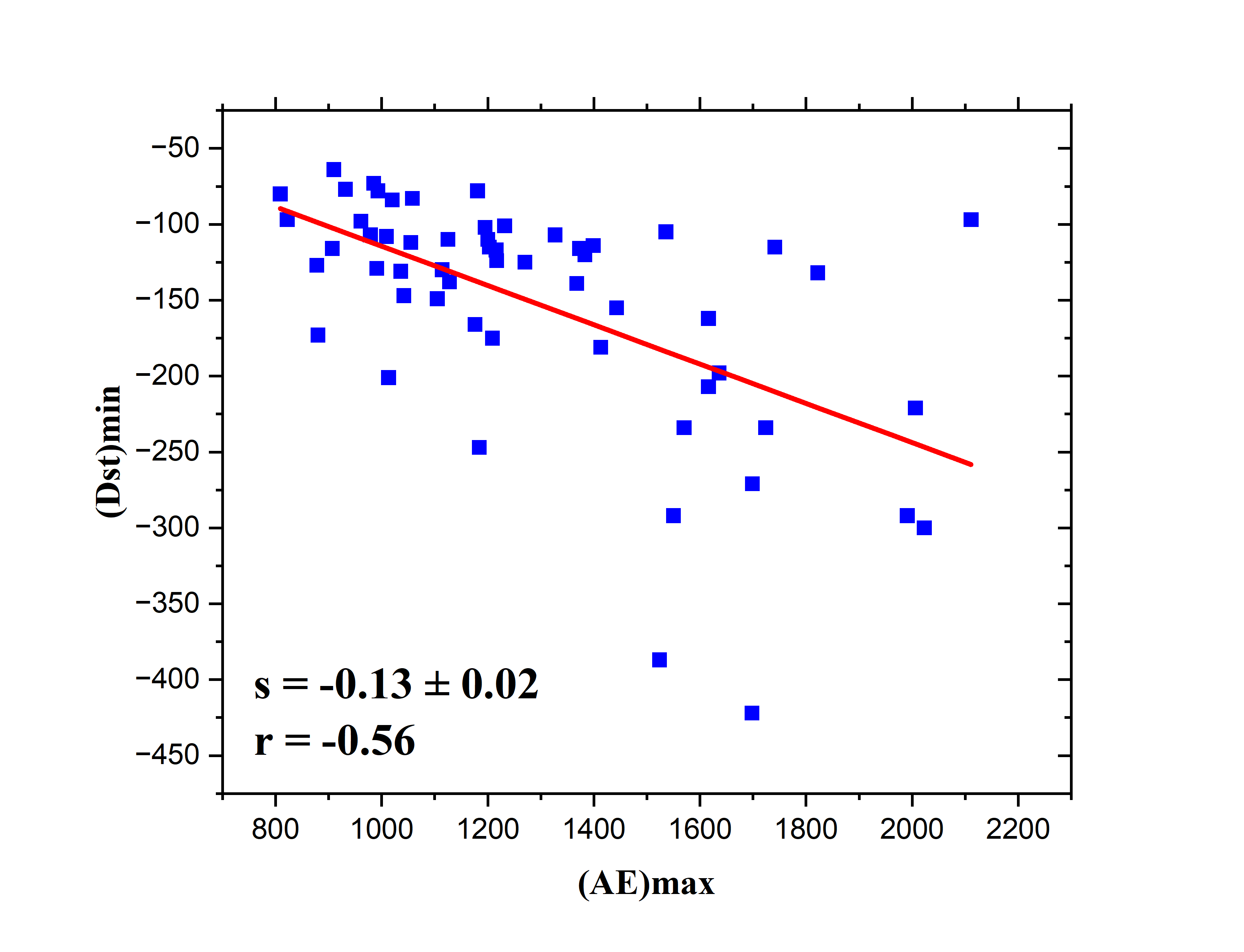}
		\includegraphics[height=4.2cm, width=5.9cm]{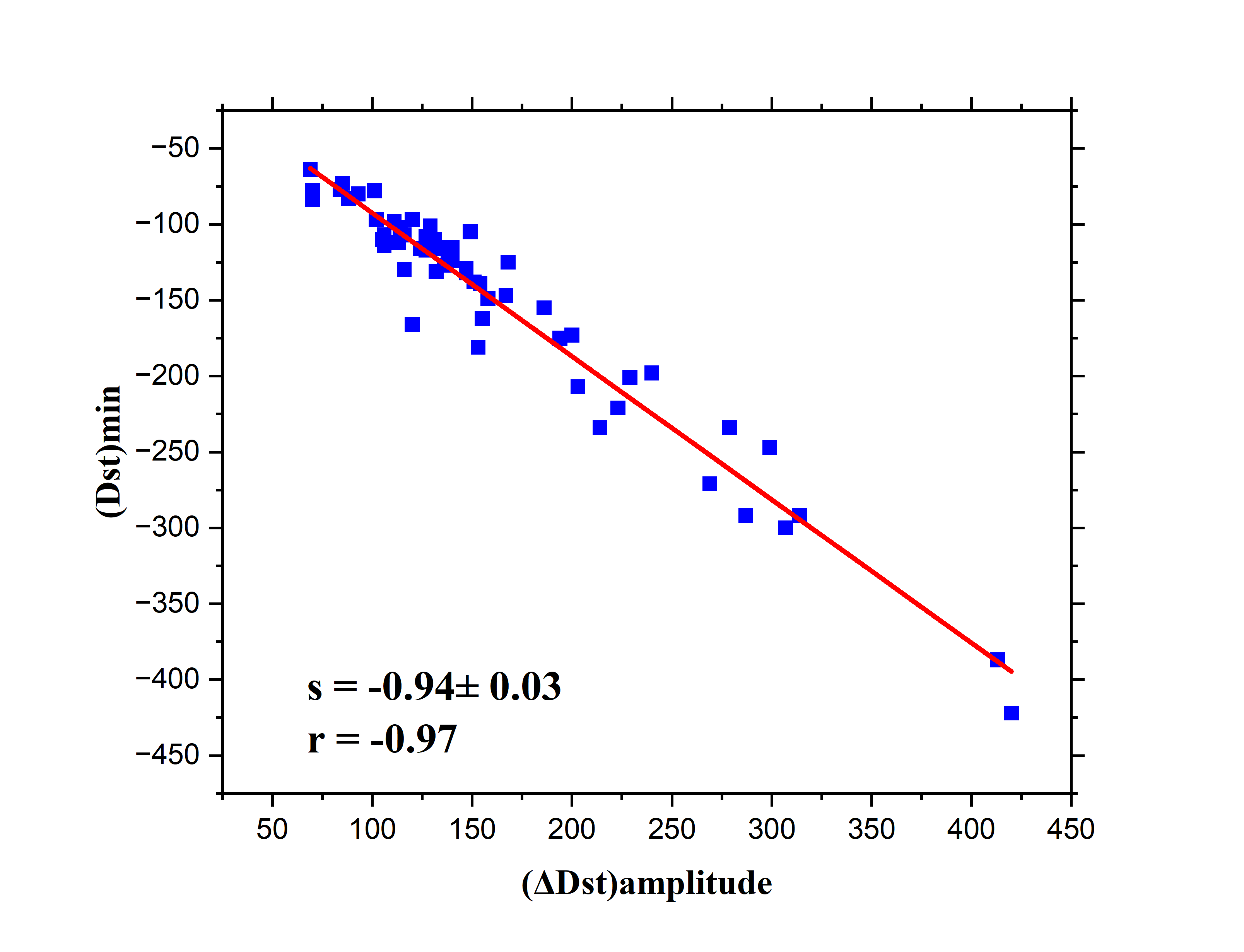}	
 		\\[\smallskipamount]	
 		\includegraphics[height=4.2cm, width=5.9cm]{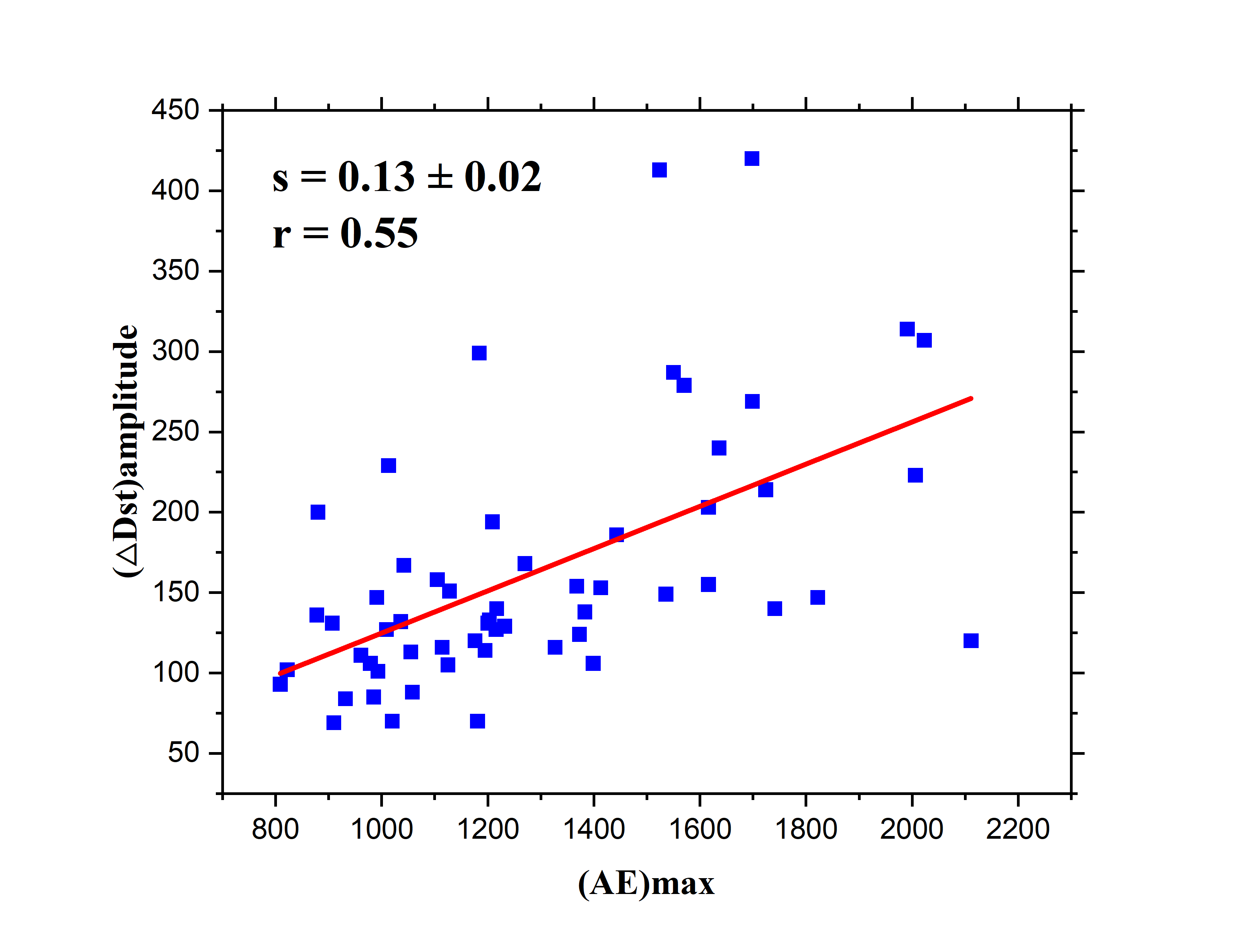}	
 		\includegraphics[height=4.2cm, width=5.9cm]{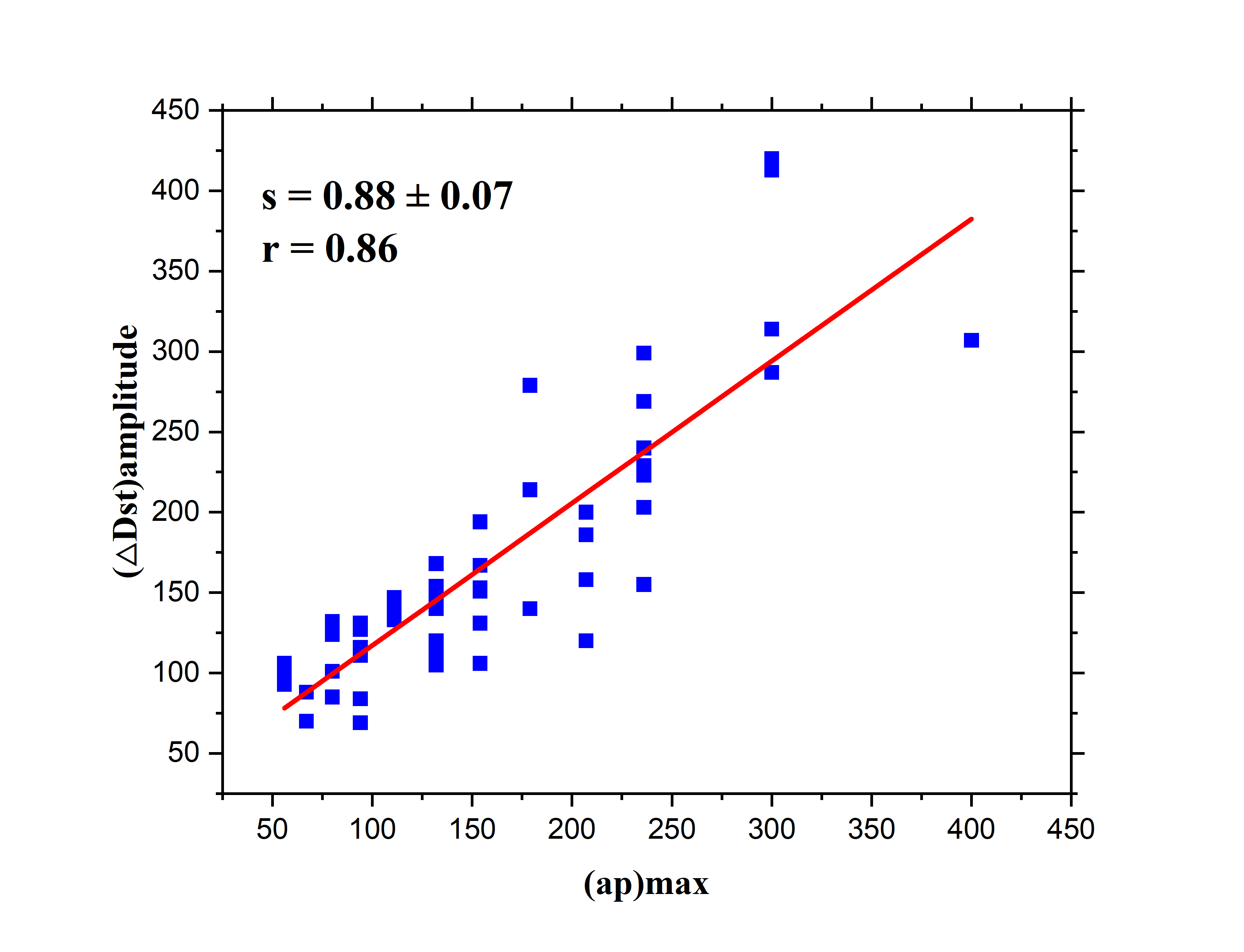}
		\includegraphics[height=4.2cm, width=5.9cm]{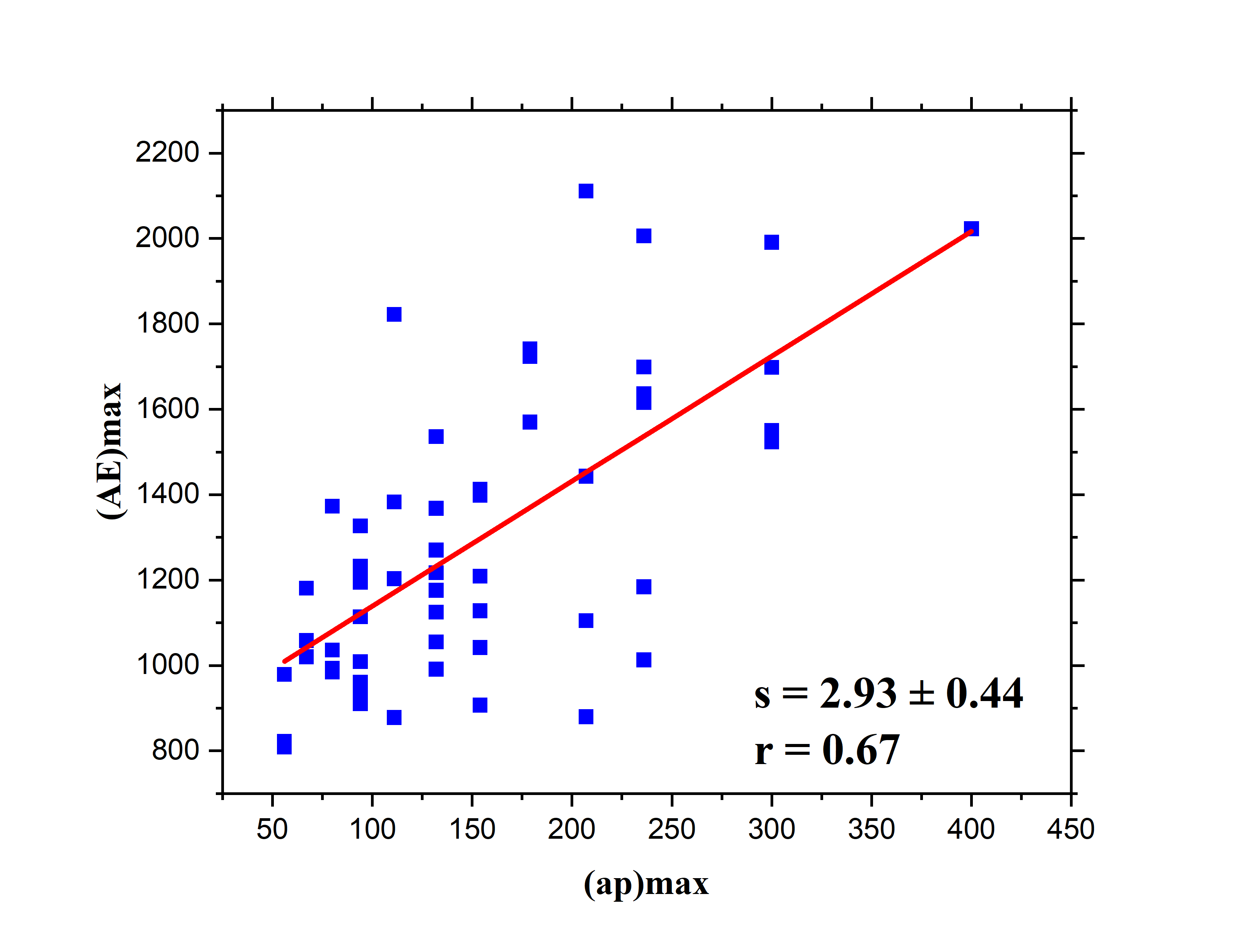} 		
		\caption{\small The scatter plots depict the best linear fit between the geomagnetic indices' peak values. The relationship between Dst with $\Delta$Dst, ap and AE is shown (upper), and the relationship between $\Delta$Dst with ap and AE, as well as,the relationship between AE and ap is shown (bottom).}\label{fig:gscorr}
	\end{figure*}

We use three geomagnetic indices: Dst, ap, and AE measurements.
The relationship between the geomagnetic indices is illustrated in Figure \ref{fig:gscorr}, which depicts scatter plots representing the linear relationship between them.\\	
During large geomagnetic disturbance, we have correlated the peak of Dst (Dst$_\mathrm{min}$) with AE (AE$_\mathrm{max}$) and ap (ap$_\mathrm{max}$) (upper left and upper middle) respectively, while (upper right) depicts the relationship between Dst$_\mathrm{min}$ and Dst amplitude ($\Delta$Dst) which are strongly related with the Pearson's correlation coefficient $-0.97$. During the strong geomagnetic disturbance Dst is strongly related with ap with its best fit equation Dst$_\mathrm{min}=(-0.85\pm0.07)$(ap$_\mathrm{max})+(-22.03\pm11.84)$, with Pearson's correlation coefficient $r =-0.85$.
However, Dst has a relatively weak correlation to AE (with a Pearson's correlation coefficient of -0.56) among the three geomagnetic indices during the period of maximum geomagnetic disturbances (i.e peak of GS). Our findings support those of \citep{joshi2011relationship}, who found that, Dst and AE have a weak linear relationship (r$ =-0.55 $) when compared to the pair of other indices. \cite{2017E3SWC..2001010S} reported that Dst has a poor relationship with AE for intense GSs which is consistent with our result. The reason for this could be due to two assumptions. First, during the peak of intense GSs, the polar auroral oval extends to mid-latitude regions of the Earth, and mid-latitude magnetometers record a reasonable intensity of disturbance, which is strongly related to Dst, whereas polar magnetometers record weak intensity.
The second assumption is that there is a large latitudinal difference between the equator and the polar regions, which cannot share storm expansion. The maximum extension of Dst during intense GS would be up to the mid-latitude regions around 30 to 40 degrees latitude, which is the reason why ap is strongly related to Dst \citep[also see,][]{2023AdSpR..71.1137B}.\\
We have correlated the amplitude of Dst ($\Delta$Dst) with AE and ap (bottom of left and middle) respectively of Figure \ref{fig:gscorr}. The best fit equation for the amplitude of Dst is $\Delta$Dst$_\mathrm{amplitude}=(0.88\pm0.07)$(ap$_\mathrm{max})+(28.65\pm12.09)$, and the Pearson correlation coefficient for this relationship is $0.86$.
The duration of the \m\ was taken as the time interval between the onset of the Dst index Dst$_0$ and the minimum value of Dst, (Dst$_\mathrm{min}$). We used the formula $\arrowvert\Delta\mathrm{Dst}\arrowvert = \arrowvert \mathrm{Dst}_\mathrm{min}-\mathrm{Dst}_0\arrowvert$ to calculate $\arrowvert\Delta \mathrm{Dst}\arrowvert$ (see column 5 from Table \ref{tabl1}), which is recorded for each event. When we take the peak of Dst, we are referring to Dst value at the time when GS is at its maximum, whereas $\Delta$Dst is the magnitude of the difference between Dst at the time of peak and onset Dst over the GS \m\ duration.\\
Figure \ref{fig:gscorr} (bottom right) depicts another interesting relationship between mid-latitude and high-latitude geomagnetic disturbance during GS.
With a Pearson's correlation coefficient of $0.61$, AE has a good relationship with ap but not as good as between Dst and ap \citep[see,][]{fares1997relationships, adebesin2016investigation}. In general, the mid-latitude geomagnetic index ap has a better relationship with Dst than AE. \cite{2018AdSpR..61..348B} has observed that at least for ICME events, the Kp index compared to AE, correlate somewhat better with the magnitude of the minimum value of the Dst index. Despite sharing a boundary with both polar and equatorial GSs, ap has a strong relationship with Dst during the peak of intense GSs. Our finding suggests that during the \m\ of intense storms Dst expand to mid-latitude region and polar storm index AE expand to the mid-latitude region as well but cannot reach to the equator \citep[see][]{2023AdSpR..71.1137B}. 
 \cite{2016STP.....2d..11B} reported that the strength of the auroral currents and their movement to lower latitudes during magnetic storms both affect AE's value.\\
Thus, the mid-latitude geomagnetic index ap is correlated with both polar and equatorial storm indices during MPs of the storm, but in a better way with equatorial Dst at least during intense GS.

	\subsection{SSC of Dst and IMF Bz}\label{SSC}
	
\begin{figure*}
		\centering
		\includegraphics[height=3.9cm, width=5.4cm]{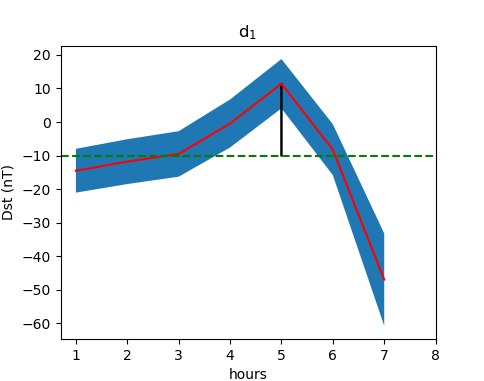}	
		\includegraphics[height=3.9cm, width=5.4cm]{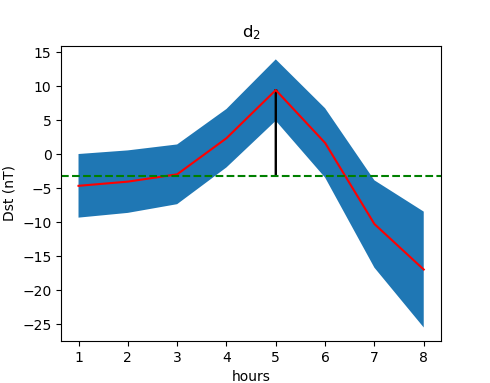}
		\includegraphics[height=3.9cm, width=5.4cm]{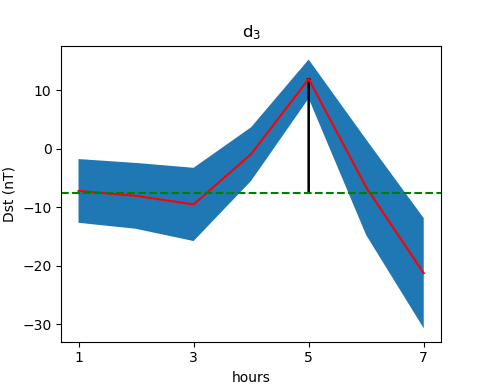}
		\\[\smallskipamount]
		\includegraphics[height=3.9cm, width=5.4cm]{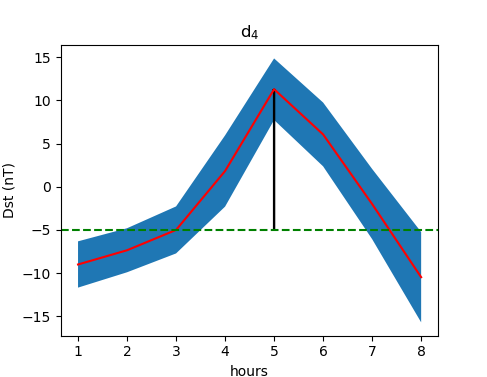}	
		\includegraphics[height=3.9cm, width=5.4cm]{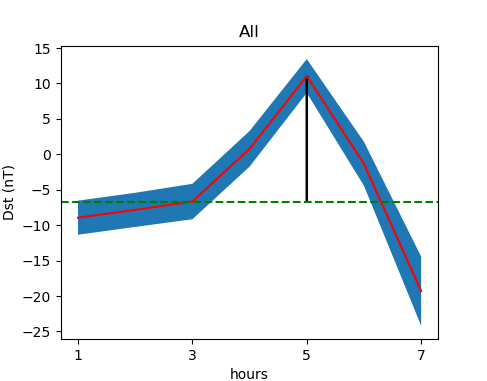}
		\caption{\small The figure reveals Dst SSC profiles for the four groups: one step decrease of group d$_1$ (top left), two step decrease of group d$_2$ (top middle), three step decrease of group d$_3$ (top right), multiple step decrease of group d$_4$ (bottom left) and group of all events at the (bottom right). Zoom of the sudden enhancement from the superposed analyses for the GS before starting the \m\ development. The red line shows the GS data and the cyan shaded areas are the calculated errors as explained in Section \ref{sec:res}. The vertical black thick line represent the amplitude of SSC, and from the horizontal dashed green line to the peak. The horizontal green dashed line represents the SSC time span in hours.}
		\label{pics:sscd}
	\end{figure*}

\begin{figure*}
		\centering
		\includegraphics[height=3.9cm, width=5.4cm]{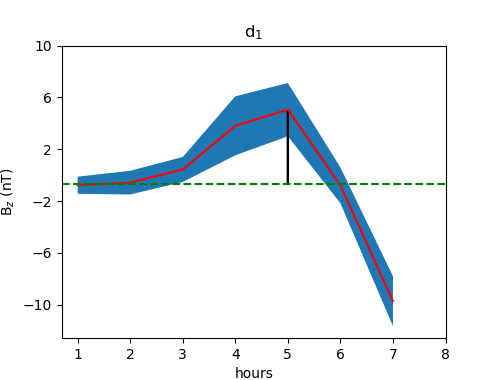}	
		\includegraphics[height=3.9cm, width=5.4cm]{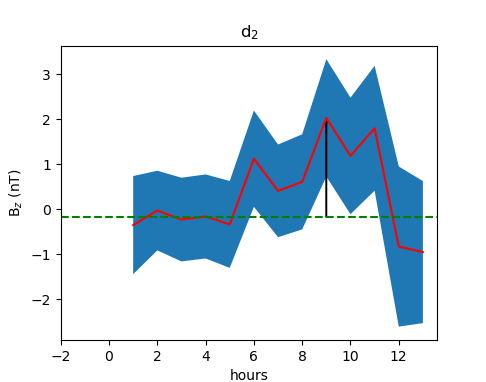}
		\includegraphics[height=3.9cm, width=5.4cm]{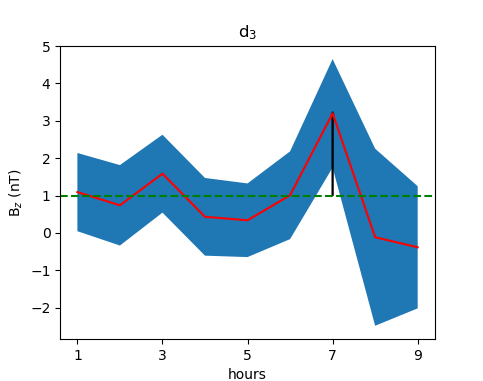}
		\\[\smallskipamount]
		\includegraphics[height=3.9cm, width=5.4cm]{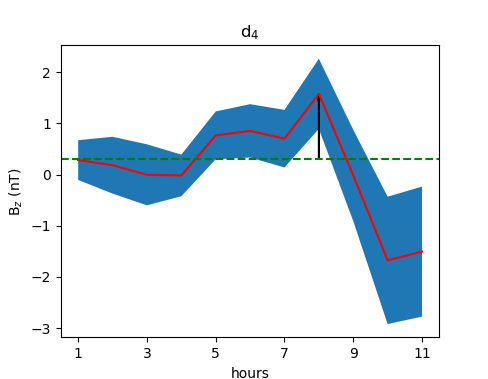}	
		\includegraphics[height=3.9cm, width=5.4cm]{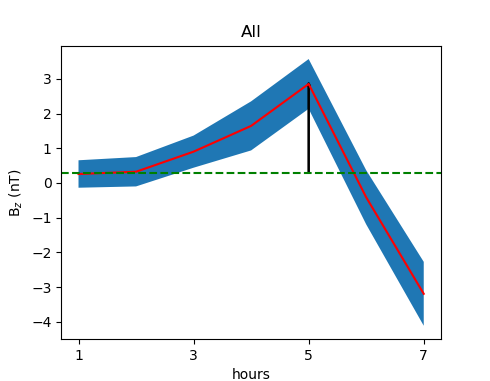}
		\caption{The figures follow similar fashion as Figure \ref{pics:sscd} and represent Bz \textbf{\textit{profiles during}} SSC for the four groups and the entire data points (the bottom right).}
		\label{pics:sscb}
	\end{figure*}

The majority of moderate and intense storms begin abruptly due to strong IP shocks that compress the magnetosphere \citep{Russell1992TheEO}. Fast CMEs commonly serve as the primary driver of SSCs \citep[e.g.,][]{Taylor1994ASE} whereas CIR-driven GSs do not typically have SSCs \citep{Venkatesan1991OnTH}.
Quick rises in the surface geomagnetic field's northward component are associated with SSCs, as are abrupt changes in the dynamic pressure of the \sw. The magnetopause compresses as the \sw\ approaches the magnetosphere's bow, which causes a quick rise in the magnetic field on the Earth's day-side which lasts only a few minutes \citep[e.g.,][]{oyedokun2018geomagnetically}. The SSC onset generally coincides with a sharp increase of \sw\ pressure and velocity as seen in Figures (\ref{d1} - \ref{d5}). \sw\ pressure increases are typically linked to IP shocks \citep{1980P&SS...28..381A}.\\	
Figures \ref{pics:sscd} and \ref{pics:sscb} reveals rapid increases called SSC of geomagnetic index Dst and the IMF Bz ahead of the \m\ of disturbance respectively. We considered at least two hours before the start time of the SSC to gather detailed information. In Figure \ref{pics:sscd}, the SSC for GS is represented by a horizontal green dashed line that takes 3--4 hours. The vertical black thick line represents the amplitude of SSC from its normal position to its peak. We plotted a zoom of the region from Figures (\ref{d1}--\ref{d5}) average SEA profile a few hours before GS development to gain a more in-depth understanding of our interest in SSC regions. The first group, one-step decrease group d$_1$ (top left), takes 3 hours to return to its pre-level position and has a peak amplitude of 21.4 nT. The second group, two-step decrease group d$_2$ (top middle), takes more than 3 hours to return to its pre-level position and has a peak amplitude of 12.5 nT. The third group, three-step decrease group d$_3$ (top right), also takes more than 3 hours to return to its pre-level position and has a peak amplitude of 19.5 nT. The fourth group, multi step decrease group d$_4$ (bottom left), takes more than 4 hours to return to its pre-level position and has a peak amplitude of 16.2 nT. The entire set of events (bottom right) takes more than 3 hours to return to its pre-level position and has an amplitude of 17.7 nT at its peak.\\	
The peak value of Bz associated with the shock was found to be the most effective predictive parameter for SSCs \citep{2020AGUFMSM015..02S}. Figure \ref{pics:sscb} depicts the enhanced IMF Bz in the same manner as Figure \ref{pics:sscd}, which takes 2--6 hours. The first group, one-step decrease group d$_1$ (top left), Bz returns to its pre-level position after 4 hours and has a peak amplitude of 5.6 nT. The second group,  two-step decrease group d$_2$ (top middle), Bz returns to its pre-level position after more than 6 hours and has a peak amplitude of 2.1 nT. The third group, three-step decrease group d$_3$ (top right), Bz has a peak amplitude of 2.2 nT and takes more than 2 hours to return to its pre-level position. The fourth group, multi-step decrease group d$_4$ (bottom left), Bz has a peak amplitude of 1.2 nT and takes more than 2 hours to return to its pre-level position. The average value for the total set of 57 events (bottom right) takes over 3 hours to return to its pre-level position and has a peak amplitude of 2.5 nT.\\	
As shown in Table \ref{tabl3}, the peak of the geomagnetic index (Dst$_\mathrm{peak}$) appears to show a direct relationship with the amplitude of SSC and an inverse relationship with the duration of SSC ($\tau$). We may conclude that IP structure generating high amplitude SSC produces more geoeffective storms, whereas the one that generating long lived SSC produces less geoeffective storms. Storms of group d$_1$ are more geoeffective in general, with a Dst$_\mathrm{peak} = -155$ nT, and the accompanying down-dusk electric field Ey peaks at 8.9 mVm$^{-1}$, consistent with earlier findings by \citep[e.g.,][]{2022AdSpR..70.2830E} $\mathrm{Ey}>5$ mVm$^{-1}$ for one step severe storms.	
The third group d$_3$ is the second higher geoeffective with a Dst$_\mathrm{peak} = -124$ nT, group d$_4$ is the third with a Dst$_\mathrm{peak} = -110$ nT, and group d$_2$ is somewhat less with a Dst$_\mathrm{peak} = -102$ nT.
As represented in Table \ref{tab8} the peak values of all parameters alongside with the corresponding groups	are shown. \cite{2017AdSpR..59.1425S} reported that SSC rising time is largely determined by IP shock speed and \sw\ dynamic pressure. 
The time interval between SSC and the start of the MP for both Dst and Bz ranges from $\sim$2 -- 6 hours as shown in Table \ref{tabl3} is almost similar to the previous study by \citep{pandey2009characteristic} who reported 1 -- 6 hours.

	\begin{table*}
	\centering
			\footnotesize
		\caption{\footnotesize Intensity (amplitude) of SSC for geomagnetic index Dst and southward IMF Bz as a result of SEA. The duration of SSC ($\tau$) represented by horizontal dashed lines, while amplitude of SSC is represented by the vertical solid black lines. The peak values of Dst and Bz during GS recorded for comparison purposes.\label{tabl3}}
		\centering
		\begin{tabular}{l cccccc}
			\hline
			Groups&\multicolumn{3}{c}{\underline{\textbf{Dst (nT)}}} &  \multicolumn{3}{c}{\underline{\textbf{Bz (nT)}}} \\
			(events) &$\tau$ (hour)&\textit{\textbf{SSC}} amplitude (nT)& \textit{\textbf{GS strength}} Dst$_\mathrm{peak}$ & $\tau$ (hour)& \textit{\textbf{during SSC}} amplitude (nT)&Bz$_\mathrm{peak}$\textit{(\textbf{during MP})} \\
			\hline
			all&3.3 & 17.7&-94 &3.8&2.5&5.75\\
			d$_4$&4.3 & 16.2& -110&2.2&1.2&10.02\\
			d$_3$&3.2 & 19.5& -124&1.8&2.2&13.26\\
			d$_2$&3.5 & 12.5& -102&6.3&2.1&8.76\\
			d$_1$&3.1 & 21.4& -155&4.0&5.6&14.16\\
			\hline
		\end{tabular}
	\end{table*}
	
	\begin{table*}
		\small 	 
		\caption{\small The average peak values of all four groups of GSs. \label{tab8}}
		\centering		
		\begin{tabular}{l cccc}
			\hline
			\textbf{Parameters}  &\multicolumn{4}{c}{\underline{\textbf{Peak values}}}\\
			&\textbf{d$_1$}&\textbf{d$_2$}&\textbf{d$_3$}&\textbf{d$_4$}\\
			\hline
			B &25.8 &15.1 &21.8 &16.8\\ 			
			$\sigma$B$_{\mathrm{m}}$ &3 &1.9 &3.3 & 1.8\\ 		
			$\sigma$B$_{\mathrm{v}}$ &10.8 & 6.8& 9.9& 6.0\\  		
			$\rho$ &20.7 &18.8 &23.2 &13.8 \\ 
			v &574 &487 &548 &526 \\
			P &12.8 &7.9 & 12.6& 8.4\\  		
			Ey &8.9 &4.2 &6.8 & 4.8\\ 		
			$\beta$ &0.28 &0.4 &0.4 & 0.4 \\ 		
			ap &94 &87 &123 & 87 \\
			AE &860 &711 &822 & 717 \\ 		 		  		
vB &$16.34\times 10^{3}$ &$7.14\times 10^{3}$ &$12.68\times 10^{3}$ &$8.76\times 10^{3}$\\ 		
vEy &$7.01\times 10^{3}$ &$2.09\times 10^{3}$ &$5.99\times 10^{3}$ &$2.76\times 10^{3}$ \\ 		
vBz$^2$ &$25.81\times 10^{4}$ &$5.05\times 10^{4}$ &$22.98\times 10^{4}$ &$14.28\times 10^{4}$ \\
v$^{\frac{4}{3}}$Bz &-$7.81\times 10^{4}$ &-$3.32\times 10^{4}$ &-$6.47\times 10^{4}$ &-$4.01\times 10^{4}$\\  		
v$^{\frac{4}{3}}$Bz$\rho$ &-$11.22\times 10^{5}$ &-$5.10\times 10^{5}$ &-$6.39\times 10^{5}$ &-$4.65\times 10^{5}$ \\ 		
v$^{\frac{4}{3}}$Bz$\rho^{\frac{1}{2}}$ &-$2.80\times 10^{5}$ &-$1.25\times 10^{5}$ &-$1.91\times 10^{5}$ & -$1.23\times 10^{5}$ \\ 			
v$^{\frac{4}{3}}$B$\rho^{\frac{1}{2}}$ &$5.10\times 10^{5}$ & $2.19\times 10^{5}$&$4.52\times 10^{5}$ & $2.20\times 10^{5}$\\
			\hline
		\end{tabular}
	\end{table*}
	

	\section{Conclusion}\label{sec:conc}
	
In this study we have extracted geomagnetic indices, SW plasma and field parameters and their derived CFs using hourly averaged resolution data. Based on the criteria explained in Section \ref{sec:data} we have retrieved 57 GS events ($\mathrm{Dst}\leq-64$ nT) from 1995 to 2022. We have also utilized IMF parameters (magnitude B, Bz, Sigma in magnitude $\sigma$B$_{\mathrm{m}}$ and Sigma in IMF vector $\sigma$B$_{\mathrm{v}}$), SW plasma parameters (proton density $\rho$, and SW speed v) geomagnetic activity indices (Dst, ap, AE) and derivatives [P, Ey, $\beta$, CFs (vB, vBz$^2$, vEy, v$^{\frac{4}{3}}$Bz, v$^{\frac{4}{3}}$Bz$\rho$, v$^{\frac{4}{3}}$Bz$\rho^{\frac{1}{2}}$, v$^{\frac{4}{3}}$B$\rho^{\frac{1}{2}}$)]. \\
All the selected and retrieved geomagnetic indices, SW plasma and field parameters and their derived CFs have grouped into four groups based on the criteria stated on Section \ref{subsec:dec}. All the selected 19 SW plasma and field parameters/functions together with geomagnetic indices and various derivatives have subjected to SEA for each groups of events into three panels (see Figures from \ref{d1} - \ref{d5} for details in Section \ref{sec:res}) seven parameters analyzed in each panel. 
Almost all SW plasma and field parameters and derived functions begin a few hours before the onset of MP. In most cases, a high density turbulent IMF shock/sheath structure formed a few hours before Dst onset and passed through during Dst onset. GSs with the shorter MP duration, are in general more intense, while storms with the longer MP duration are less intense.  \\
In addition to SEA, we employed a linear correlation between the peak of geomagnetic activity indices and SW plasma and field parameters as well as their functions. To that end, we fitted the peak values of all parameters linearly. After fitting all parameters with the three geomagnetic indices, we arrived at the following conclusions.
Although individual SW field-plasma parameters that show a little difference with individual geomagnetic indices magnitude during MP of GS; it is Bz/B magnitude with Dst$_\mathrm{min}$, $\Delta$Dst and ap$_\mathrm{max}$ however, it is v that relates much better than B/Bz with AE$_\mathrm{max}$. \\
Overall, the most significant function influencing the MP of the GS for two geomagnetic indices Dst$_\mathrm{min}$ and ap$_\mathrm{max}$ is a triple parameter CF (v$^{\frac{4}{3}}$Bz$\rho^{\frac{1}{2}}$), which consists of a combination of viscous term ($\rho^{\frac{1}{2}}$) and terms related to the electric field/merging term (v$^{\frac{4}{3}}$Bz).\\
However, the function (v$^{\frac{4}{3}}$B$\rho^{\frac{1}{2}}$) represents slightly better the peak of AE$_\mathrm{max}$ during the storms compared to (v$^{\frac{4}{3}}$Bz$\rho^{\frac{1}{2}}$). 
In conclusion, we can invoke the magnitude of the function (v$^{\frac{4}{3}}$Bz$\rho^{\frac{1}{2}}$) to represent reasonably well the strength of GS (Dst$_\mathrm{min}$ and $\Delta$Dst) but also the intensification of geomagnetic activity in mid-latitude region (ap$_\mathrm{max}$) and polar region (AE$_\mathrm{max}$) during most intense period of GS activity (as observed by Dst$_\mathrm{min}$ in near-equatorial latitude). However, search for better coupling should continue in the future studies.\\
As discussed in Section \ref{sub:DstBz}, the time lag between the peak of geomagnetic index Dst and Bz was calculated. The majority of events (58\%) are delayed by one to four hours (with an average of one hour and thirty minutes). According to our findings, one and two step decrease GS events have a shorter time lag on average than three and multiple step GSs. \\
As discussed in Section \ref{SSC}, the peak of Dst has a direct relationship with the amplitude of SSC and an inverse relationship with the duration of SSC. We can draw the conclusion that short-lived, high-intensity transient SSC is associated with more severe GSs, whereas long-lived SSC is linked to less severe GSs. In contrast, the duration and intensity of Bz have no linear relationship with its peak value. \\
We also looked at how geomagnetic indices correlated with each other during the most intense period in GS.
The relationship between the three geomagnetic indices; Dst, ap, and AE measurements have been tested (discussed the detail in Section \ref{sub:correndex}). Dst is better correlated with mid-latitude geomagnetic index ap than the polar index AE. 
Almost all of the merging/electric field related CFs (including vB and Ey) that we utilized in our study have a stronger association with ap$_\mathrm{max}$ during strong GSs than with AE$_\mathrm{max}$. These findings support the \cite{2023AdSpR..71.1137B} result, who found that, during the MP of GSs, specifically those caused by Sheath/ICME events, there was a stronger association between the SW electric field and Kp than with AE. A substantial linear relationship exists between ap and Dst during the peak of intense GSs, despite the fact that these two GS types share a boundary.\\
In summary, we proposed a suitable function (a combination of viscous and electric field related terms) whose amplitude represents not only the magnitude of GS (Dst$_\mathrm{min}$), but also the intensity of geomagnetic disturbances observed in the mid-latitude (ap$_\mathrm{max}$) and polar (AE$_\mathrm{max}$) domains during the occurrence of strong geomagnetic disturbances. 
The time lag between the onset of GSs and the characteristics and functions of the SW has been determined. The estimated time lag between the onset of GSs and the suggested functions (and their component parameters) can be used to estimate the predicted onset/time of maximum geomagnetic disturbance in different latitude zones using SW observations.
The best-fit inter-relationship between geomagnetic indices' peak values was found. These interrelationships between peak geomagnetic index values in three separate domains can be used to estimate disturbance levels in other latitude zones based on observations in a certain latitude zone.


\section*{Acknowledgements}
This study made use of data from the GSFC/SPDF OMNIWeb interface, which can be
found at https://omniweb.gsfc.nasa.gov.
Reviewers'  suggestions were helpful in improving the scientific level of the paper.

\bibliographystyle{jasr-model5-names}
\biboptions{authoryear}
\bibliography{bib}

\end{document}